\documentclass[aps,amsmath,amssymb,twocolumn,10pt]{revtex4-1}
\usepackage[T1]{fontenc}
\usepackage{graphicx}
\usepackage{hyperref} 
\hypersetup{colorlinks=true,
            breaklinks=true,
            bookmarks=true,
            urlcolor=magenta,
            linkcolor=blue,
            citecolor=red, 
            frenchlinks=true} 
\usepackage{charter}  
  
\begin{document}   
   
\title{One-dimensional mixtures  of several ultracold atoms: a review}
\author{Tomasz Sowi\'nski$^{1}$ and Miguel \'Angel Garc\'ia-March$^{2}$} 
\affiliation{  
\mbox{$^1$ Institute of Physics, Polish Academy of Sciences Aleja Lotnik\'ow 32/46, PL-02668 Warsaw, Poland}\\
\mbox{$^2$ ICFO -- Institut de Ciencies Fotoniques,  Av. Carl Friedrich Gauss 3, 08860 Castelldefels (Barcelona), Spain}}
\date{\today} 

\begin{abstract} 
Recent theoretical and experimental progress on studying one-dimensional systems of bosonic, fermionic, and Bose-Fermi mixtures of a few ultracold atoms confined in traps is reviewed in the broad context of mesoscopic quantum physics. We pay special attention to limiting cases of very strong or very weak interactions and transitions between them. For bosonic mixtures, we describe the developments in systems of three and four atoms as well as different extensions to larger numbers of particles. We also briefly review progress in the case of spinor Bose gases of a few atoms.  For fermionic mixtures, we discuss a special role of spin and present a detailed discussion of the two- and three-atom cases. We discuss the advantages and disadvantages of different computation methods applied to systems with intermediate interactions. In the case of very strong repulsion, close to the infinite limit, we discuss approaches based on effective spin chain descriptions. We also report on recent studies on higher-spin mixtures and inter-component attractive forces. For both statistics, we pay particular attention to impurity problems and mass imbalance cases. Finally, we describe the recent advances on trapped Bose-Fermi mixtures, which allow for a theoretical combination of previous concepts, well illustrating the importance of quantum statistics and inter-particle interactions. Lastly, we report on fundamental questions related to the subject which we believe will inspire further theoretical developments and experimental verification.
\end{abstract} 
\maketitle   
\tableofcontents 
    
\section{Introduction}

\subsection{Few-body physics of ultracold atoms}

Quantum engineering is a rapidly developing field of modern physics. Its successes in the last three decades originate in the deep progress of the experimental control of matter on subatomic scales interacting with the electromagnetic field. Currently, quantum engineering is typically identified with a broadly defined field of photonics (quantum informatics, interferometry, nonclassical correlations between photons) and with physics of ultracold atoms \cite{PethickBook}. Typically in the second case, the main objective is to study the macroscopic behavior of many quantum particles in optical lattices -- periodic potentials formed by standing waves of spatially arranged laser beams. This path is inspired by the idea of quantum simulators for condensed matter systems, {\it i.e.}, preparing realistic and fully controllable quantum systems described by simple toy models of condensed matter physics, like the Hubbard model or spin-chain models, {\it etc.} \cite{LewensteinBook}.

Importantly, in parallel to this very fashionable direction of lattice models, an equally fascinating path of theoretical and experimental exploration is present in the field -- {\it the physics of few-body ultracold systems}. Systems of a few quantum particles form a natural link between one-, two-body physics and the many-body physics which has spectacular consequences of collective properties originating in inter-particle interactions and quantum statistics \cite{2010BlumePhysics,Blume2012Rev}. Therefore their quantum simulation is a fundamental and very interdisciplinary milestone for building our understanding of physical quantum systems. Up to a few years ago, engineering of such systems, {\it i.e.}, their coherent control and manipulation, was not experimentally possible. However, due to recent progress in the field of ultracold gases, it became feasible to prepare on demand few-particle interacting systems of a well-defined number of particles. In this way, a completely new era in experimental studies of mesoscopic quantum systems started, {\it i.e.}, systems too large to be reduced to simple two- and three-body problems and too small to be described with the whole sophisticated machinery of the quantum statistical mechanics. In this review we want to focus on the most intriguing subset of {\it one-dimensional systems} having many unique properties forced by strongly reduced dimensionality.

Obviously, it is a very demanding task to experimentally achieve ultracold one-dimensional few-body systems. It can be done only if one can control atomic systems on different levels with tremendous accuracy. The crucial experimental landmark is a set of experiments in which strongly interacting ultracold bosons forming the Tonks-Girardeau gas was obtained~\cite{2004KinoshitaScience,paredes2004tonks}. Then a very striking experiment reported in \cite{2005KinoshitaPRL} showed how the famous fermionization of bosons occurs in a strongly interacting system. In general, trapping a few bosons in the ultracold regime shows a larger difficulty than for fermions, due to losses associated with three-body recombination. To overcome this difficulty, perfect control of interactions is needed. Fortunately, it is facilitated when bosons are loaded to appropriately prepared optical lattice. For example, in \cite{2008CheinetPRL,2010WillNature} it was shown that with appropriate manipulation of optical double-well confinement it is possible to fill one of its sites with a successive number of bosons. At the same time, it was shown that with an appropriate reshaping of a microscopic optical trap it is possible to load exactly two atoms with a very high efficiency \cite{2010HeOE}. On the other hand, dipole traps can be loaded via evaporating cooling with tens of bosons~\cite{2013BourgainPRA}. In the case of fermions, the first experimental preparation of a one-dimensional two-component mixture of $^{40}$K atoms was reported in \cite{2005MoritzPRL} where the creation of two-particle bound states was examined. Later, in \cite{2010LiaoNATURE} preparation of a one-dimensional imbalanced system of $^6$Li atoms was announced. A completely different concept of preparation of one-dimensional few-fermion systems was presented in a groundbreaking series of experiments performed in the J. Selim's group in Heilderberg \cite{2011SerwaneScience,2013WenzScience,2015MurmannPRL,2015MurmannPRLb}. In these experiments (by imposing a very deep one-dimensional trap to previously confined three-dimensional system) it was proven that quasi-one-dimensional systems of a small, well-defined number of particles can be deterministically prepared, controlled, and measured  \cite{2011SerwaneScience}. During a whole experiment, the strength of inter-particle interactions together with the shape of the external potential can be changed almost adiabatically or instantaneously without losing coherence in the system. Moreover, by adding particles to the system one by one, it was shown how the Fermi sea of interacting particles is built in the  system \cite{2013WenzScience}. Then the few-fermion system in the limit of very strong interactions was examined  experimentally \cite{2015MurmannPRL} proving that in fact the system can be effectively described with the spin-chain Hamiltonian in this limit. The setup is so flexible that even the multi-well confinements can be seriously considered \cite{2015MurmannPRLb}. All these experiments pointed evidently that the unexplored field of few-body problems can now be studied and examined with high-precision experiments. In consequence, many theoretical aspects of corresponding problems and completely new questions have been addressed (for example: the impurity immersed in the Fermi sea problem \cite{1964McGuireJMP,2013WenzScience}, the 1D Cooper pairing problem \cite{2012ZurnPRL,2013ZurnPRL}, self-formation of fermionic chains problem \cite{2015MurmannPRL}, the correlated tunneling to the open space problem \cite{2005ChuuPRL,2012RontaniPRL,2012LodePNAS}, {\it etc.}). 

\subsection{Review plan}

Our review should be considered as a specific continuation of previous attempts for obtaining a comprehensive view of the problems of a mesoscopic number of interacting particles. Therefore at this moment, we want to recall other recent reviews which can be very helpful to the reader. First, there are a few comprehensive descriptions of the many-body ultracold system \cite{2008BlochRevModPhys,2010ChinRevModPhys} which give an appropriate background for a better understanding of the most important results in the field. Since our review is devoted to one-dimensional systems, we should definitely point out here two reviews devoted to these kinds of many-body systems, {\it i.e.}, \cite{2011CazalillaRMP} and \cite{2013GuanRevModPhys} for bosons and fermions, respectively. From the other side, few-body limit of ultracold systems is adequately described in  \cite{Blume2012Rev}. In this work, however, the discussion is oriented mostly on two- and three-dimensional confinements. Having in mind all these comprehensive presentations, our aim is to fill a gap between them and focus on a detailed description of ultracold mixtures of several atoms confined in one-dimensional traps. Although this subject was already partially covered by the mini-review \cite{2016ZinnerRev}, here we would like to give an extensive discussion of different issues related to the subject.

Before we start our story, we would like to mention some topics strictly related to one-dimensional few-body systems which we intentionally do not discuss or discuss only briefly. First, we do not discuss any results related to the whole branch of few-body problems connected with the Efimov physics. An interested reader can find a comprehensive description of these problems in \cite{2006BraatenPR} and \cite{2017NaidonRPP}. Second, in this review, we mainly focus on few-body systems confined in a single parabolic trap. Therefore, the discussion on multi-well and/or periodic confinements is only mentioned when it is essential for keeping the context. Finally, we limit ourselves to the static problems and we are mostly oriented to the ground-state properties. Therefore, we do not elaborate on the dynamical problems related to different initial states being out-of-equilibrium, different quench scenarios, or periodic modulations of the system's parameters. These paths of explorations, although very important, interesting, and appropriately justified in the case of large number of particles (see \cite{2011PolkovnikovRevModPhys,2015EisertNatPhys} for review) just started to gain interest recently in the case of a few-particle problems. Therefore, we believe that it is too early to include these considerations in our review. However, we mention appropriate works, whenever dynamical properties of the system are crucial to giving a route for a better understanding of statical properties of interacting few-body systems.

Keeping all the above constraints our review has the following structure. In Sec.~\ref{Sec:Bosons} we introduce the Hamiltonian for a two-component mixture of bosonic particles and we identify eight different interesting limits of repulsive interactions, which we discuss in the Section. We devote a subsection to succinctly review developments in the case of attractive interactions, and focus in the rest of the section in the most studied case of repulsive interactions. To get a better understanding of few-bosons systems, we start with a brief discussion of the seminal Girardeau observation that infinitely repulsive bosons may be directly mapped to the system of non-interacting fermions. Then we discuss with all details the problem of three and four bosons and show how these studies can be extended to the problem of a larger number of particles. We also discuss the developments in the study of few-atom spinor bose mixtures. In Sec.~\ref{Sec:Fermions} similar discussion is provided for fermionic mixtures. Here, however, we strongly focus on the role of particles' spin and correlations forced by the Pauli exclusion principle for identical fermions. We discuss a very fresh idea of the spin-chain description of the system being close to infinite repulsions and we briefly overview different methods for intermediate interactions. Inspired by recent experiments, we also report the progress in our understanding of attractively interacting particles and different mass mixtures. In Sec.~\ref{Sec:BoseFermi} we merge both previous attempts and discuss properties of Bose-Fermi mixtures. In Sec.~\ref{Sec:OtherExt} we briefly discuss different possible extensions of problems discussed in previous sections. We focus on those we believe can give rise for further exploration and may bring many interesting results. Finally, in Sec.~\ref{Sec:Final}, we summarize the review and address some relevant and open questions which in our opinion may bring a fundamental breakthrough in our understanding of one-dimensional few-body systems and their links to the many-body world.

\subsection{Two particles in a harmonic trap}
\begin{figure}[t]
\includegraphics[width=\linewidth]{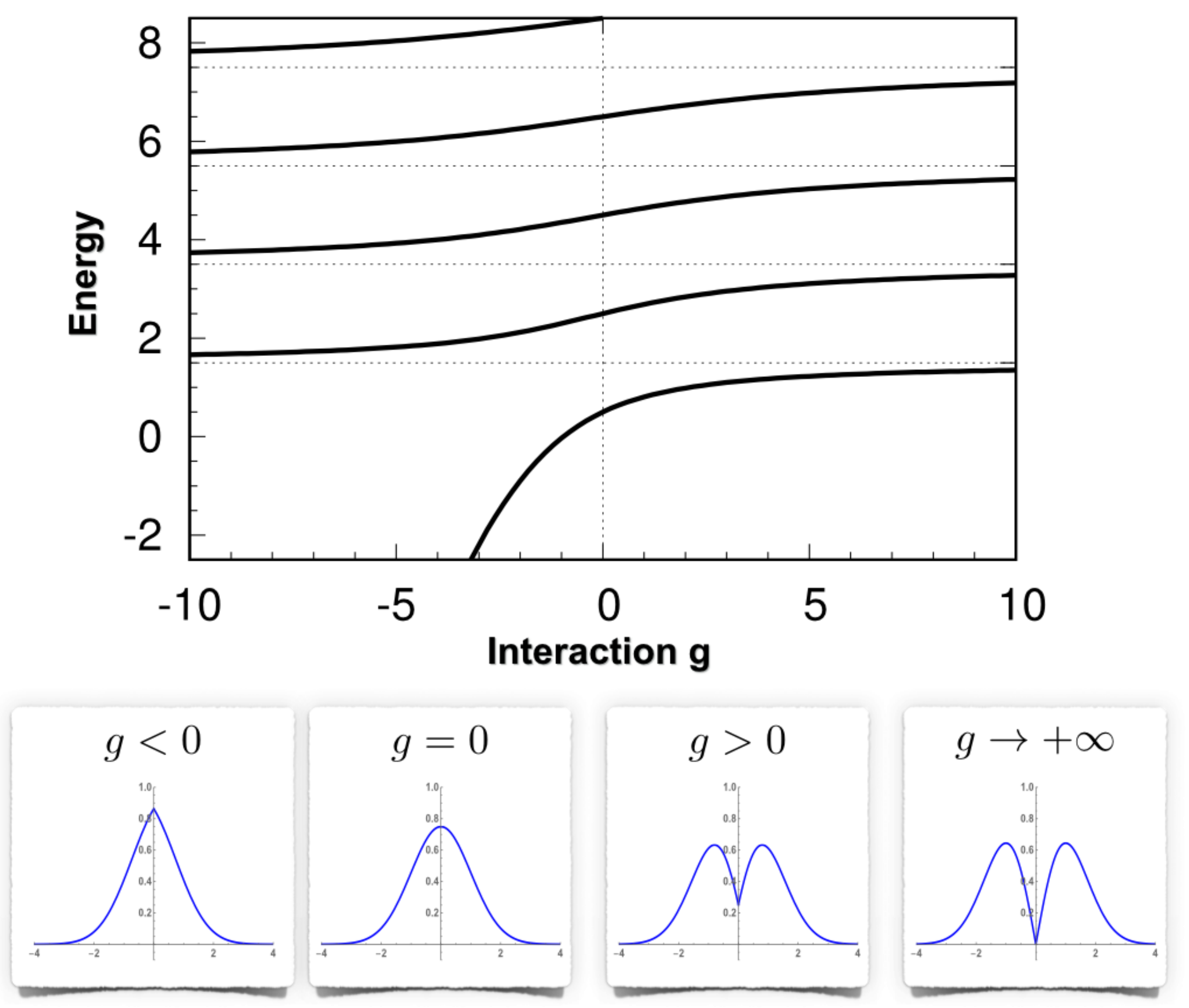}
\caption{Energy spectrum of the relative motion Hamiltonian \eqref{BuschRel} in the subspace of even wave functions (bosons). Horizontal dashed lines represent values achieved in infinite (attractive as well as repulsive) interactions. In the bottom panel we present the ground state wave function $\Psi_0(X)$ for different interactions. Note characteristic cusps at $r=0$ forced by contact interactions. \label{_Fig-BuschSolution}}
\end{figure}

Many theoretical and experimental considerations described in this review were inspired by the seminal paper of Busch {\it et al.}  \cite{1998BuschFoundPhys} where the exact analytical solution of the eigenproblem for two ultracold bosons confined in a harmonic trap (of any dimension) was presented. At that times the paper was treated only as an interesting theoretical curiosity since there were no experimental ways to validate its predictions. However, along with experimental progress in controlling and manipulating quantum systems with a small number of particles, the paper turned out to be one of the milestones in our understanding of properties of a small number of quantum particles.

Since many of the upcoming discussions are based or inspired by the two-body solution of Busch {\it et al.}, let us make (following detailed argumentation presented in \cite{2009WeiIJMPB}) a brief overview of the original problem in the one-dimensional case. The Hamiltonian of the considered system of two bosons of mass $m$ moving in a one-dimensional harmonic trap of frequency $\omega$ has the form
\begin{equation} \label{HamBuschFull}
\hat{\cal H} = -\frac{\hbar}{2m}\left(\frac{\partial^2}{\partial x^2}+\frac{\partial^2}{\partial y^2}\right)+\frac{m\omega^2}{2}\left(x^2+y^2\right)+g\delta(x-y),
\end{equation}
where $x$ and $y$ are positions of particles interacting via contact interactions with strength $g$. Whenever one deals with harmonic confinement, it is extremely convenient to express all quantities in natural units of a harmonic oscillator, {\it i.e.}, to measure energy, positions, and momenta in units of $\hbar\omega$, $\sqrt{\hbar/m\omega}$, and $\sqrt{\hbar m\omega}$, respectively. Then the Hamiltonian \eqref{HamBuschFull} becomes dimensionless and it has the form
\begin{equation} \label{HamBusch}
\hat{\cal H} = -\frac{1}{2}\left(\frac{\partial^2}{\partial x^2}+\frac{\partial^2}{\partial y^2}\right)+\frac{1}{2}\left(x^2+y^2\right)+g\delta(x-y),
\end{equation}
provided that the interaction strength $g$ is measured in its natural unit $\sqrt{\hbar^3\omega/m}$.

At this point let us mention that in fact the interaction coupling $g$ can be expressed by the corresponding effective one-dimensional scattering length $a$
\begin{align} \label{gexpress}
g=-\frac{2\hbar^{2}}{m}a^{-1}.
\end{align}
One-dimensional scattering length is however directly related to the three-dimensional $s$-wave scattering lenght $a_s$ as follows
\begin{equation} \label{aexpress}
a=-\frac{a_{\perp}^2}{2a_s}\left(1-C\frac{a_s}{a_{\perp}}\right),
\end{equation}
where  $a_\perp=\sqrt{2\hbar/m\omega_\perp}$ is the natural with of the ground-state of perpendicular confinement of frequency $\omega_\perp$ and $C=1.4603$ \cite{1998OlshaniiPRL}. It means that interactions in the one-dimensional confinement are controlled by three-dimensional $s$-wave scattering as well as the shape of the external perpendicular confinement.

To diagonalize the Hamiltonian \eqref{HamBusch} one changes variables to the center-of-mass and relative motion positions. It is very convenient to make this transformation in rescaled form 
\begin{equation} \label{BuschTransform}
R = (x+y)/\sqrt{2}, \qquad X=(x-y)/\sqrt{2}.
\end{equation}
After this transformation the Hamiltonian decouples to two independent single-particle Hamiltonians $\hat{\cal H}=\hat{\cal H}_R+\hat{\cal H}_X$ describing the center-of-mass motion and the relative motion of particles
\begin{subequations}
\begin{align}
{\cal H}_R &= -\frac{1}{2}\frac{\mathrm{d}^2}{\mathrm{d}R^2}+\frac{1}{2}R^2, \\
{\cal H}_X &= -\frac{1}{2}\frac{\mathrm{d}^2}{\mathrm{d}X^2}+\frac{1}{2}X^2 + \frac{g}{\sqrt{2}}\delta(X). \label{BuschRel}
\end{align}
\end{subequations}
As shown in \cite{1998BuschFoundPhys} the relative motion Hamiltonian \eqref{BuschRel} can be analytically diagonalized. In the subspace of odd wave functions the diagonalization is trivial since $\delta$ does not affect solutions vanishing at $X=0$. In the subspace of even (bosonic) wave functions the eigenenergies $E_k$ are given by roots of the transcendental equation 
\begin{equation}
-g\,\Gamma\left(\frac{1-2E_k}{4}\right)=2\sqrt{2}\,\Gamma\left(\frac{3-2E_k}{4}\right),
\end{equation}
and the corresponding eigenfunctions are expressed in terms of the Tricomi confluent hypergeometric function
\begin{equation}
\Psi_k(X) = N_k\, \mathrm{e}^{-X^2/2}\, \mathrm{U}\left(\frac{1-2E_k}{4},\frac{1}{2},X^2\right).
\end{equation}
Having these solutions one can show straightforwardly that, in the limit of infinite repulsions, the ground-state wave function has the form $\Psi_0(X)\sim|X|\mathrm{exp}(-X^2/2)$ with eigenenergy $E_0=1.5$. It means that it is degenerated with the odd ground-state wave function $\Psi_{odd}(X)\sim X\,\mathrm{exp}(-X^2/2)$.
The spectrum of the relative motion Hamiltonian \eqref{BuschRel} and shapes of the ground-state wave functions (in the even subspace) as functions of interaction strength are presented in Fig.~\ref{_Fig-BuschSolution}.

Having analytical solutions of a two-body problem in hand one can study different dynamical properties of the system \cite{2006IdziaszekPRA,2010SowinskiPRA,2016EbertAnnPhys,2018LedesmaARX,2019BudewigMolPhys}. Note that, although the Hamiltonian is separable into the centre-of-mass and the relative motion coordinates \eqref{BuschTransform}, this is not the case in the configuration space of particles' position, {\it i.e.}, interactions induce quantum correlations during the dynamics. From the other hand, one should also have in mind that the separation of the Hamiltonian in the relative motion coordinates is the immanent feature of the harmonic confinement. Any anharmonicity present in a trapping shape leads directly to a coupling between the center-of-mass and the relative positions. In consequence, it gives rise to transfer excitations between these two degrees of freedom and, as proved experimentally \cite{2013SalaPRL}, may be very helpful for the formation of bound pairs.

Since an existence of the exact analytical solutions of many-body problems is rather rare, we want here also to mention a few other examples of exactly solvable models. First, we want to mention the Moshinsky model, {\it i.e.}, an exactly solvable model of two particles confined in a harmonic trap and interacting via harmonic forces \cite{1968MoshinskyAJP}. This model was extended to many interacting particles \cite{1985BialynickiLetMPhys,2000ZaluskaPRA,2013KoscikFBS} and also many components \cite{2017KlaimanChemPhys,2017AlonJPhysA}. Second, the Busch {\it et al.} solution for two particles can be extended to cases of anisotropic harmonic traps \cite{2005IdziaszekPRA}. Third, in the case of a four-body problem and contact forces, neat analytical solutions associated with the symmetries of the
three-dimensional and four-dimensional icosahedra were discussed in \cite{2016ScoquartSciPost} while a very specific system of $N$ hard-sphere particles having special mass ratios was solved in \cite{2018OlshaniiPRA}. Finally, different exact solutions of the two-body problem with other than contact interactions were also announced: an attractive $1/r^6$ interaction in \cite{1998GaoPRA}, a repulsive $1/r^3$ interaction in \cite{1999GaoPRA,2016JieJPhysB}, and a finite-range (repulsive and attractive) interaction modeled by a step function in \cite{DeuretzbacherPhD,2018KoscikSciRep,2019KoscikARX}. Of course, we should also mention here two other seminal many-body solutions, {\it i.e.}, the Lieb--Liniger model of $N$ bosons \cite{1963LiebPR,1963LiebPRb} and the the Calogero--Sutherland model of interacting particles via inversely quadratic potentials \cite{1971CalogeroJMP,1971SutherlandJMP} and its extensions \cite{2016CampoNJP,2017PittmanPRB}.

 \section{Bosonic mixtures} \label{Sec:Bosons}

In this section we discuss the properties of bosonic mixtures with a small number of atoms. Unless otherwise clarified, for simplicity we consider that the two atomic components  consist of different hyperfine states of the same atomic species and therefore they have identical  mass and they are trapped in a one-dimensional parabolic external potential with the same oscillator frequency. The trapping in the two other directions is sufficiently tight to effectively {\it freeze} the dynamics in these directions, {\it i.e.}, all excitations in perpendicular directions are very unfavorable energetically. Since the number of atoms is small  and conserved in each component, it is often possible to work within the first quantization formalism and write the Hamiltonian being a straightforward extension of the two-boson Hamiltonian \eqref{HamBusch}. Therefore the Hamiltonian describing mixture of $N_\mathrm{A}$  identical bosons of kind $A$, with coordinates $x_i$, and $N_\mathrm{B}$ atoms
of kind $B$, with coordinates $y_i$ has a form
\begin{align}
\label{Eq:Hamiltonian1stQ}
\hat{\cal H} &=\sum_{i=1}^ {N_\mathrm{A}}\left[-\frac {1}{2} \frac {\partial^2}{\partial x_i^2}  + \frac {1}{2} x_i^2\right] + g_\mathrm{A} \sum_{i=1}^{N_\mathrm{A}}\sum_{j=i+1}^{N_\mathrm{A}}\delta (x_i -x_j) \nonumber \\
&+ \sum_{i=1}^ {N_\mathrm{B}}\left[-\frac {1}{2} \frac {\partial^2}{\partial y_i^2} + \frac {1}{2} y_i^2\right]+g_\mathrm{B}\! \sum_{i=1}^{N_\mathrm{B}}\sum_{j=i+1}^{N_\mathrm{B}}\delta (y_i -y_j) \nonumber \\
&+ g_\mathrm{AB} \sum_{i=1}^{N_\mathrm{A}}\sum_{j=1}^{N_\mathrm{B}}\delta (x_i -y_j), 
\end{align}
All mutual contact interactions are modeled by delta functions. In general one deals with three independent interactions strengths $g_\mathrm{A}$, $g_\mathrm{B}$, and $g_\mathrm{AB}$ for intra-component interactions in species $\mathrm{A}$, $\mathrm{B}$ and inter-component interactions, respectively. 

 At this point let us note that equivalently the Hamiltonian of the system \eqref{Eq:Hamiltonian1stQ} can also be written in the second quantization formalism by introducing the bosonic field operators $\hat\Phi_\sigma(x)$ annihilating a boson from the component $\sigma\in\{\mathrm{A},\mathrm{B}\}$ at position $x$. Bosonic nature of particles is encoded in the commutation relations which must be fulfilled by these operators
\begin{subequations}
\begin{align}
\left[\hat\Phi_\sigma(x),\hat\Phi_{\sigma'}^\dagger(x')\right]&=\delta_{\sigma\sigma'}\delta(x-x'), \\
\left[\hat\Phi_\sigma(x),\hat\Phi_{\sigma'}(x')\right]&=0.
\end{align}
\end{subequations}
With this notation the Hamiltonian \eqref{Eq:Hamiltonian1stQ} is transformed to the form
\begin{align} \label{Eq:Hamiltonian2ndQTS}
\hat{\cal H} &= \sum_\sigma\int\!\!\mathrm{d}{x}\,\hat\Phi^\dagger_\sigma(x)\left[-\frac{1}{2}\frac{\mathrm{d}^2}{\mathrm{d}x^2}+\frac{1}{2}x^2\right]\hat\Phi_\sigma(x) \nonumber \\
&+ \sum_\sigma\frac{g_\sigma}{2}\int\!\!\mathrm{d}{x}\,\hat\Phi^\dagger_\sigma(x)\hat\Phi_\sigma^\dagger(x)\hat\Phi_\sigma(x)\hat\Phi_\sigma({x}) \nonumber \\
&+g_{\mathrm{AB}}\int\!\!\mathrm{d}{x}\, \hat\Phi^\dagger_\mathrm{A}({x})\hat\Phi^\dagger_\mathrm{B}({x})\hat\Phi_\mathrm{B}({x})\hat\Phi_\mathrm{A}({x}).
\end{align}
In fact, the Hamiltonian \eqref{Eq:Hamiltonian2ndQTS} describes the system with arbitrary number of particles $N_\mathrm{A}$ and $N_\mathrm{B}$. However, since it commutes with the number operators $\hat{N}_\sigma=\int\!\mathrm{d}x\,\hat\Phi_\sigma^\dagger(x)\hat\Phi_\sigma(x)$ it can be analyzed in each subspace of given number of particles independently. In each of these subspaces it has the form of the Hamiltonian \eqref{Eq:Hamiltonian1stQ} with fixed particle numbers $N_\mathrm{A}$ and $N_\mathrm{B}$. 

\begin{figure}[t]
\includegraphics[width=0.98\columnwidth]{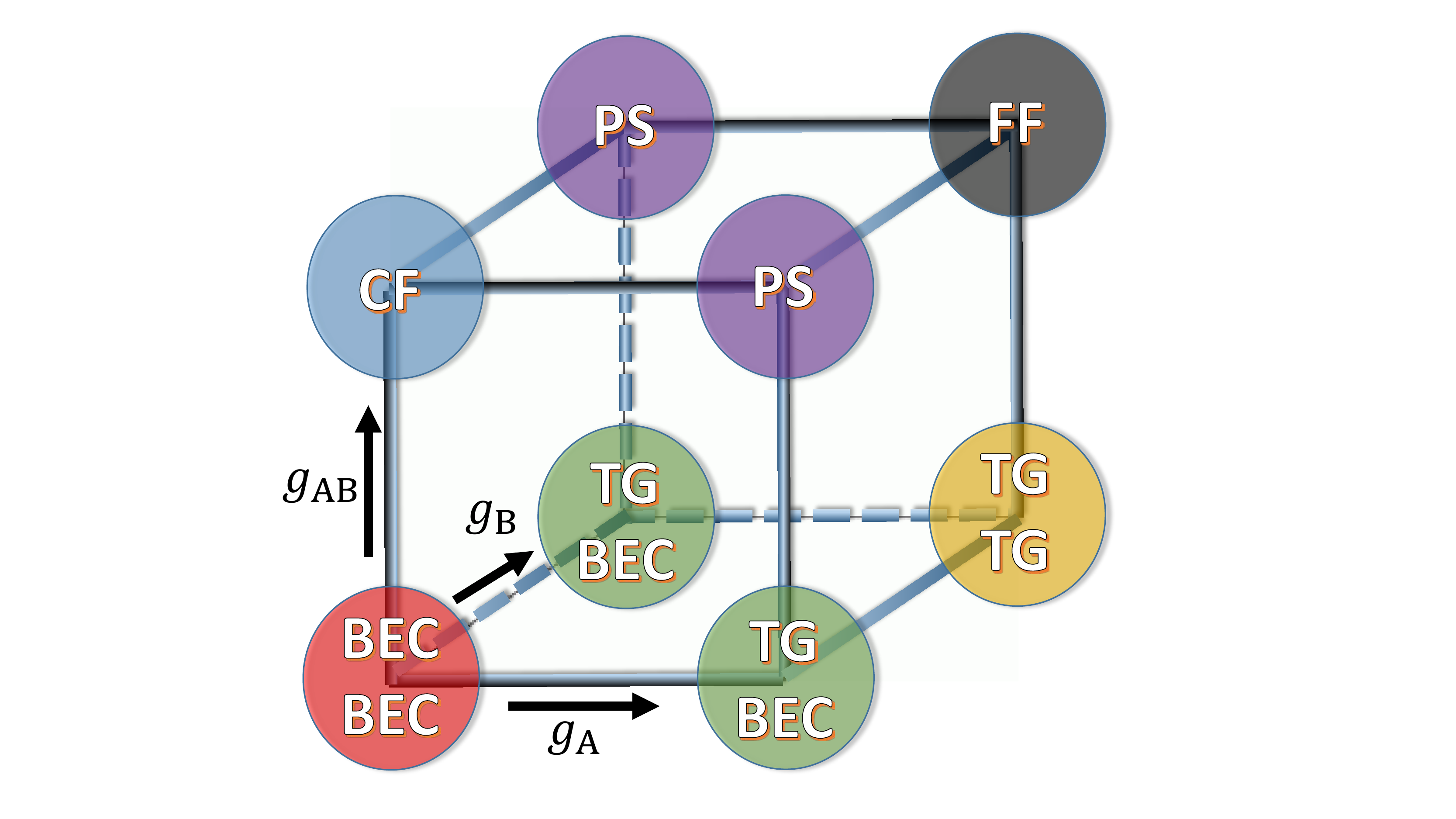}
\caption{Schematic view of the eight limits for two-component bosonic mixtures. Figure adapted  from~\cite{2014GarciaMarchNJP}. Copyright (2014) by the IOP Publishing.   \label{Fig:GarciaMarch_Cube}}
\end{figure}

In the following we consider the general Hamiltonian~\eqref{Eq:Hamiltonian1stQ}. Therefore there are two intra-component coupling constants $g_\mathrm{A}$ and $g_\mathrm{B}$ and one inter-component coupling constant $g_\mathrm{AB}$. For repulsive interactions ($g_\mathrm{A},g_\mathrm{B},g_\mathrm{AB}\geq 0$), there are eight natural limiting cases (see cube representation in Fig.~\ref{Fig:GarciaMarch_Cube}). These limits are:
\begin{itemize}
\item BEC-BEC limit, when all interactions are zero, \\ $g_\mathrm{A}=g_\mathrm{B}=g_\mathrm{AB}=0$.  
\item BEC-TG limit, when one of the intra-component interaction tends to infinite, while remaining ones are zero, \\
$g_\mathrm{AB}=g_\mathrm{A (B)}=0$ and $g_\mathrm{B (A)}\to\infty$. 
\item TG-TG limit, when inter-component interactions vanish but both intra-component interactions tend to infinite, \\ $g_\mathrm{AB}=0$, $g_\mathrm{A}\to\infty$, and $g_\mathrm{B}\to\infty$.
\item Composite fermionization (CF), when the inter-component interaction tends to infinite while intra-component interactions vanish, \\
$g_\mathrm{AB}\to\infty$, $g_\mathrm{A}=g_\mathrm{B}=0$. 
\item Phase separation (PS), when the inter-component together with one of the intra-component interactions tend to infinite while remaining one  vanishes, \\
$g_\mathrm{AB}\to\infty$, $g_\mathrm{A(B)}\to\infty$, and $g_\mathrm{B(A)}=0$.
\item Full fermionization (FF), when  all interactions tend to infinity \\
$g_\mathrm{AB}\to\infty$, $g_\mathrm{A}\to\infty$, and $g_\mathrm{B}\to\infty$. 
\end{itemize}

Note that, in the case of the first three limits, the inter-component interactions vanish. Therefore the problem is substantially simplified since, in these cases, both components are completely independent and the system can be treated as a simple composition of one-component bosonic gases (see for example \cite{2019MarchukovAnnPhys} for a detailed discussion of single-component systems). Conversely, in all other cases the inter-component interactions are very strong and they induce non-trivial correlations between particles belonging to opposite species. In these cases, the particular components cannot be treated as independent.

 Whenever few-atom bosonic mixtures are studied it is very helpful to have in mind a clear idea of some cornerstones originating in the limit of a large number of particles. These considerations lead directly to criteria for the four famous concepts: {\it (i)} the phase separation, {\it (ii)} the Tonks-Girardeau limit, {\it (iii)} the composite fermionization of the mixture, {\it (iv)} the full fermionization. Before discussing these four concepts in details (in subsections II\,B-E) let us first make a short overview on bosonic mixtures when attractive interactions between particles are present.

\subsection{Attractive forces -- a brief overview} 
\label{sec:attract}
The discussion of attractive forces is started from considering only a single component of bosonic gas. Then as shown in \cite{1964McGuireJMP}, in the limit $g\to -\infty$, the $N$-particle state forms a very exotic many-body state called the super-Tonks-Girardeau gas. Its properties were deeply studied theoretically~\cite{2004AstrakharchikPRL,2005AstrakharchikPRL,2005BatchelorJSM} and later it was also observed experimentally, as reported in~\cite{2009HallerScience}. When bosonic mixtures are considered, a very rich scenario opens, as one can distinguish different relative signs of different couplings, {\it i.e.}, the case in which both intra- and inter-component coupling have the same (attractive) sign, or cases when sings are opposite (one or more attractive, the rest repulsive). For the case when intra-component interactions are strongly repulsive while the inter-component interactions are attractive ($g_\mathrm{AB}<0$ and $g_\mathrm{A},g_\mathrm{B}\to +\infty$), a detailed theoretical study in~\cite{2003CazalillaPRL}  convinces that two different scenarios (depending on the density) are possible: the system is collapsing or pairs of particles are created. In contrast, when intra-component interactions are not necessarily very repulsive or they are attractive, numerical studies in the framework of the Multi-Configuration-Hartree-Fock techniques presented in \cite{2009TempfliNJP} show a very rich variety of phenomena -- different mechanisms of pairing, collapses, states with loosely bound particles, {\it etc}. Recently, this direction of research was additionally triggered by the theoretical proposal \cite{2015PetrovPRL,2018ZinPRA,2018ChiquilloPRA} and experimental confirmation~\cite{2018CabreraScience,2018SemeghiniPRL,2018CheineyPRL} of the existence of quantum liquid droplets in a two-component bosonic gas. In consequence, a number of works have explored this scenario in the limit of small number of atoms. A key ingredient for this liquid droplets to exist is the role played by three-body interactions. Therefore, initial works studied the effect of considering both two-body and three-body interactions in a single-component bosonic gas~\cite{2018NishidaPRA}. Then, using diffusion quantum Monte Carlo numerical computations and analytical predictions, a number of works studied bosonic mixtures with inter-component attractive and intra-component repulsive   interactions~\cite{2018PricoupenkoPRA,2019CikojevicPRA,2019ParisiPRL}. Particularly, in ~\cite{2018GuijarroPRA} the problem of three interacting bosons was considered.  

Previously, three-boson interactions were considered rather in the case of optical lattice systems to mimic influence of higher bands \cite{2011TiesingaPRA,2011SilvaPRA,2012SowinskiPRA,2016HincapiePRA}. Recently, this direction was reduced to problems of a few bosons confined in a one-dimensional double-well \cite{2018DobrzynieckiPRA}. We believe that the problem of effective three-body interactions and their competition with two-body forces is not well explored yet. Since it is an extremely interesting and blooming topic, it will attract a great attention in upcoming years.

\subsection{Phase separation}
  
 There is a long tradition on literature on phase segregation, which also is rooted in the study of other superfluidic systems, such as $^3$He-$^4$He~\cite{2006BarrancoJLP}. For Bose-Einstein condensates  of alkali atoms the first theoretical study~\cite{1996HoPRL} was shortly before the experimental realization~\cite{1996MyattPRL}. Then many important contributions to the understanding of binary bosonic mixtures came in the next few years~\cite{1997EsryPRL,1997BuschPRA,1998AoPRA,1998PuPRL,1998HallPRL,1998GordonPRA,1997GoldsteinPRA,1998OhbergPRA}. To make further discussion as simple as possible let us now present simple mean-field argumentation that in the limit of large number of particles the phase separation may appear in the system.
In this presentation we follow arguments presented in~\cite{1998AoPRA} for three-dimensional system. 

For clearness of the argumentation let us consider a mixture of $N_\mathrm{A}$ and $N_\mathrm{B}$ confined in a box potential with total length $L$, {\it i.e.}, it is described by the Hamiltonian \eqref{Eq:Hamiltonian2ndQTS} with omitted parabolic confinement and all integrals are over a region of length $L$. The mean-field description is based on the assumption that all bosons of a given component occupy only one single-particle orbital represented by the wave function $\phi_\sigma(x)$. Consequently, the corresponding field operators can be written as
\begin{equation}
\hat\Phi_\sigma(x) \approx \phi_\sigma(x) \hat{a}_\sigma,
\end{equation} 
where $\hat{a}_\sigma$ is the operator that annihilates an atom from component $\sigma$ being in a state $\phi_\sigma(x)$. With this notation one immediately writes the mean-field approximation of the ground-state of the system as
\begin{equation}
| \psi_{\rm MF}\rangle=\frac{\left(\hat{a}_{\rm{A}}^\dagger\right)^{N_{\rm{A}}}}{\sqrt{N_{\rm{A}}!}}\frac{\left(\hat{a}_{\rm{B}}^\dagger\right)^{N_{\rm{B}}}}{\sqrt{N_{\rm{B}}!}}| \mathtt{vac}\rangle,
\end{equation}
provided that the mean-field wave functions $\phi_\mathrm{A}(x)$ and $\phi_\mathrm{B}(x)$ are chosen in such a way that the mean-field energy $E_{\rm{MF}}=\langle \psi_{\rm MF}|\hat{\cal H}|\psi_{\rm MF}\rangle $ is minimal. 

In the problem studied, there are two conserved quantities ($N_\mathrm{A}$ and $N_\mathrm{B}$) and therefore the minimization is done with two constraints encoded in two Lagrange multipliers $\mu_\sigma$ (chemical potentials), {\it i.e.}, the condition for minimization reads $\delta E_{\rm{MF}} -\mu_{\rm{A}}\delta N_{\rm{A}} -\mu_{\rm{B}}\delta N_{\rm{B}}=0$.  This procedure gives rise to the set of coupled Gross-Pitaevskii equations of the form
\begin{align}
\left[-\frac{1}{2}\!\frac{\mathrm{d}^2}{\mathrm{d}x^2}+g_{\rm{A}}|\phi_{\rm{A}}(x)|^2+g_{\rm{AB}}|\phi_{\rm{B}}(x)|^2-\mu_{\rm{A}}\right]\phi_{\rm{A}}({x})=0,\nonumber\\
\left[-\frac{1}{2}\!\frac{\mathrm{d}^2}{\mathrm{d}x^2}+g_{\rm{B}}|\phi_{\rm{B}}(x)|^2+g_{\rm{AB}}|\phi_{\rm{A}}(x)|^2-\mu_{\rm{B}}\right]\phi_{\rm{B}}({x})=0.\label{Eq:GPE}
\end{align}

For a homogeneous solution of equations~\eqref{Eq:GPE} the kinetic term vanishes and the densities $n_\sigma=N_\sigma/L$ are position independent. Consequently, chemical potentials can be expressed as $\mu_\sigma=g_\sigma n_\sigma+g_{\mathrm{AB}}n_{\sigma'}$. Therefore,   the total energy of the homogeneous state is  
\begin{equation}
E_1=\frac{1}{2}\left[g_{\rm A}\frac{N_{\rm A}^2}{L}+g_{\rm B}\frac{N_{\rm B}^2}{L}+2g_{\rm{AB}}\frac{N_{\rm A}N_{\rm B}}{L}\right]. 
\end{equation} 
Here we assumed that the number of particles in each component is very large, {\it i.e.}, one can use approximation $N_\sigma-1\approx N_\sigma$ and consequently $N_\sigma(N_\sigma-1)\approx N_\sigma^2$. 

In contrast, if we consider the inhomogeneous case in which the two components have non-overlapping densities with a sharp interface the total energy is substantially different. Indeed, if $L_\sigma$ is the volume occupied by the component $\sigma$, then the densities are $n_\sigma=N_\sigma/L_\sigma$ and the total energy of the inhomogeneous state is
\begin{equation}
E_2=\frac{1}{2}\left[g_{\rm A}\frac{N_{\rm A}^2}{L_{\rm A}}+g_{\rm B}\frac{N_{\rm B}^2}{L_{\rm B}}\right]. \label{Eq:InhomE}
\end{equation} 
After minimization of Eq.~\eqref{Eq:InhomE} with respect to $L_\sigma$ with constrain $ L_{\rm A}+L_{\rm B}=L$ one finds
\begin{subequations}
\begin{align}
 {L}_\sigma={L}\left[1+\sqrt{\frac{g_{\sigma'}}{g_\sigma}}\frac{N_{\sigma'}}{N_\sigma}\right]^{-1}, \\
 \rho_\sigma=\left[1+\sqrt{\frac{g_{\sigma'}}{g_\sigma}}\frac{N_{\sigma'}}{N_\sigma}\right]\frac{N_\sigma}{L},
\end{align}
\end{subequations}
with chemical potentials $\mu_\sigma=g_\sigma\rho_\sigma$. Then,  the total energy for the inhomogeneous state reads
\begin{equation}
E_2=\frac{1}{2}\left[g_{\rm A}\frac{N_{\rm A}^2}{L}+g_{\rm B}\frac{N_{\rm B}^2}{L}+2\sqrt{g_{\rm A}g_{\rm B}}\frac{N_{\rm A}N_{\rm B}}{L}\right]. 
\end{equation} 
By comparing energies $E_1$ and $E_2$ one finds the condition that the inhomogeneous state has lower energy
\begin{equation}
 E_2-E_1=-\left(g_{\rm{AB}}-\sqrt{g_{\rm A}g_{\rm B}}\right)\frac{N_{\rm A}N_{\rm B}}{L}<0.
\end{equation}
This implies  that whenever 
\begin{equation} 
 g_{\rm{AB}}>\sqrt{g_{\rm A}g_{\rm B}},\label{Eq:PScrit}
\end{equation}
the homogenous state is not energetically favorable and the phase separation occurs in the many-body system. The derivation of the criterion \eqref{Eq:PScrit}, introduced in~\cite{1998AoPRA}, is given for illustrative purposes. More sophisticated derivations take into account the presence of an external trap as well as corrections from the finite particles' number and the thickness of the overlapping region. Properties of the phase separation at finite temperatures can also be examined \cite{2014RoyPRA,2015RoyPRA}. For our purposes, it is useful to have in mind that the physics of mixtures of a few bosons will  show the footprints of the phase separation phenomena that would appear at the large atom limit. Finally, we highlight a recent  interesting  study in the three-dimensional case, which aimed to compare the thermodynamic predictions with the results from numerical Monte Carlo simulations of smaller number of atoms, of the order of a few hundreds~\cite{2018CikojeviNJP}. 
 
\subsection{Tonks-Girardeau gas in a parabolic trap} 

The initial Girardeau papers on the strong repulsions limit were originated from the observation that the eigenstates $\phi_i(x_1,\ldots,x_N)$ of the first quantized Hamiltonian for interacting bosons
\begin{align}
\label{Eq:Hamiltonian1stQ1comp}
&\hat{\cal H} =\sum_{i=1}^{N}\left[-\frac {1}{2} \frac {\partial^2}{\partial x_i^2}  + \frac {1}{2} x_i^2\right] + g \sum_{i<j}\delta (x_i -x_j), 
\end{align}
in the limit $g\to \infty$ are exactly the same as those of the Hamiltonian
\begin{align} \label{BoseFermiHamEq}
&\hat{\cal H} =\sum_{i=1}^{N}\left[-\frac {1}{2} \frac {\partial^2}{\partial x_i^2}  + \frac {1}{2} x_i^2\right] , 
\end{align}
provided that the boundary condition $\phi|_{x_i=x_j}=0$ is applied to the many-body wave function. In fact this condition is exactly equivalent to the condition forced by the Pauli exclusion principle and therefore one can formulate the famous Bose-Fermi mapping for one-dimensional systems: eigenstates $\phi$ of the Hamiltonian \eqref{Eq:Hamiltonian1stQ1comp} for infinitely strong repelling bosons are in one-to-one correspondence with the eigenstates $\psi_\mathrm{F}$ of the Hamiltonian \eqref{BoseFermiHamEq} of non-interacting fermions  and can be constructed by appropriate symmetrizations, which turns into the simple relation $\phi=|\psi_{F}|$ for the many-body ground state~\cite{1960GirardeauJMP}. This was, in turn, the extension of the classical Tonks theory of hard-spheres~\cite{1936TonksPhysRev} to the quantum realm (named the Tonks-Girardeau gas). Shorty after the original paper of Girardeau~\cite{1960GirardeauJMP}, the calculation of the solution at all interaction strengths for the homogeneous potential was obtained with the Lieb--Liniger approach~\cite{1963LiebPR,1963LiebPRb}. This was further reduced to an eigenvalue problem of matrices of the same sizes as the irreducible representations of the permutation group $S_N$ for $N$ atoms~\cite{1967YangPRL}. In fact, all these analytical solutions were possible since, for the homogeneous external potential one can solve corresponding problems within the famous Bethe ansatz approach~\cite{1931BetheZTM,2014GaudinBook}. These initial works were followed by a thorough and rich study of bosonic systems in one dimension.  The crossover from BEC to TG was studied in several works~\cite{2000PetrovPRL,2001DjunkoPRL,2001GirardeauPRL,2002BlumePRA,2003GangardtPRL,2003KheruntsyanPRL,2004BrandJPhysB}. It was shown that   a Bose-Einstein condensate in a thin cigar-shaped trap has dynamics that approach those of a 1D TG gas, for large interaction strength and low  temperatures and  densities~\cite{1998OlshaniiPRL,2000PetrovPRL}. Thus, for ultracold quantum gases, the most relevant set-up includes a parabolic trap, which is not analytically solvable for the whole range of interactions by the Bethe ansatz. Before studying the parabolic trap case in detail, let us briefly mention that many works have studied the TG gas in different external trapping potentials, a non-comprehensive list includes potentials such as split traps~\cite{2003BuschJPhysB,2008MurphyPRA,2008GooldPRA,2008YinPRA,2010GooldNJP,2011GuoJPhysB}, optical lattices~\cite{2005AlonPRL,2008ChenEPL}, hard wall boxes~\cite{2005BatchelorJPhysA,2006HaoPRA,2009HaoPRAb,2015OlshaniiNJP}, ring potentials~\cite{2005SakmannPRA}, double wells~\cite{2006ZollnerPRA,2007ZollnerPRA,2008ZollnerPRL,2012ChatterjeePRA,2016DobrzynieckiEPJD,2018DobrzynieckiPLA,2018DobrzynieckiPRA,2009OkopinskaFBS,2015GarciaMarchPRA,2017HarshmanFBS,2017HarshmanPRA}, and anharmonic potentials~\cite{2007MatthiesPRA}.  The experimental realization of a TG gas was first reported in~\cite{2004KinoshitaScience,paredes2004tonks}.   

\begin{figure}[t]
\includegraphics[width=0.85\columnwidth]{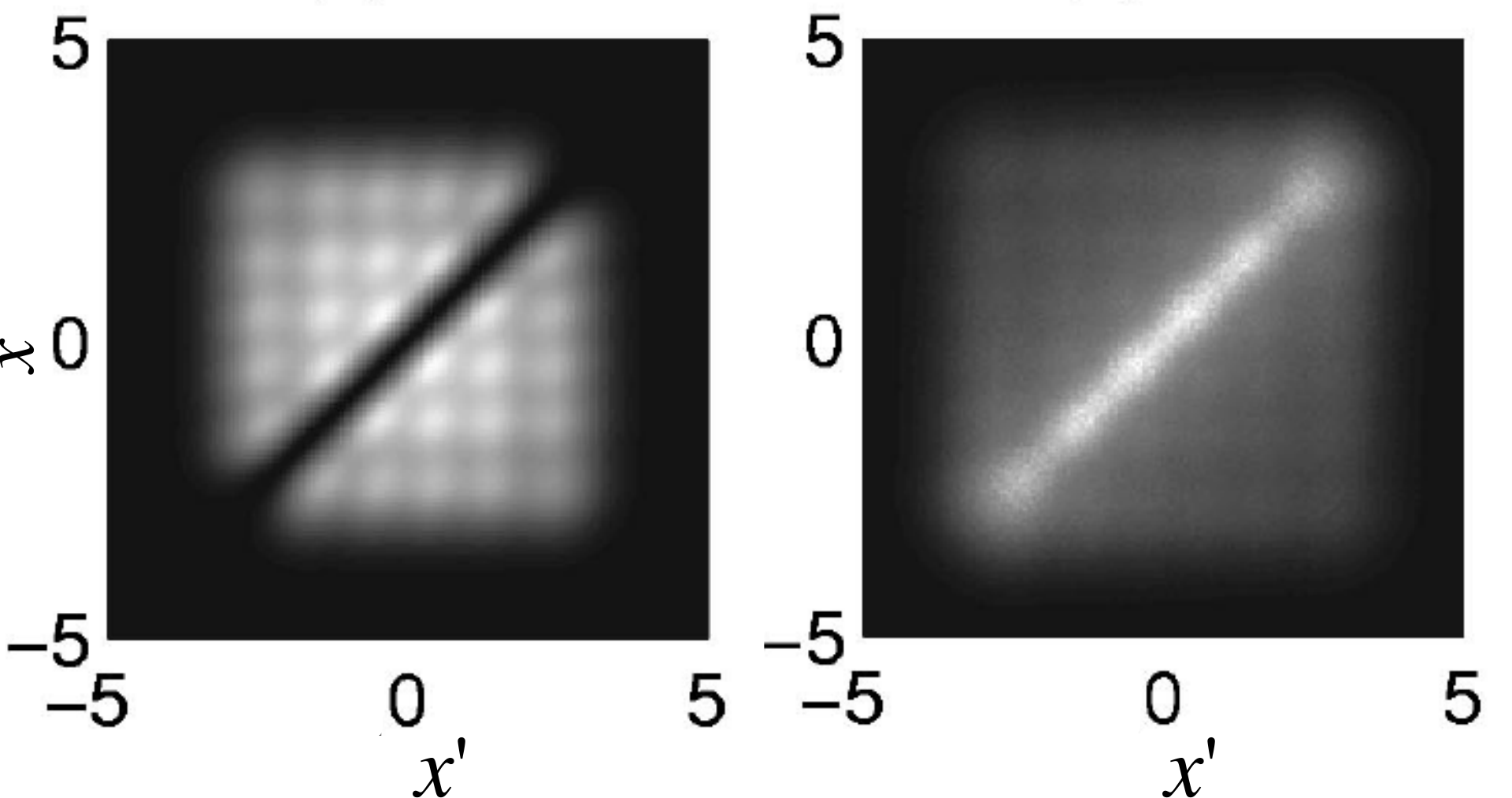}
\caption{(left panel) Single-particle reduced density matrix \eqref{Eq:OBDM} for a Tonks-Girardeau with $N=5$ atoms. (right panel) Two-particle density profile \eqref{Twoparticledensity} for the same system, showing a zero in the diagonal $x_1=x_2$. Figure adapted  from~\cite{2001GirardeauPRA}. Copyright (2001) by the American Physical Society. \label{Fig:GirardeauPRA}}
\end{figure}

In the following, we discuss the parabolic trap case in detail. Let us start with the solution in the TG limit in the presence of an external trap potential~\cite{2001GirardeauPRA}. The ground state wave function for the ideal gas of fermions with $N$ atoms can be expressed as the Slater determinant of the lowest $N$ single-particle eigenfunctions of the external confinement $\varphi_i(x)$
\begin{equation}
 \psi_{F}(\boldsymbol{r})=\frac{1}{\sqrt{N!}}\mathrm{det}\left[\varphi_i(x_j)\right]_{i=0,\ldots,N-1}^{j=0,\ldots,N-1}, 
\end{equation}
where the positions vector $\boldsymbol{r}=(x_1,\ldots,x_N)$. In the case of harmonic confinement one finds with $\varphi_i(x)=N_i\,\mathrm{H}_i(x)\,\mathrm{e}^{-x^2/2}$ with $\mathrm{H}_i(\cdot)$ being the Hermite polynomials. After applying appropriate symmetrization between particles' positions and a bit of algebra one finds the ground-state wave function for $N$ bosons in the Jastrow form
\begin{subequations}
\begin{equation}
\label{eq:Jastrow}
 \phi(\boldsymbol{r})=C_N \left(\prod_{i=1}^{N}\prod_{j=i+1}^{N}|x_i-x_j|\right)\mathrm{e}^{-\sum_i x_i^2/2} , 
\end{equation}
 with 
 \begin{equation}
  C_N=2^{N(N-1)/4}\left(N!\prod_{n=0}^{N-1}n!\sqrt{\pi}\right)^{-1/2}. 
 \end{equation}
\end{subequations}
Different properties of the system are encoded in the single-particle reduced density matrix usually defined as:
 \begin{equation}
 \label{Eq:OBDM}
 \rho^{(1)}(x,x')\!=\int\!\!\! \phi(x,\ldots,x_N) \phi(x',\ldots,x_N){\rm d} x_2 \dots{\rm d} x_N,
 \end{equation}
 In the TG limit an expression for $\rho^{(1)}$ in terms of $N-1$ integrals can be obtained (see~\cite{2001GirardeauPRA}).   The diagonal part of $\rho^{(1)}$ is the single-particle density profile $n(x)=\rho^{(1)}(x,x)$ which can be written explicitly as $n(x)=\sum_{j=0}^{N-1}|\varphi_j(x)|^2$~\cite{2001GirardeauPRA}. In Fig.~\ref{Fig:GirardeauPRA}a we show the single-particle reduced density matrix for $N=5$ bosons. 
Relevant information is also encoded in the single-particle momentum distribution, {\it i.e.}, diagonal part of the Fourier transform of the single-particle reduced density matrix
 \begin{equation}
 n(k)=\frac{1}{2\pi}\int  {\rm d}  x \int {\rm d} x'  \rho(x,x') \exp[-ik(x-x')], 
 \end{equation}
as well as in the two-particle density profile 
 \begin{equation} \label{Twoparticledensity}
 n_2(x_1,x_2)\!=\int\!\! |\phi (x_1,x_2,\dots,x_N)|^2 {\rm d} x_3 \dots{\rm d} x_N.
 \end{equation}
Since any two particles cannot be found at the same position, this density vanishes at the diagonal ($x_1=x_2$), see Fig.~\ref{Fig:GirardeauPRA}b. 

According to the Penrose-Onsager criterion of condensation an occurrence of the dominant eigenvalue in the spectral decomposition of the single-particle reduced density matrix $\rho^{(1)}(x_1,x_2)$ indicates condensation in the corresponding dominant orbital.  Shortly after first Girardeau paper, in a series of paper Lenard studied the momentum distribution and gave a bound for the dominant eigenvalue in the uniform Tonks-Girardeau gas, a topic with was open anyhow in the subsequent years~\cite{1964LenardJMP,1966LenardJMP,1979VaidyaPRL}. It took many years for a generalization to the trapped case~\cite{2003ForresterPRA}. The occupation of the dominant natural orbital grows with the number of particles like $\sim N^{0.5}$ showing that bosons have a natural tendency to condense into a single orbital even in this strongly repelling fermionized limit. We note that only recently a beautiful generalization to any trapping potential has been provided~\cite{2017SBrunScipost}. 
 
 \begin{figure}[t]
\includegraphics[width=0.92\columnwidth]{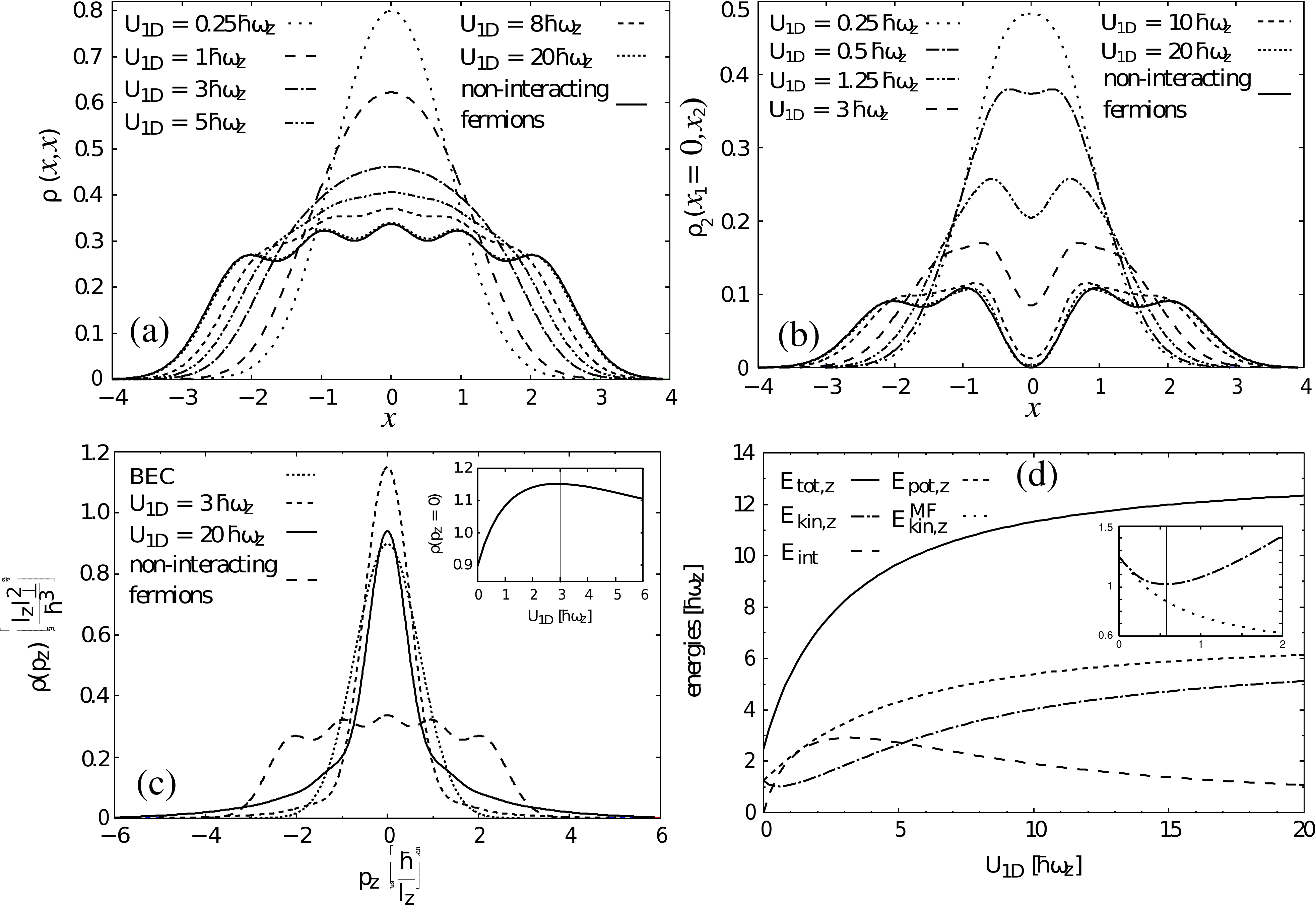}
\caption{(a) Density profile (diagonal  part of the single-particle reduced density matrix) for a Tonks-Girardeau with $N=5$ atoms as the interactions are increased. (b) two-particle density matrix profile (two-particle reduced density matrix at the origin for one of the variables) for the same system, as the interactions are incrteased. It developes a zero at the origin, {\it i.e.}, when $x_1=x_2$. (c) Density profile in the momentum domain for the same system and different interaction compared to the distribution of $N=5$ non-interacting fermions, (d) Total energy and its components: the kinetic, potential, and interaction energy for $N=5$. Here, the coupling constant $g$ in Eq.~\eqref{Eq:hamExactdiag} is termed as $U_{\rm{1D}}$, as in~\cite{2007DeuretzbacherPRA}. Figure adapted  from~\cite{2007DeuretzbacherPRA}. Copyright (2007) by the American Physical Society. \label{Fig:DeuretzbacherPRA}}
\end{figure} 

An important question is how a condensed system with $N$ bosons fermionizes as the interactions are increased (that is how it reaches he TG limit). This study has been attempted with different techniques, such as Multi-Configuration Hartree-Fock techniques (MCTDH), which are borrowed from chemistry~\cite{2005AlonPRL,2006ZollnerPRAb,2011ErnstPRA,2018GwakArxiv}, Monte Carlo numerical methods~\cite{2004AstrakharchikPRL}, semi-analytical methods~\cite{2012BrouzosPRL,2014WilsonPLA},  and the exact diagonalization~\cite{2007DeuretzbacherPRA} (also some studies have attempted the exact diagonalization when the delta, contact interactions are approximated by a thin Gaussian~\cite{2012KoscikFBS,2009ChristenssonPRA}).    To illustrate the process of fermionization we will discuss the exact diagonalization method, due to its simplicity. This is based on an expansion of the field operator $\hat\Phi(x)=\sum_i \hat{a}_i\varphi_i(x)$ in the basis of the eigenstates of corresponding single-particle Hamiltonian. After substitution of this expansion to the Hamiltonian \eqref{Eq:Hamiltonian2ndQTS} one finds
\begin{equation} 
\hat{\cal H}=\sum_{j} E_j \hat{n}_j+\frac{g}{2}\sum_{ijkl} U_{ijkl}\hat{a}_i^\dagger\hat{a}_j^\dagger\hat{a}_k\hat{a}_l
\label{Eq:hamExactdiag}
\end{equation} 
where $\hat{n}_j=\hat{a}_j^\dagger\hat{a}_j$ are the number operators, and $g$ is the coupling constant, accounting for the strength of the interactions. In the case of harmonic confinement a single-particle energies read $E_j=\hbar\omega(1/2+j)$. Interaction integrals can be calculated straightforwardly knowing shapes of single-particle states $U_{ijkl}=\int{\rm d}x\,\varphi_i(x) \varphi_j (x) \varphi_k (x) \varphi_l(x)$. The drawback of this method is that one has to truncate this basis to a maximum number of modes $M$. Then, one constructs the Fock basis $\{F_i\}$ for $N$ particles build from these $M$ modes, calculate all possible matrix elements of the Hamiltonian in this basis $H_{ij}=\langle F_i|\hat{\cal H}|F_j\rangle$ and diagonalize the resulting matrix. The dimension of the matrix to be diagonalized grows with the number of particles and the number of single-particle states taken into account as $(N+M-1)!/[N!(M-1)!]$. Thus this is restricted to a small number of atoms, to have reasonably big matrices and sufficient accuracy (a recent study shows how to accelerate its convergence~\cite{2018KoscikPhysLettA,2018JeszenszkiPRA}). This method allows anyhow to illustrate many aspects of the transition from a small interacting gas of bosons to a TG gas. In Fig.~\ref{Fig:DeuretzbacherPRA}(a) we show the single-particle density profile $n(x)$ (the diagonal part of the single-particle reduced density matrix),  as $g$ is increased for $N=5$ bosons (in the figure, the nomenclature from~\cite{2007DeuretzbacherPRA} is used, that is, $g=U_{\rm{1D}}$). As shown the density profile evolves from a Gaussian form to the characteristic profile for strongly repelling bosons, with a number of oscillations equal to the number of atoms.  These specific oscillations of the single-particle density profile can be viewed as counterparts of the famous Friedel oscillations known from solid-state physics \cite{1952FriedelPhylMag,1958FriedelMetAll}. Throughout the fermionization process, the two-particle density profile $n_2(x_1,x_2)$, develops a zero at $x_1=x_2$, and gets very close to the one for fermions (Fig.~\ref{Fig:DeuretzbacherPRA}(b)). Also, the momentum distribution $n(k)$  develops a peak and is rather different from that of fermions (Fig.~\ref{Fig:DeuretzbacherPRA}(c)). Particularly, high-momentum tails have the predicted behavior,  $n(k)\propto 1/k^4$~\cite{2002MinguzziPLA,2002LapeyrePRA}. 
This figure clearly illustrates that, at the TG limit, the system is different from that of ideal fermions. This information is also encoded in the natural orbitals occupations, obtained after diagonalization of the single-particle reduced density matrix, which shows that the largest value is significantly big (see discussion above), showing some degree of condensation, contrarily to fermions. 

Finally, it is interesting to study different contributions to the total energy $E$ of the ground-state:  the kinetic part $E_{\rm{kin}}$,  the potential part $E_{\rm{pot}}$, and the interaction part $E_{\rm{int}}$. In  Fig.~\ref{Fig:DeuretzbacherPRA}(d)  we plot these three components together with the total energy $E$ as functions of interactions for $N=5$ particles. Naturally, the total energy asymptotically tends to the energy of non-interacting fermions $E\rightarrow\hbar\omega\sum_{j=1}^N(1/2+j)=12.5$ (for $g=0$ all five atoms have energy 1/2 so the total energy of the non-interacting system is $E_0=2.5$). It is quite obvious that in the limit of strong repulsions ($g\rightarrow\infty$), the interacting energy should go to zero. The fact that the calculated $E_{\rm{int}}$ plotted in Fig.~\ref{Fig:DeuretzbacherPRA}(c) gets small but not zero shows that the exact diagonalization method fails to describe the TG limit accurately, due to the truncation in the number of basis modes used. 
Very interestingly, in~\cite{2015GudymaPRA}, Monte Carlo methods together with Local density approximation calculations were used to show the differences in this transition from non-interacting (ideal gas) limit to the TG limit when evaluated at a small and large number of atoms. Particularly, the study of excitations and the breathing mode showed that they behave differently for a very small number of atoms.

 \begin{figure}[t]
\includegraphics[width=0.92\columnwidth]{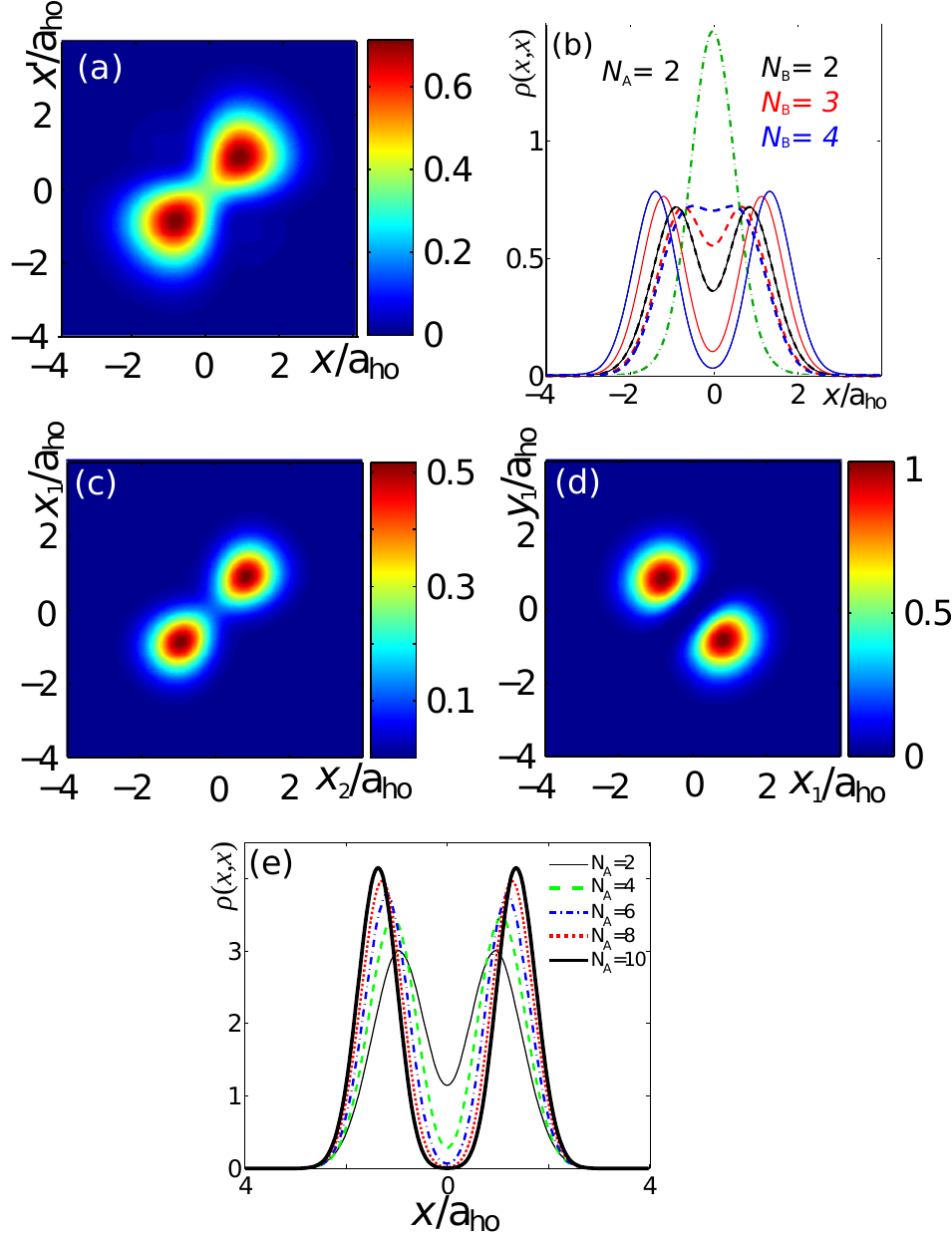}
\caption{(a)  Single-particle density matrix for a mixture with $N_{\rm{A}}=N_{\rm{B}}= 2$ bosons in the Composite fermionization limit, for $g_{\rm{A}}=g_{\rm{B}}=0$ and $g_{\rm{AB}}=20\hbar\omega a_{\rm{ho}}$. (b) Density profiles  $n(x) $ for increasing number of atoms in component B,  $N_{\rm{B}}=2,3,4 $. Dashed lines represents the profiles of B, solid lines for A. Green dashed line represents the limiting profile for B for very large $N_{\rm{B}}$. (c) Two-particle  density profile $n_2(x_1,x_2)$ for two atoms in species A. (d) Two-particle density profile $n_2(x_1,x_2)$ for one atom  in species A and one in B. (e) Density profiles  $n(x)$ for increasing number of atoms, but equal in both species, $N_{\rm{A}}=N_{\rm{B}}$.  Figure adapted  from~\cite{2014GarciaMarchNJP} and from~\cite{2013GarciaMarchPRA} (last paper, only panel (b)). Copyright (2014) by the IOP Publishing and by the American Physical Society.
 \label{Fig:Smelcher}}
\end{figure}

 \subsection{Composite Fermionization}

An important limit for the analysis is that of the composite fermionization of the bosonic mixtures introduced in~\cite{2008ZollnerPRA,2009HaoEPJD}. This limit occurs when in the Hamiltonian~\eqref{Eq:Hamiltonian1stQ} we neglect intra-component interactions ($g_\mathrm{A}=g_\mathrm{B}=0$) keeping inter-component interaction very large. The limit strictly occurs when $g_\mathrm{AB}\to\infty$. The Hamiltonian for finite inter-component interactions reads  
\begin{align}
\label{Eq:HamiltonianCF}
\hat{\cal H} &= \sum_{i=1}^ {N_\mathrm{A}}\left[-\frac {1}{2} \frac {\partial^2}{\partial x_i^2}  + \frac {1}{2} x_i^2\right] + \sum_{i=1}^ {N_\mathrm{B}}\left[-\frac {1}{2} \frac {\partial^2}{\partial y_i^2} + \frac {1}{2} y_i^2\right]    \nonumber\\
&+ g_\mathrm{AB} \sum_{i=1}^{N_\mathrm{A}}\sum_{j=1}^{N_\mathrm{B}}\delta (x_i -y_j). 
\end{align}
In the case $g_\mathrm{AB}\to\infty$, the wave function should vanish whenever $x_i -y_j=0$. Therefore, inspired by the Bose-Fermi mapping and the wave function for a single component \eqref{eq:Jastrow}, we find that the many-body wave function of the system has a form
\begin{equation}
\label{eq:JastrowCF}
 \phi(\boldsymbol{r}_\mathrm{A},\boldsymbol{r}_\mathrm{B})\propto \!\left(\prod_{i=1}^{N_\mathrm{A}}\prod_{j=1}^{N_\mathrm{B}}\!|x_i-y_j|\right)\mathrm{e}^{-\sum_i x_i^2/2} \mathrm{e}^{-\sum_j y_j^2/2}. 
\end{equation}
where $\boldsymbol{r}_\mathrm{A}=(x_1,\ldots,x_{N_\mathrm{A}})$ and $\boldsymbol{r}_\mathrm{B}=(y_1,\ldots,y_{N_\mathrm{B}})$ are just shortcuts for atoms' positions in components A and B respectively.

The main features of a composite fermionized gas can be illustrated in terms of the single-particle reduced density matrices $\rho_\mathrm{A}(x,x')$ and $\rho_\mathrm{B}(y,y')$ together with the two-particle density profiles $n^{(A)}_2(x_1,x_2)$, $n^{(B)}_2(y_1,y_2)$  and $n^{(AB)}_2(x,y)$ of particles from the same or opposite components, respectively. First two are defined in analogy to \eqref{Twoparticledensity}, while the latter is defined as 
\begin{align}
 n^{(AB)}_2(x,y)=\int\!\!|\phi(\boldsymbol{r}_\mathrm{A},\boldsymbol{r}_\mathrm{B})|^2 {\rm d} x_2{\rm d} y_2 \cdots{\rm d} x_{N_\mathrm{A}}{\rm d} y_{N_\mathrm{B}}.
 \end{align}
The single-particle reduced density matrix of a system where $N_\mathrm{A}=N_\mathrm{B}=2$ (in this case it is the same for both components) is presented in Fig.~\ref{Fig:Smelcher}a. One notices two distinct peaks showing that there is equally probable to find the atoms on the left or on the right side of the trap. The two-particle density profiles for two atoms belonging to the same and opposite components $n_2^{(\sigma)}(x_1,x_2)$ $n_2^{(AB)}(x_1,x_2)$ are shown in Fig.~\ref{Fig:Smelcher}c and Fig.~\ref{Fig:Smelcher}d respectively. In the latter case, a characteristic separation (vanishing of the density) along with $x_1=y_1$ line is clearly visible.  This allows one to make comprehensive interpretation of the result: the system manifests a density separation, if a boson from the component A is found in one of the maxima the others particles from the same component will be located nearby. At the same time, bosons from the remaining component will be localized around the second peak. The largest occupation of the natural orbitals of the single-particle reduced density matrix is $\lambda_0^{\rm{A,B}}\approx 0.55$~\cite{2013GarciaMarchPRA}, showing that though there is some kind of fermionization in the system,  there is also a strong tendency to condense all indistinguishable bosons. For completeness, in Fig.~\ref{Fig:Smelcher}b we plot the density profile as the number of atoms in the B component is increased, while there are only two atoms in A. As shown, species B has a larger tendency to occupy the center of the trap, while species B tends to split to two fragments located in the edges of the system. Indeed, $\lambda_0^{\rm{B}}\approx 0.62$ and  $\lambda_0^{\rm{B}}\approx 0.68$ for $N_\mathrm{B}=3$ and $N_\mathrm{B}=4$, respectively, while $\lambda_0^{\rm{A}}$ tends to 0.5~\cite{2013GarciaMarchPRA}. In the limit of  $N_\mathrm{B}\ll N_\mathrm{A}$, the species B condenses in a Gaussian profile in the center of the trap (schematically represented as a green line in Fig.~\ref{Fig:Smelcher}b) while species A fragments in two incoherent peaks with one atom at each side of B. Therefore, in this limit,  $\lambda_0^{\rm{B}}\to 1$  and $\lambda_0^{\rm{A}}\to 1/2$, being similar to a phase separated limit~\cite{2013GarciaMarchPRA}. In Fig.~\ref{Fig:Smelcher}e we illustrate how the two peaks move further away from the center of the trap as $N_\mathrm{A}=N_\mathrm{B}$ is increased (calculated with Diffusion Monte Carlo  in~\cite{2014GarciaMarchNJP}). The extreme limit in which one of the species has only one atom connects with the impurity problem, discussed in the subsection~\ref{sec:larger}.

\begin{figure}
\includegraphics[width=0.94\columnwidth]{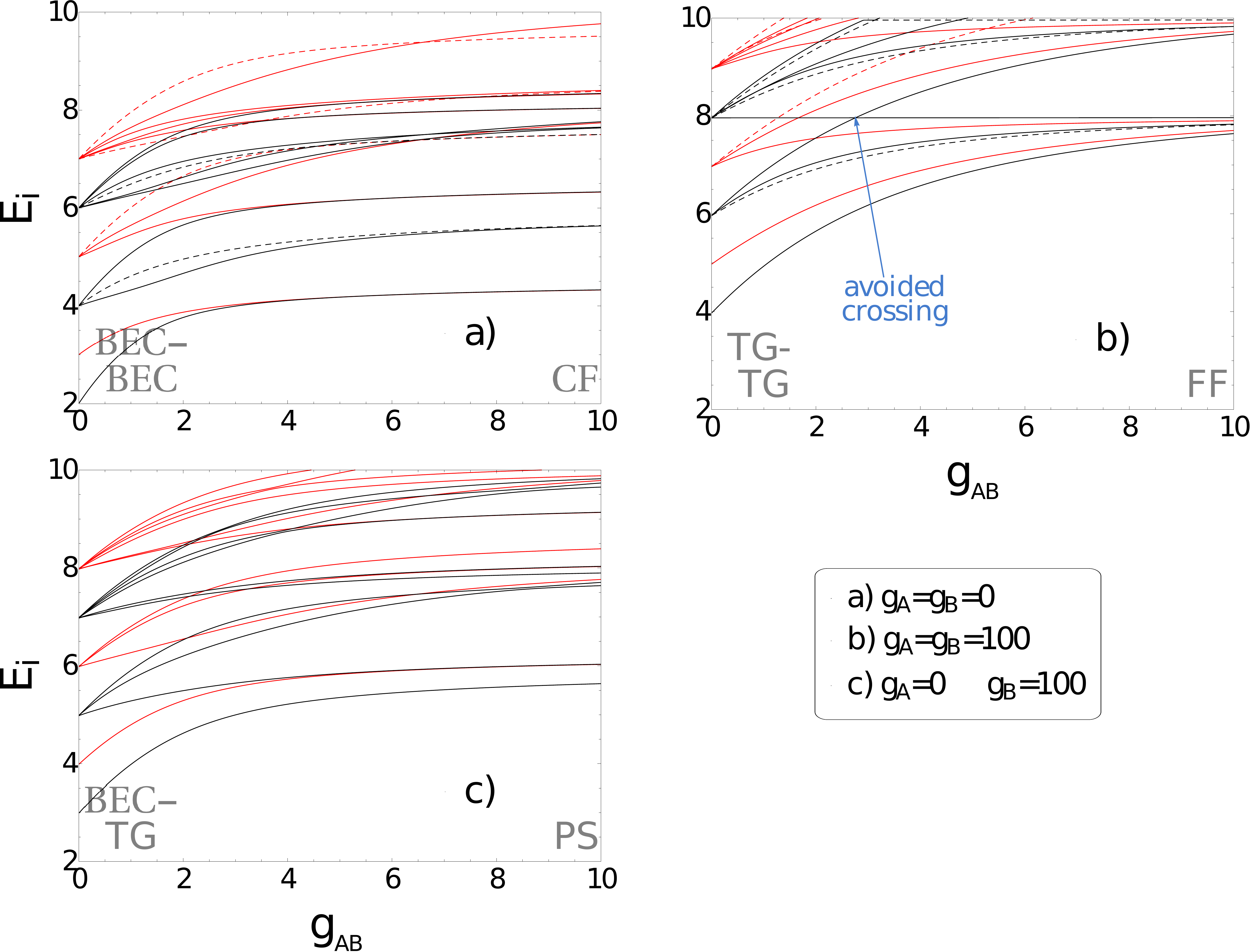}
\caption{ Energy spectrum for $N_\mathrm{A}=N_\mathrm{B}=2 $ as a function of $g_\mathrm{AB}$ for different fixed 
$g_\mathrm{A}$ and $g_\mathrm{B}$. The decoupled total CM is assumed to be in its ground state (see discussion on subsection~\ref{sec:four}).  The total parity is thus determined solely by the $R_{AB}$ parity and is marked by  black lines (even states) and red lines (odd states).  Solid curves correspond to symmetric ($+1$) and dashed to antisymmetric ($-1$)  eigenstates under the $S_{r}$ operation. The indicated (avoided) crossings are exemplary and simply outline specific features.  
All quantities are given in HO units. Figure adapted  from~\cite{2018PyzhNJP}. Copyright (2014) by  IOP Publishing.   \label{Fig_Smelcher_2}}
\end{figure}

We show the energy spectra as $g_\mathrm{AB}$ is increased with $g_\mathrm{A}=g_\mathrm{B}=0$ in Fig.~\ref{Fig_Smelcher_2}a. As observed, the ground state is doubly degenerate when $g_\mathrm{AB}\to\infty$~\cite{2018PyzhNJP}. Therefore, the double-peaked ground state is doubly degenerated (see density profiles in Fig.~\ref{Fig:Smelcher}). A small perturbation breaks the symmetry giving rise to the density profiles in which one peak separates from the other one, each localized at one side of the trap so that the density separation is evident. This figure is discussed in detail in subsection~\ref{sec:four}. 

 \subsection{Full Fermionization}
 
The isotropic limit corresponds to the case in which all coupling constants are equal, $g_{\rm{A}}=g_{\rm{B}}=g_{\rm{AB}}$~\cite{2003LiEPL}. In such case the system is integrable for the homogeneous case (see e.g.~\cite{2003LiEPL} for the solution via the Bethe ansatz) but not in the trapped case. However, the problem is analytically solvable even in the trapped case in the Full Fermionization limit, {\it i.e.}, the limit where all coupling constants tend to infinity $g_{\rm{A}}$, $g_{\rm{B}}$, and $g_{\rm{AB}}\to\infty$~\cite{2007GirardeauPRL}.  This solution is obtained by an extension of the Bose-Fermi mapping theorem to mixtures: for a system with $N_{\sigma}$ atoms in individual components, one constructs the fermionic many-body ground state with the Slater determinant for $N=N_{\rm{A}}+N_{\rm{B}}$ particles in $N$ the lowest single-particle states. Then, appropriate symmetrization is implemented in each component independently
\begin{equation}
\phi_\mathrm{B}(\boldsymbol{r}_\mathrm{A},\boldsymbol{r}_\mathrm{B}) = {\cal S}_\mathrm{A}{\cal S}_\mathrm{B} \left(\mathrm{det}\left[\varphi_i(r_j)\right]^{i=0,\ldots,N-1}_{j=0,\ldots,N-1}\right),
\end{equation}
where ${\cal S}_{\sigma}$ introduces appropriate bosonic symmetrization with respect to the permutations of particles in individual components. In the case of harmonic confinement the ground-state wave function takes the Jastrow form
\begin{multline}
 \phi(\boldsymbol{r}_\mathrm{A},\boldsymbol{r}_\mathrm{B})\propto \\\left(\prod_{i=1}^{N_\mathrm{A}}\prod_{j=i+1}^{N_\mathrm{A}}\!|x_i-x_j|\prod_{i=1}^{N_\mathrm{B}}\prod_{j=i+1}^{N_\mathrm{B}}\!|y_i-y_j|\prod_{i=1}^{N_\mathrm{A}}\prod_{j=1}^{N_\mathrm{B}}\!|x_i-y_j|\right)\\\times\mathrm{e}^{-\sum_i x_i^2/2} \mathrm{e}^{-\sum_j y_j^2/2}. 
 \end{multline}
 Note that in the mixed term we have intentionally selected a positive symmetry whenever positions of A and B atoms are exchanged. In fact, this choice is arbitrary in the limit of infinite repulsions. In consequence, it leads to the degeneracy of the ground-state manifold. In the next section, we will describe all the possibilities for the systems of three and four atoms. For a system of $N$ distinguishable atoms, with infinite interactions, there would be $N!$ degenerate states. For a mixture of $N_{\rm{A}}$ and $N_{\rm{B}}$ atoms, there are instead $N!/(N_\mathrm{A} ! N_\mathrm{B} !)$ degenerate states (for a detailed study on the degeneracies see~\cite{2008DeuretzbacherPRL,2011FangPRA}). To prove this, one has to rely on the symmetries of the system. To this end, it is usual to use the Young Tableaux associated with the system, as is a combinatorial object that permits for a convenient way to describe the group representations~\cite{2011FangPRA}. For very strong but finite interactions the degeneracy is lifted and the state with the lowest energy (the true ground-state) is the appropriate superposition of states with different symmetries. 

\subsection{Minimal mixture: Three atoms}  
\label{sec:three}

The minimal mixture of bosons in which the quantum statistics plays any role consists of two bosons of species A and one atom of a different species B. In the case of harmonic confinement the system is described by the Hamiltonian
\begin{align}
    \label{model31b}
\hat{\cal H}&= \sum_{i=1}^2 \frac{1}{2}\left( -\frac{\partial^2}{\partial x_i ^2}+  x_i^2 \right) + g_{\rm{A}}\delta(x_1 - x_2) \\
&+\frac{1}{2}\left(-\frac{\partial^2}{\partial y ^2}+ y^2\right) + g_{\rm{AB}}\left[\delta(x_1 - y)+\delta(x_2 - y)\right], \nonumber
\end{align}
where $x_i$ and $y$ are positions of the bosons in appropriate components.  This Hamiltonian, for $\omega\rightarrow 0$,  (no trapping potential) was  discussed in seminal papers~\cite{1964McGuireJMP,1975GaudinnJP}. Particularly,  the case with $g_\mathrm{A}=0$ and $g_\mathrm{AB}\neq 0$ is the Faddev-solvable Gaudin-Derrida model~\cite{1975GaudinnJP}.  

In this case there are four meaningful limits. Namely:
\begin{itemize}
\item BEC-BEC limit ($g_\mathrm{A}=g_\mathrm{AB}=0$),  
\item BEC-TG limit ($g_\mathrm{A} \to\infty$, $g_\mathrm{AB}=0$), 
\item CF limit ($g_\mathrm{A}=0$, $g_\mathrm{AB}\to\infty$), 
\item FF limit ($g_\mathrm{A}\to\infty$, $g_\mathrm{AB}\to\infty$). 
\end{itemize}
In all these limits  analytical solutions can be found~\cite{2012HarshmanPRA,2014ZinnerEPL} by introducing convenient transformation to the Jacobi coordinates 
\begin{subequations}
\begin{align}
    \label{model33}
    R&=\frac{x_1+ x_2 + x_3}{3}\,, \\
    X&=\frac{x_1-x_2}{\sqrt{2}}\,,\\
    Y&=\frac{x_1+x_2}{\sqrt{6}}-\sqrt{\frac{2}{3}}x_3.
    \end{align}
\end{subequations}
In these variables the three-particle Hamiltonian~\eqref{model31b} decouples to the Hamiltonian of the center-of-mass motion in coordinate $R$ and the relative motion encoded in the two remaining variables $(X,Y)$ (see appendix~\ref{app:threeatoms}). Obviously, the center-of-mass Hamiltonian does not include the interaction term and it is simply equivalent to the single-particle problem in a harmonic confinement.  On the other hand, the contact interactions define six lines  in the $(X,Y)$ plane (see Fig.~\ref{Fig_distinguishability1_app} in appendix~\ref{app:threeatoms} and Ref.~\cite{2012HarshmanPRA}). They correspond to the the locus of points where two particles meet.  These lines are located at $X=0$ and $X=\pm \sqrt{3}Y$.  Thus, they delimit six regions in the plane (note that Jacobi transformation breaks a symmetry between $X$ and $Y$ coordinates and they cannot be interchanged). By introducing another transformation to the polar coordinates
\begin{equation}
\rho=\sqrt{X^2+Y^2} \qquad \theta=\arctan(Y/X)
\end{equation}
the three lines  correspond to three angles $\theta=\{-\pi/6,0,\pi/6\}$. As discussed in   Appendix~\ref{app:threeatoms} this fact directly leads to a six-fold symmetry in the $(X,Y)$ plane in the case $g_\mathrm{A}=g_\mathrm{AB}$. In the case $g_\mathrm{A}\neq g_\mathrm{AB}$ the symmetry is reduced to two-fold one. This gives important hints to construct Ansatz functions for different systems, either three indistinguishable bosons, fermions, or mixtures of two bosons or fermions with an additional distinguishable particle~\cite{2012HarshmanPRA,2014ZinnerEPL}. Let us first discuss the case of two bosons interacting with one additional impurity when $g=g_\mathrm{A}=g_\mathrm{AB}$. In such case, a useful basis for representing functions for every $g>0$ is 
 \begin{equation}
 \label{eq:m,nrho,ntheta}
  \phi_{m,n_\rho,n_\theta}\!(\rho,\theta)\propto\mathrm{U}\left(-n_\rho,n_\theta+1,\rho^ 2\right)\mathrm{F}_{m,n_\rho}\!(\theta)\rho^{n_\theta}\mathrm{e}^{-\rho^2/2},
 \end{equation}
where $\mathrm{U}(a,b,x)$ is the Tricomi confluent hypergeometric function and the angular function $\mathrm{F}_{m,n_\theta}(\theta)$ is defined in each of the sextants of the plane as 
\begin{equation}
\label{eq:Angularpart}
 {\mathrm F}_{m,n_\rho}(\theta)=\alpha\sin[n_\rho(\theta-j\pi/3)]\exp[-im(j-1)\pi/3],
\end{equation}
when $\theta\in[(j-1)\pi/3,j\pi/3[\,$, with $j=1,\dots,6$. Here $n_\rho,n_\theta$ are positive integers and $m=0,\pm1,\pm2$ and $3$ due to the six-fold symmetry. We call the index $m$ orbital angular pseudo-momentum (OAPM), as it was introduced in the context of vortex solitons~\cite{2009GarciaMarchPhysD,2019GarciaMarchArxiv}.  The OAPM is associated to discrete rotations of multiples of $\pi/3$ in the polar plane. It provides a useful tool to express in an easily interpretable way the results from~\cite{2012HarshmanPRA} in terms of the hyperspherical coordinates (called here polar coordinates). The OAPM identifies how the state transforms under a rotation by $\pi/3$  and  it  gives  the  charge  of  the  central  singularity~\cite{2009GarciaMarchPRA}. For $m=0$ or $3$ the solution belongs to a one-dimensional representation of the point group  ${\mathcal{C}_{6\nu}}$, either $\mathcal{A}_1$,$\mathcal{A}_2$, $\mathcal{B}_1$, or $\mathcal{B}_2$ (see appendix~\ref{app:threeatoms}). For OAPM $m=\pm1$ or $\pm2$  the function belongs to the two dimensional representation of $\mathcal{E}_1$ or $\mathcal{E}_2$, respectively. Excitations due to nodes   in the radial direction are given by $n_\rho$ and the number of nodes within a sextant in the angular direction is given by $n_\theta$. 
 For finite non-zero $g$, the functions have to be continuous at the boundary between the sextants where we defined the functions  $\mathrm{F}_{m,n_\rho}(\theta)$. These boundaries are the lines at the angles $\theta_j=j\pi/3$, with $j=1,\dots,6$. The interactions are thus implemented by matching solutions using the condition
\begin{align}
\label{eq:matchingcondition}
 \frac{-1}{2\rho^ 2}\left(\left.\frac{\mathrm{d}\phi(\rho,\theta)}{\mathrm{d}\theta}\right|_{\theta_j+\epsilon}-\left.\frac{\mathrm{d}\phi(\rho,\theta)}{\mathrm{d}\theta}\right|_{\theta_j-\epsilon} \right)=\frac{g}{\sqrt{2}\rho}\phi(\rho,\theta),
\end{align}
at the angles $\theta_j$. Writing  $g$ as $\hat{g}=\sqrt{2}\rho g$,  Eq.~\eqref{eq:matchingcondition} becomes independent  of $\rho$, and therefore of $n_\rho$. As a consequence, $n_\rho$ will label the solutions. 
The functions~\eqref{eq:m,nrho,ntheta} are exact eigenfunctions when $g\to\infty$.
The condition~\eqref{eq:matchingcondition} will give two types of solutions, even-parity solutions with $n_\theta=3+6n$ and odd-parity solutions with $n_\theta=6n$, with $n=0,1,\dots$. The energies are $E=2n_\rho+n_\theta+1$. We remark that the six solutions with same values of $n_\rho$ and $n_\theta$ have the same energy independently of  $m=0,\pm1,\pm2,3$.  This strong degeneracy is broken for finite $g$. 

 \begin{figure}[t]
\includegraphics[width=0.98\columnwidth]{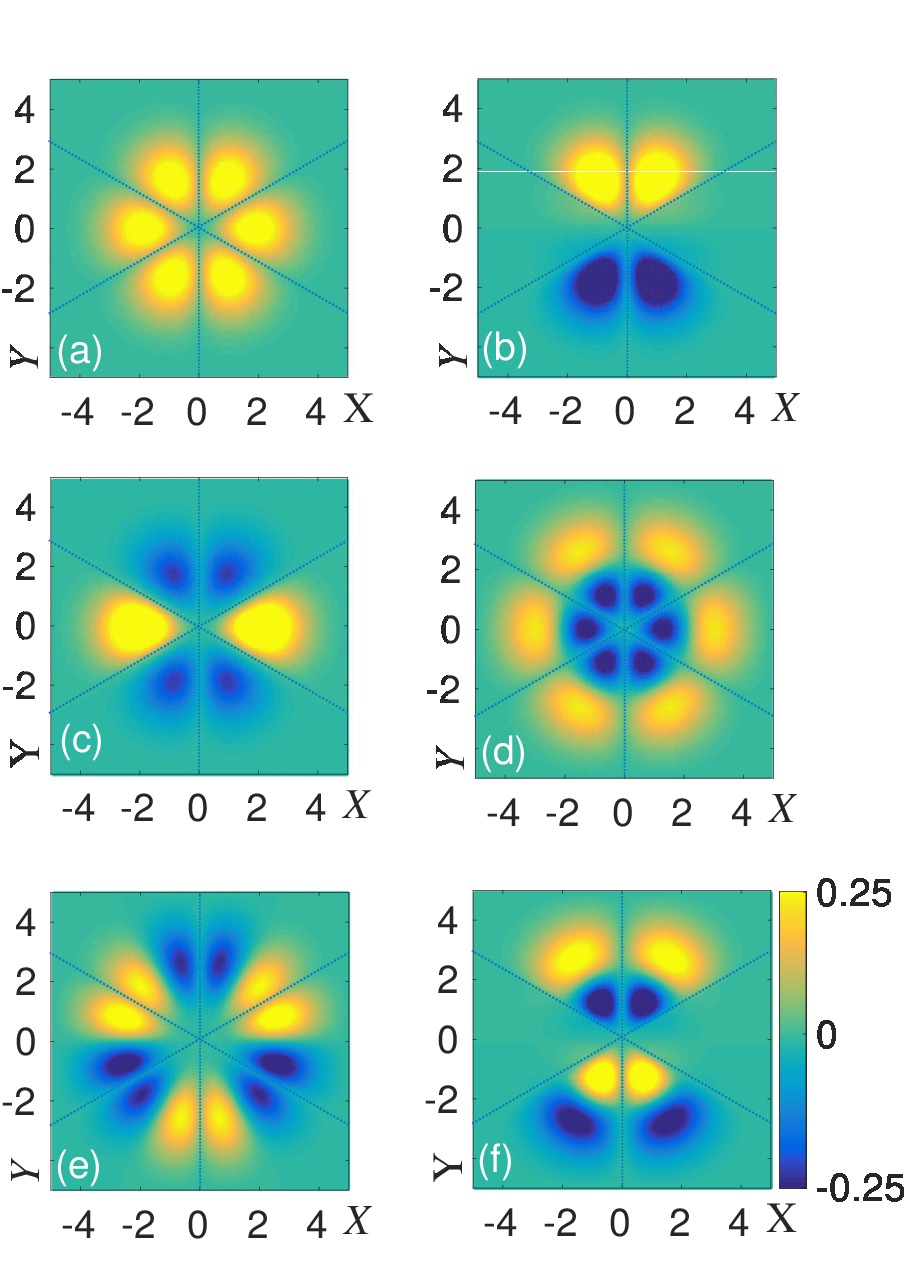}
\caption{Eigenfunctions when $g_\mathrm{AB}=g_\mathrm{A}=20$ .  (a) For finite but large $g$, the ground state  belongs to $\mathcal{A}_1$.  (b) first and (c) second excitations  for finite $g$; These states have a reduced $\mathcal{C}_{2\nu}$ symmetry. As discussed in the text,  they are obtained combinations of the two (vortex like) eigensates associated to the  $\mathcal{E}_1$ and $\mathcal{E}_2$  irreducible representation, respectively.  (d) a radial and (e) an angular excitation of the ground state; (f) a radial excitation; of the first excited state. Figure adapted  from~\cite{2019GarciaMarchArxiv}. \label{Fig_distinguishability1}}
\end{figure} 

In Fig.~\ref{Fig_distinguishability1} we show the three quasi-degenerate solutions for  zero $n_\rho$ and $n_\theta$ and different $m$ for large, but finite $g=g_\mathrm{A}= g_\mathrm{AB}$.
Not all solutions with any $m$ (which will be the six solutions with $m=0,\pm1,\pm2,3$ -- see appendix~\ref{app:threeatoms}) can be realized in a system with two indistinguishable bosons. To understand this, let us first denote the 2-cycle operation as $\hat{\sigma}_{ij}$, which is the operation that permutes particles $i$ and $j$. These permutations correspond to reflections or rotations in the $X-Y$ plane: the interchange of particles 1 and 2 correspond to a reflection with respect to the $X=0$ axis, which can also be written as a transformation in the angular variable as $\theta\to-\theta+\pi$, while $\hat{\sigma}_{23}$ and $\hat{\sigma}_{31}$ correspond to  $\theta\to-\theta+\pi/3$ and $\theta\to-\theta-\pi/3$, respectively. Let us label the two indistinguishable bosons as 1 and 2. Then all eigenfunctions have to be symmetric under $\hat{\sigma}_{12}$ for the mixture of two bosons and a distinguishable particle. Under the 2-cycle transformations $\hat{\sigma}_{23}$ and  $\hat{\sigma}_{31}$ they do not show any specified symmetry. The solutions with $m=0$ obey this condition, while the solutions with $m=3$ do not, and therefore they do not appear in the spectra.   The eigenfunction as written in~\eqref{eq:m,nrho,ntheta} with $m=\pm 1$ and $\pm2$  do not have any defined symmetry under $\hat{\sigma}_{12}$. This imposes a  new condition, and thus these wave functions have to show a reduced $\mathcal{C}_{2\nu}$ symmetry. For the two bosons plus distinguishable particle,  the combinations  $ |1,n_{\rho},n_\theta\rangle+ i |1,n_{\rho},n_\theta\rangle$ and $ |2,n_{\rho},n_\theta\rangle- i |2,n_{\rho},n_\theta\rangle$ give the solutions with the correct permutation property (which now do not belong to ${\mathcal{C}_{6\nu}}$ but to its subgroup ${\mathcal{C}_{2\nu}}$). In  Fig.~\ref{Fig_distinguishability1}, panel (a) corresponds to the solution with $m=0$, which is the one with the lowest energy in case $g$ is finite but large. Panels (b) and (c) give the first and second excited state when $g$ is large but finite. We finally note that $n_\rho$ labels radial excitations, as the one in Fig.~\ref{Fig_distinguishability1} (d) which has $m=0$, $n_\rho=1$ and $n_\theta=0$ and in Fig.~\ref{Fig_distinguishability1} (f), which is a radial excitation of the first excited state. Also, $n_\theta$ labels angular excitations on site, as the one in Fig.~\ref{Fig_distinguishability1} (e) which has $m=0$, $n_\rho=0$ and $n_\theta=1$. 

 \begin{figure}[t]
\includegraphics[width=0.98\columnwidth]{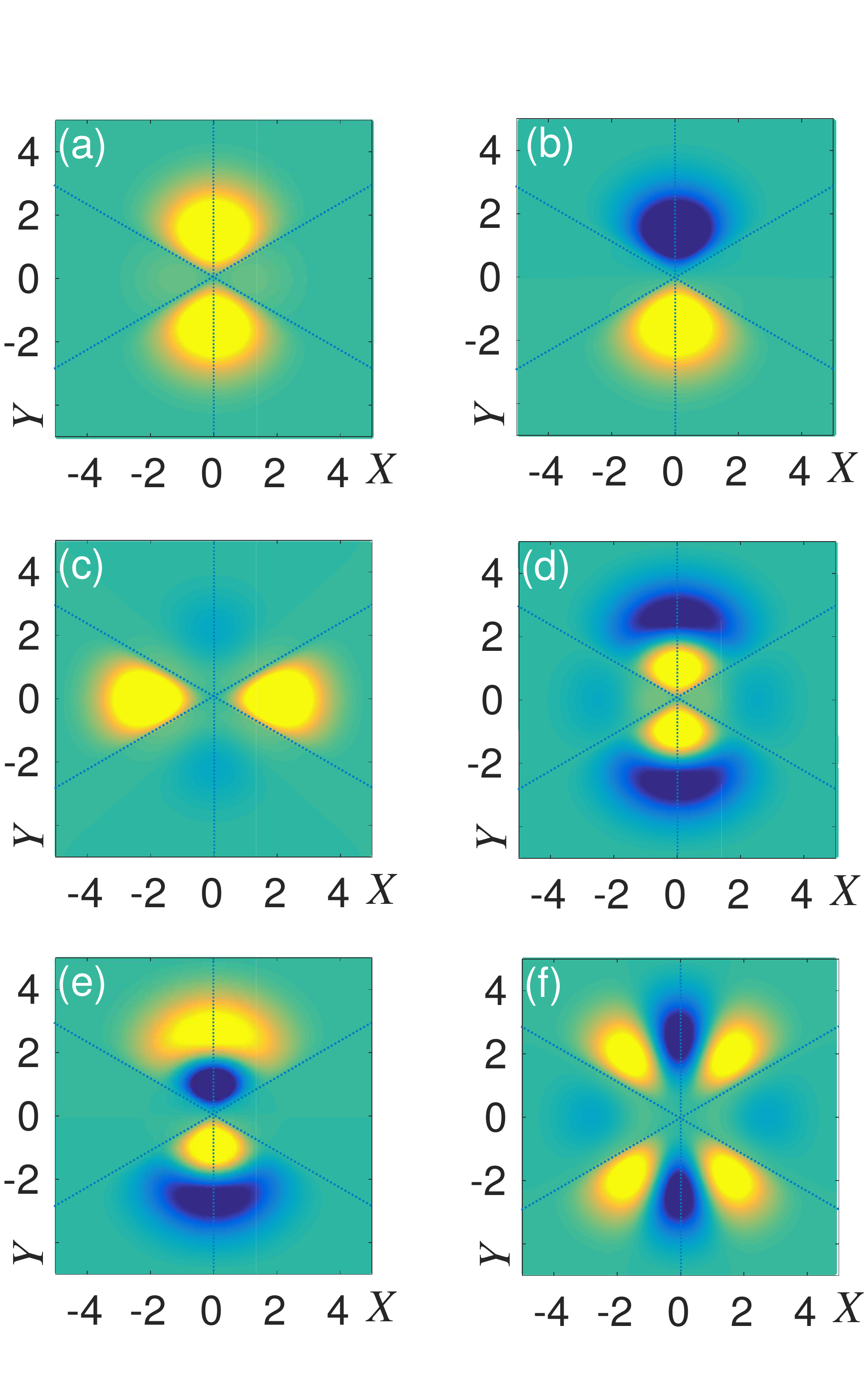}
\caption{Eigenfunctions when $g_\mathrm{AB}=20$ and $g_\mathrm{A}=0$. (a) and (b) are degenerate in energy while the first excitation is plotted in (c). All these three state have integer energy. The third excitation, plotted in (d) has half integer energy.    \label{Fig_distinguishability3}}\end{figure}

\begin{figure}
\includegraphics[width=0.94\columnwidth]{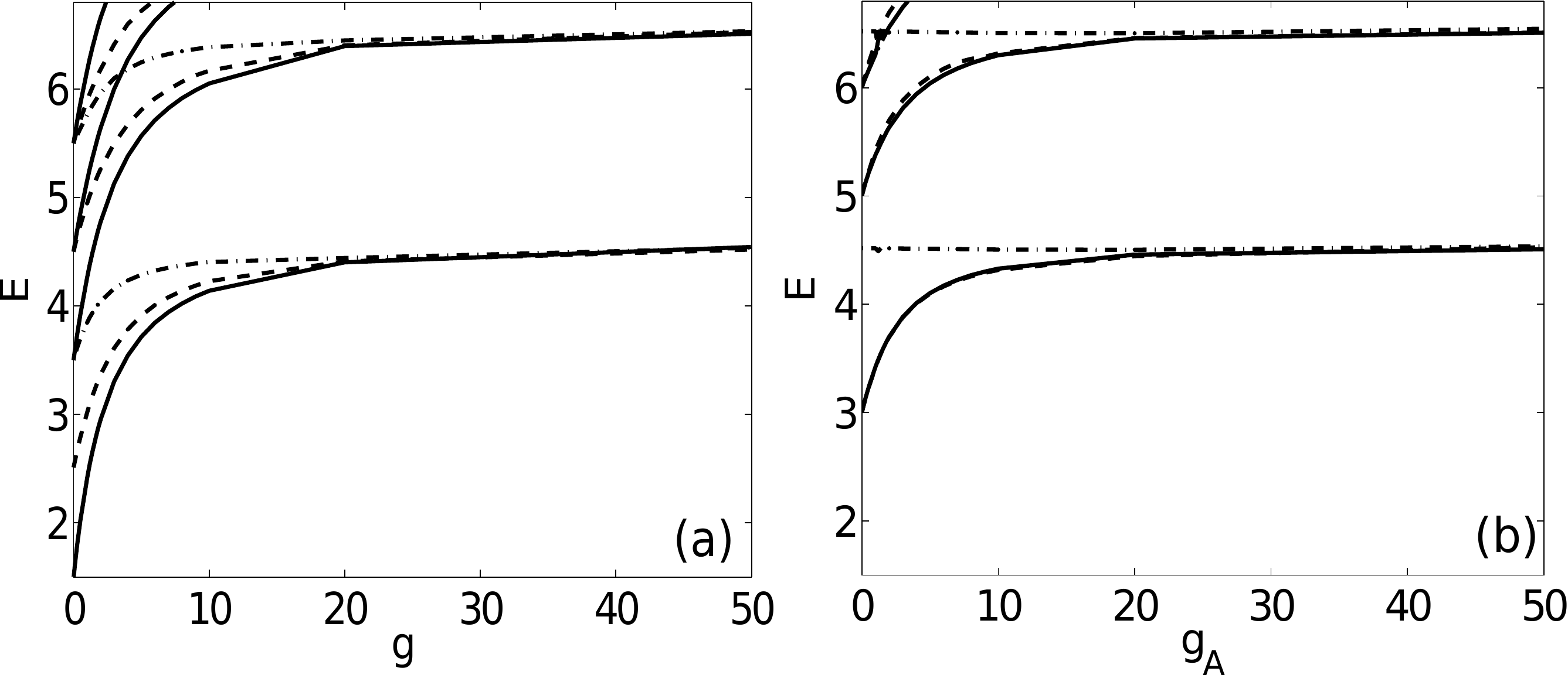}
\caption{ (a) Energy eigenspectrum as a function of $g_{\mathrm{AB}} =g_{\mathrm{A}}=g$. Different line styles are used to help identify the different excited states. (b) Energy eigenspectrum as a function of $ g_{\mathrm{A}}$ for  $g_{\mathrm{AB}} =50$. Harmonic oscillator units are used for the energies and distances. Figure adapted  from~\cite{2014GarciaMarchPRA}. Copyright (2014) by the American Physical Society.    \label{Fig_distinguishability4}}
\end{figure}

Finally, we emphasize an important set of solutions that emerge for $g_\mathrm{A}\neq g_\mathrm{AB} $. First, as   discussed in~\cite{2014ZinnerEPL} and numerically calculated in~\cite{2014GarciaMarchPRA}, when  $g_{\rm{A}}=0$ and $g_{\rm{AB}}\to\infty$ the eigenfunctions are exact again, and the conditions~\eqref{eq:matchingcondition} are also satisfied with half-integer $E$ given by $E=3/2+3n$ (for the expression of the eigenfunctions valid also in this limit see~\cite{2014ZinnerEPL}). In Fig.~\ref{Fig_distinguishability3} we plot the eigenfunctions for finite $g_\mathrm{AB}=50$ and $g_\mathrm{A}=0$. The one in panel (d) is the half-integer solutions. It is degenerated with the one in panel (e).   The ground state is doubly degenerated [panels (a) and (b)]. We plot the energies as $g_{\rm{A}}$ is increased in Fig.~\ref{Fig_distinguishability4} for various values of $g_{\rm{AB}}$. The eigenfunction plotted in  Fig.~\ref{Fig_distinguishability3} panel (f) shows that on-site radial excitations also occur here, but now due to symmetry under the  $\hat{\sigma}_{12}$, only those with odd number of nodes are excited.     For large $g_{\rm{AB}}$ and varying $g_{\rm{A}}$ we find that there is a non-interacting solution, which is the one in panel (c).  We finally note that we do not discuss here the limit TG-BEC [ $g_{\rm{AB}}=0$ and $g_{\rm{A}}\to\infty$] because in this case the third particle does not play a relevant role, and the system behaves as the two-atom system (discussed in~\cite{1998BuschFoundPhys}). 
For further details in case $g_\mathrm{A}\neq g_\mathrm{AB} $ we refer appendix~\ref{app:threeatoms} and Refs.~\cite{2012HarshmanPRA,2014ZinnerEPL,2014GarciaMarchPRA}. 

\subsection{A fruitful example: four atoms}
\label{sec:four}

A mixture of four atoms represents the minimal system in which the eight limits discussed above are meaningful, as both the PS and the TG-BEC limits occur in this case. The PS limit corresponds to $g_{\rm{AB}}\to\infty$ and $g_{\rm{A }}$ (or $g_{\rm{B }} $) tend to infinity with $g_{\rm{B}}$ (or  $g_{\rm{A}}$) vanishing (see Fig.~\ref{Fig:GarciaMarch_Cube}). Let us show how this simple mixture with $N_{\rm{A }}=N_{\rm{B }}=2$  atoms unravels a great amount of physics, which include not only the usual fermionization when $g_{\rm{B}}=g_{\rm{A}}$ is increased~\cite{2009HaoEPJD}, but also  sharp crossovers between the different limits~\cite{2013GarciaMarchPRA,2014GarciaMarchNJP}, the first hints of quantum magnetism~\cite{2016DehkharghaniSciRep} and a very rich energy spectra~\cite{2018PyzhNJP}. 

The first meaningful limit of {\it composite fermionization} occurs when $g_{\rm{AB}}\to\infty$ with vanishing $g_{\rm{A }}$ and $g_{\rm{B }} $. In such case, the two-particle density profile of opposite-component particles $n_2^{(AB)}(x,y)$ identically vanishes along with the diagonal $x=y$ due to the infinitely strong repulsions. Contrarily, the two-particle densities between particles belonging of the same species $n^{(A)}_2(x_1,x_2)$ and $n^{(B)}_2(y_1,y_2)$ do not vanish along the diagonal. In  Fig.~\ref{Fig:Smelcher} we illustrate this behavior. It is also useful to calculate the single-particle reduced density matrices $\rho_\mathrm{A}(x,x')$ and $\rho_\mathrm{B}(y,y')$ for each component which is presented in Fig.~\ref{Fig:Smelcher}. The diagonalization of them gives the natural orbitals occupations, $\lambda^i_{\rm{A,B}}$, which are normalized between 0 and 1. The largest value, $\lambda^0_{\rm{A,B}}$ shows the degree of condensation in each component. In this case, for both species, one obtains the same number $\lambda^0$, which is large but significantly smaller than 1.  
The two-particle densities and the single-particle reduced density matrices give the following information: whenever a particle of one species is detected in one position with, e.g., $x>0$ all particles of the same species will be found located at the left peak, whilst the atoms of the other species will be found in the right peak. So though according to the density profiles it may seem that in the CF limit there is the overlap of densities. In practice, this limit corresponds also to phase separation.  

\begin{figure}
\includegraphics[width=0.94\columnwidth]{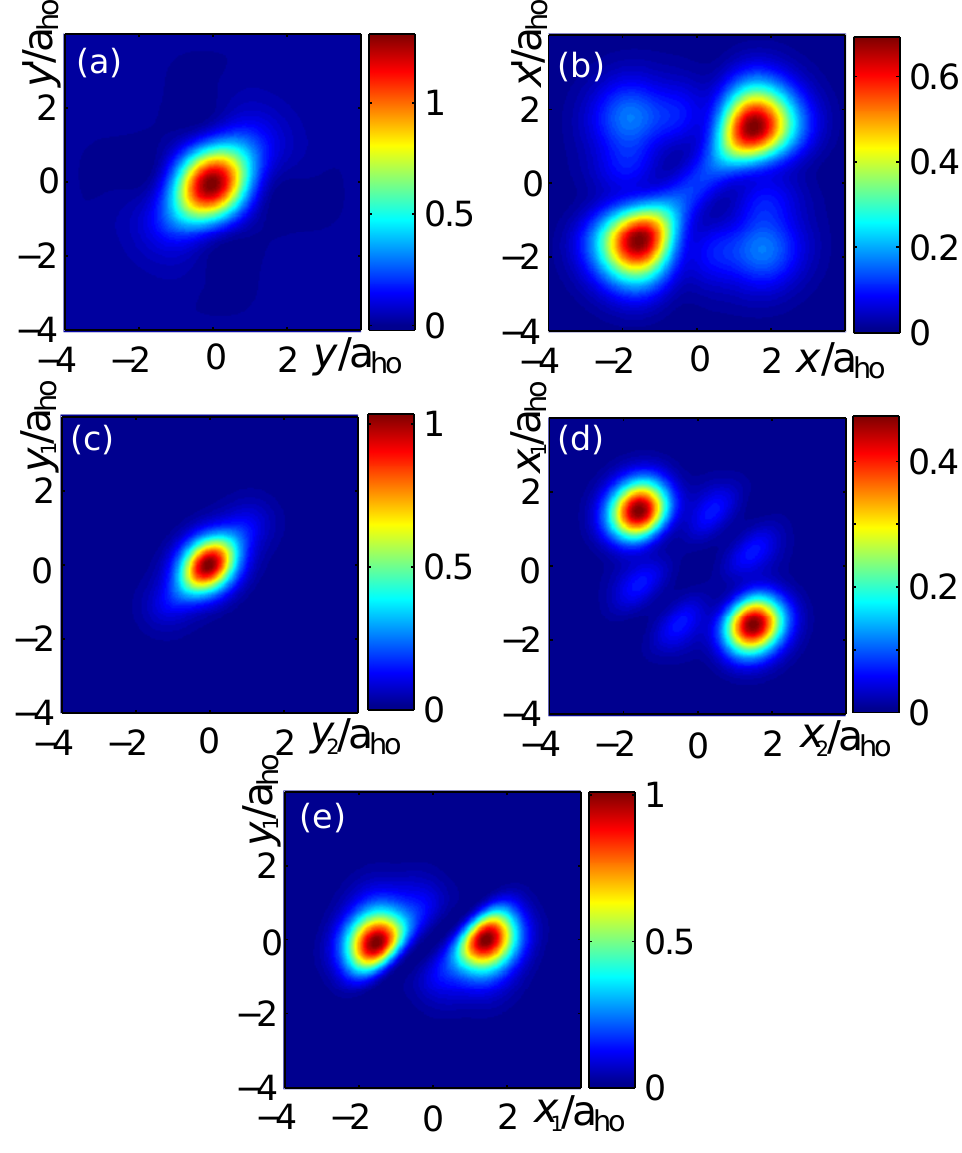}
\caption{ (a)-(b) Single-particle reduced density matrix (for species A and B, respectively) in a mixture with $N_{\rm{A}}=N_{\rm{B}}= 2$ bosons in the phase separation limit,   for $g_{\rm{A}}=g_{\rm{AB}}=20\hbar\omega a_{\rm{ho}}$ and $g_{\rm{B}}=0$.  (c)-(d) Two-particle density profile matrix for two atoms in species A and B, respectively.  (e) Two-particle density profile for one atom  in species A and one in B.  Figure adapted  from~\cite{2013GarciaMarchPRA}. Copyright (2013) by the American Physical Society.  \label{Fig_PS}}
\end{figure}

A very different situation occurs in the {\it phase separation limit} when $g_\mathrm{AB}\to\infty$, and e.g.  $g_\mathrm{A}\to\infty$, with $g_\mathrm{B}=0$. In this particular case, not only the two-particle density profile $n_2^{(AB)}(x,y)$ vanishes along the diagonal $x=y$ but also other two-particle density (related to a strong intra-component interaction $g_\mathrm{A}$) also reveal this property (see Fig.~\ref{Fig_PS}). The one for species B does not have a zero at $y_1=y_2$. The density profiles show density separation: species B occupies the center of the trap, whiles species A is moved to the edges. A key question here is that the diagonalization of the single-particle reduced density matrix for both species shows a value closer to 1 for species B and a number closer to 1/2 for species A. This means that species B, which is located in the center of the trap, is well condensed. Simultaneously, species A occupies the edges of the B atoms cloud and has very little coherence between the halves. 

In the BEC-TG limit, one has, e.g.,  $g_\mathrm{A }\to\infty$ and  $g_\mathrm{B}=g_\mathrm{AB}=0$. This case is trivial, as for A species the system behaves as a TG system and a corresponding occupation of the most occupied natural orbital of $~0.4$. For species B the atoms behave as an ideal gas, with a Gaussian density profile and an occupation of the lowest natural orbit of 1. The TG-TG limit corresponds trivially to a mixture of two independent two-atom TG gases; finally, the FF limit implies that both species fermionize as a single  4 atom species. 

\begin{figure}
\includegraphics[width=0.94\columnwidth]{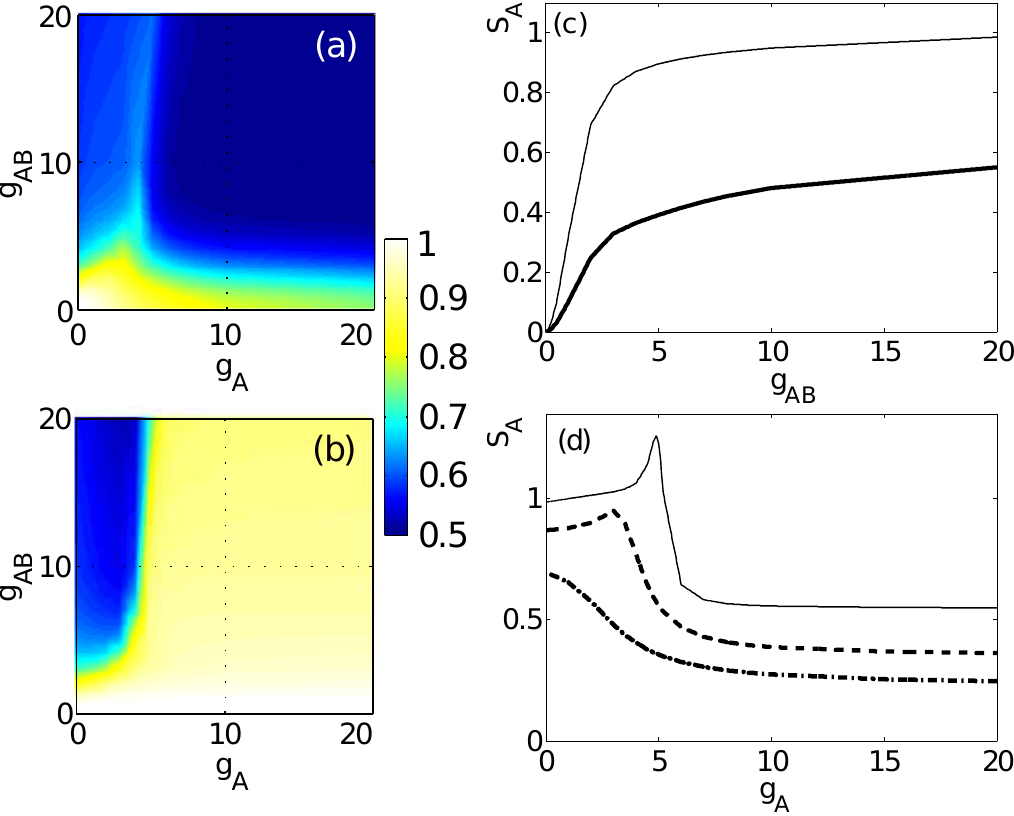}
\caption{(a)-(b) the largest occupation number of the natural orbital (for spieces A and B, respectively) in a mixture with $N_\mathrm{A}=N_\mathrm{B}=2$ bosons as a function of $g_\mathrm{AB}$ and $g_\mathrm{A}$, with $g_\mathrm{B}=0$. (c) the von Neumann entropy as a function of $g_\mathrm{AB}$ for $g_\mathrm{A}=0$ (thick line)  and $g_\mathrm{A}$ large (thin line). (d) the von Neumann entropy as a
function of $g_\mathrm{A}$ for $g_\mathrm{AB}=2$, 4, 20 (dash-dotted, dashed, and solid line, respectively). Figure adapted  from~\cite{2014GarciaMarchNJP}. Copyright (2014) by the IOP Publishing.   \label{Fig_CF_PS}}
\end{figure}

\begin{figure}
\includegraphics[width=0.8\columnwidth]{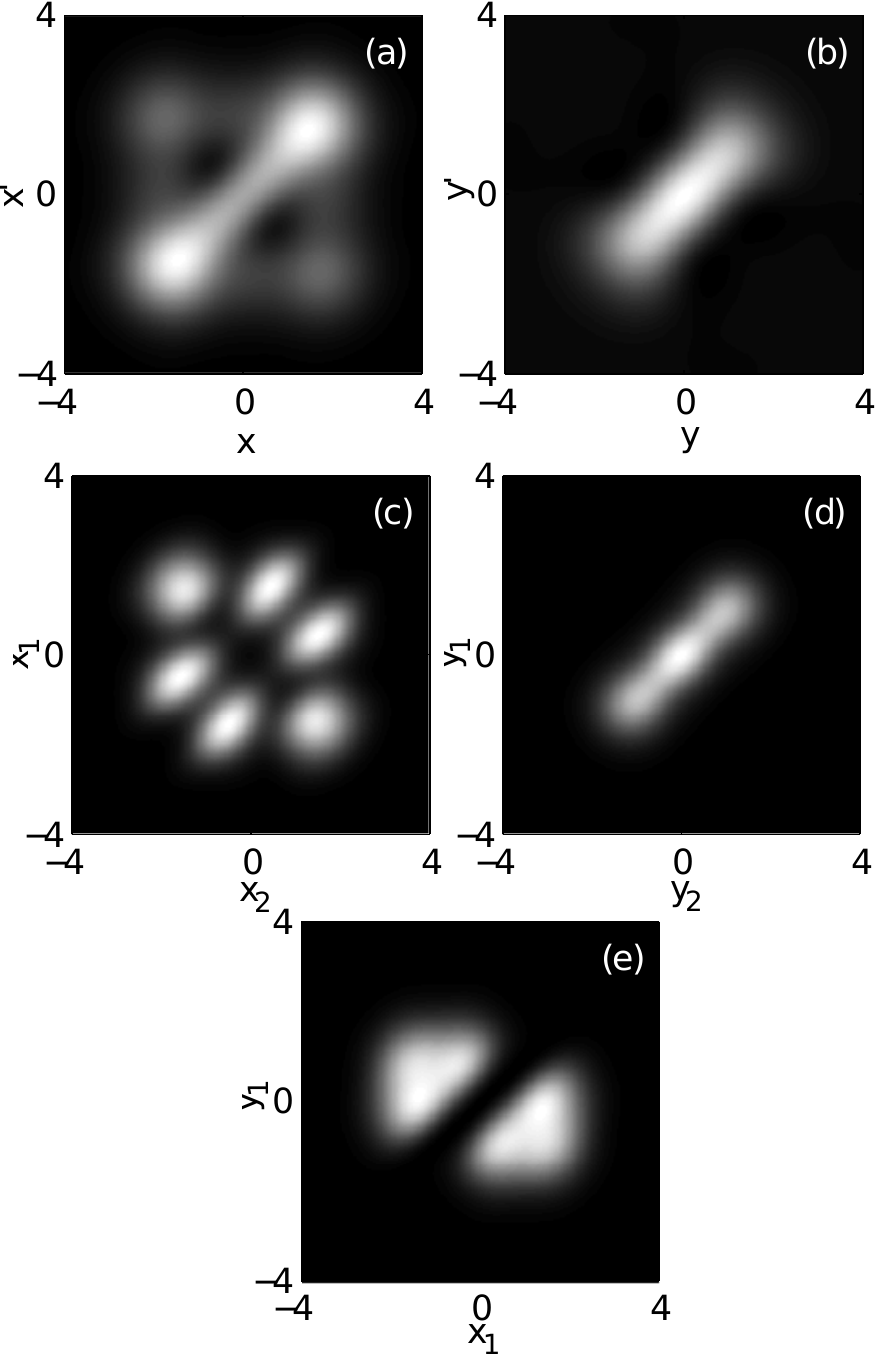}
\caption{ (a)-(b) Single-particle reduced density matrix \eqref{Eq:OBDM} for species A and B, respectively, in a mixture with $N_{\rm{A}}=N_{\rm{B}}= 2$ bosons   for $g_{\rm{A}}=4\hbar\omega a_{\rm{ho}}$ $g_{\rm{AB}}=20\hbar\omega a_{\rm{ho}}$ and $g_{\rm{B}}=0$, that is, just before the crossover. (c)-(d)  Two-particle density profile \eqref{Twoparticledensity} for two atoms in species A and B, respectively. (e) Two-particle reduced density matrix for one atom  in species A and one in B. Figure adapted  from~\cite{2014GarciaMarchNJP}. Copyright (2014) by the IOP Publishing.  \label{Fig_before_PS}} 
\end{figure}

With this description of very different limits, the transitions among them as the coupling constants are changed is a very interesting question. This has been attempted for the plane delimited by the BEC-BEC, TG-BEC, CF and PS limits, leaving aside what happens within the cube, towards the FF limit. In Fig.~\ref{Fig_CF_PS} we plot the largest occupation of a natural orbital. As seen there is a region in which there is a sharp crossover between limits, with an abrupt increase of the occupation. This transition was first discussed in~\cite{2013GarciaMarchPRA}. Just before the transition, the two-particle and single-particle reduced density matrices show large off-diagonal terms (see Fig.~\ref{Fig_before_PS}). This indicates large correlations between both species. Indeed to illustrate this one can rely on an inter-component measure of entanglement, the von Neumann entropy defined as
\begin{equation}
 S=-\mbox{Tr}[\rho_{\rm A}\mbox{log}_2\rho_{\rm A}]=\sum_j -\lambda_j \mbox{log}_2\lambda_j.
\end{equation}
where the reduced density matrix of the component $A$ is obtained by tracing-out all degrees of freedom of opposite component from the density matrix of the system, $\rho_\mathrm{A}=\mathrm{Tr}_\mathrm{B}\left[|\phi\rangle\langle\phi|\right]$ . Real positive numbers $\lambda_j$ are the eigenvalues of the reduced matrix $\rho_\mathrm{A}$.

This quantity is plotted in Fig.~\ref{Fig_CF_PS}. It shows a peak at the region where the crossover occurs, showing that indeed large correlations between both species occur. Finally, we emphasize that in~\cite{2014GarciaMarchNJP} the authors showed that the two-parameter $(\eta_1,\eta_2)$ ansatz of the form
\begin{multline}
\label{eq:JastrowCFPS}
 \phi(\boldsymbol{r}_\mathrm{A},\boldsymbol{r}_\mathrm{B}|\eta_1,\eta_2)\propto \\\left(\prod_{i=1}^{N_\mathrm{A}}\prod_{j=i+1}^{N_\mathrm{A}}\!|x_i-x_j-\eta_1|\prod_{i=1}^{N_\mathrm{A}}\prod_{j=1}^{N_\mathrm{B}}\!|x_i-y_j-\eta_2|\right)\\\times\mathrm{e}^{-\sum_i x_i^2/2} \mathrm{e}^{-\sum_j y_j^2/2}. 
\end{multline} 
reproduces faithfully the numerical results shown in Fig.~\ref{Fig_before_PS}, that is the solution  just before the crossover. 

The spectra of excitations for this system at all possible limits of interactions and as the coupling constant are varied is discussed in~\cite{2018PyzhNJP}. To understand the energy spectra it is very convenient to perform a transformation of coordinates to the center of mass $R_{CM}$, the relative center-of-mass coordinate $R_{AB}$, and the relative coordinates for each component $r_{A,B}$:
\begin{subequations}
\begin{align}
R_{\mathrm{CM}}&=(x_1+x_2+y_1+y_2)/4, \\
R_{\mathrm{AB}}&=(x_1+x_2-y_1-y_2)/4, \\
r_{\mathrm{A}}&=x_1-x_2, \\
r_{\mathrm{B}}&=y_1-y_2.
\end{align}
\end{subequations}
With this choice of variables, as usual, the center-of-mass coordinate $R_{\mathrm{CM}}$  decouples from the rest, as expected, and then we reduce dimensionality. The Hamiltonian associated with this variable is diagonalized trivially with the single-particle solutions of the harmonic oscillator. Moreover, in this framework, one can easily identify different symmetries of the system. First, the parity with respect to the $R_{\rm{AB}}$ coordinate determines the total parity
of the eigenstates. Second, for $g_\mathrm{A}=g_\mathrm{B}$ the Hamiltonian is invariant under 
$r_\mathrm{A} \leftrightarrow r_\mathrm{B}$ exchange, called  the ${\cal S}_{r}$ symmetry. Both transformations correspond to certain spatial transformations in the four-dimensional coordinate space, as in the three particle case (see appendix~\ref{app:threeatoms}). In Fig.~\ref{Fig_Smelcher_2} we show the energy spectra for this system. In the BEC-CF transition shown in Fig.~\ref{Fig_Smelcher_2}a one observes how the ground state becomes doubly degenerate. The two states have even and odd parity with respect to $R_{\rm{AB}}$. The even parity corresponds to the one shown in Fig.~\ref{Fig:Smelcher}. This two-fold degeneracy reflects the two possible configurations discussed above:  (A in left -- B in right) or (A in right -- B in left). Also, it is important to note that there are non-integer eigenvalues in the CF limit, as in the case of three atoms.  

The energy spectra along the transition between the TG-TG and PS limits are plotted in Fig.~\ref{Fig_Smelcher_2}c. The TG-TG limit corresponds to the mixture of two independent two-atom TG gases. In this case, the wave functions can be expressed analytically, they have integer-valued eigenenergies 
$E_{0,k,l,m}^{(0)} \approx k+2l+2m+4$ with equal 
spacings, and have the same degree of degeneracy
as in the non-interacting case (see spectrum for $ g_{\mathrm{AB}}=0$ in Fig.~\ref{Fig_Smelcher_2}a).  In the FF limit, when all coupling constants tend to infinity, the eigenfunctions are again analytic and the system resembles a non-interacting ensemble of four fermions with the ground state energy $\hbar\omega N^2 /2 = 8\hbar\omega$~\cite{2007GirardeauPRL}. Importantly, one has to take into account that there are two bosons in each component which are indistinguishable, but distinguishable between components. Due to this the degeneracy is  $N!/(N_\mathrm{A} ! N_\mathrm{B} !)=6$-fold. A detailed study on the degeneracies in this limit can be found in~\cite{2008DeuretzbacherPRL,2011FangPRA}, where the latter corresponds to a Bose-Fermi mixture, but the techniques apply also to this case. Finally, a detailed study where one of the species is non-interacting or strongly interacting is provided in~\cite{2016DehkharghaniJPB}, where special care is given to explain the coordinate ordering of particles in the different resulting wave functions in a relative motion expressed in hyperspherical coordinates.

\subsection{Mixtures with several atoms}
\label{sec:larger}

\begin{figure}
\includegraphics[width=0.94\columnwidth]{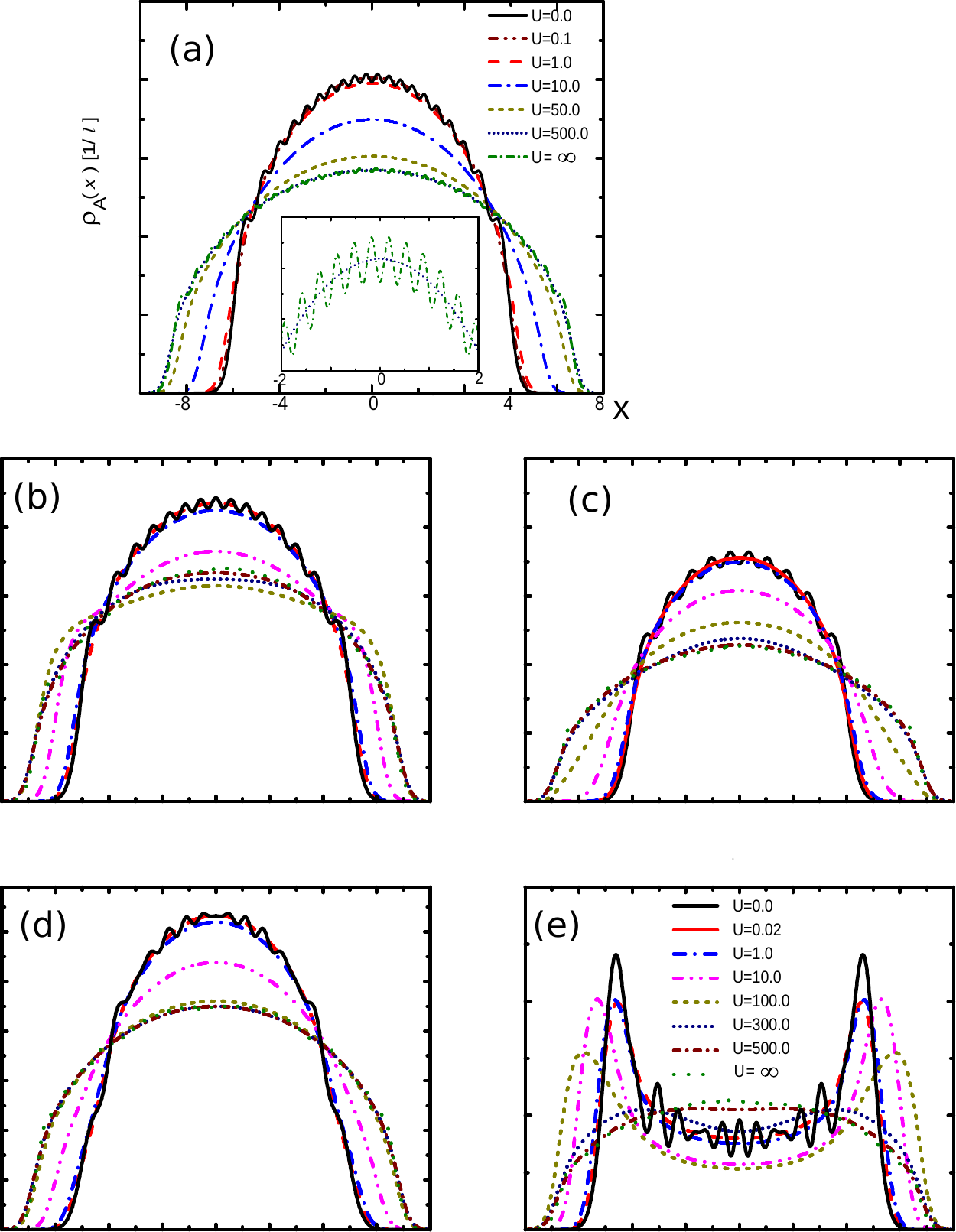}
\caption{Ground-state single-particle density profiles in the transition from TG-TG  to FF limits. (a) Density profiles $\rho_{\rm{A}}=\rho_{\rm{B}}$  as the interaction is increased for the unpolarized case,  $N_{\rm{A}}=N_{\rm{B}}=20$. (b)-(c) Density profiles $\rho_{\rm{A}}$ and $\rho_{\rm{B}}$, for each species, as the interaction is increased for a polarized case,  $N_{\rm{A}}=15$ and $N_{\rm{B}}=10$. To illustrate density separation, in (d) and (e) we plot $\rho_{\rm{A}}+\rho_{\rm{B}}$  and $\rho_{\rm{A}}-\rho_{\rm{B}}$, respectively. Figure adapted  from~\cite{2009HaoPRA}. Copyright (2009) by the American Physical Society. \label{2009HaoPRA}}
\end{figure}

As explained in the previous subsections~\ref{sec:three} and~\ref{sec:four}, a three- and four-boson mixtures allow one to explore the eight limits for the inter- and intra-species interactions (see Fig.~\ref{Fig:GarciaMarch_Cube}). Let us now discuss several theoretical works which explore these limits and the transitions between them for larger systems. 

In~\cite{2009HaoPRA} the transition from the TG-TG limit to the FF limit was studied, with a density functional approach, valid for weak harmonic traps (small trapping frequency). They consider two cases, an unpolarized mixture with $N_{\rm{A}}=N_{\rm{B}}=20$ and a polarized one, $N_{\rm{A}}=15$ and $N_{\rm{B}}=10$. For the unpolarized case, they observe a smooth transition between a mixture of two TG gases with no correlations among them, and a FF gas, with $N=40$ peaks, as corresponds with Girardeau-Minguzzi's prediction~\cite{2007GirardeauPRL}, see Fig.~\ref{2009HaoPRA}a. For the polarized case (see Fig.~\ref{2009HaoPRA}b-e) an important behavior is observed: while in the TG-TG and FF limits the behavior is as expected (two independent TGs and a FF gas with $N_{\rm{A}}+N_{\rm{B}}$ peaks), in the transition (intermediate inter-species interactions) signatures of density separation occur, see Fig.~\ref{2009HaoPRA}e. This is a {\it shell structure}. This is not evident from the plots of $\rho_{\rm{A}}$ and $\rho_{\rm{B}}$ independently (panels (b) and (c) in  Fig.~\ref{2009HaoPRA}). Therefore, this figure also presents the total density   $\rho_{\rm{A}}+\rho_{\rm{B}}$  and  the spin density distribution defined as the density difference $\rho_{\rm{A}}-\rho_{\rm{B}}$.  This latter spin  density shows, for weakly and intermediate  interactions, two peaks at the edges of the trap. So  there  is a non-polarized mixture in the centre of the trap, which is surrounded by the majority component. For larger inter-component interactions the peaks diminish and eventually disappear, and the  spin density profile is also flat, as the total density. 
This is a further indication that, while the limits are well described, the transition between them is still intriguing in many instances, particularly in polarized (imbalanced) cases. 

\begin{figure}
\includegraphics[width=0.94\columnwidth]{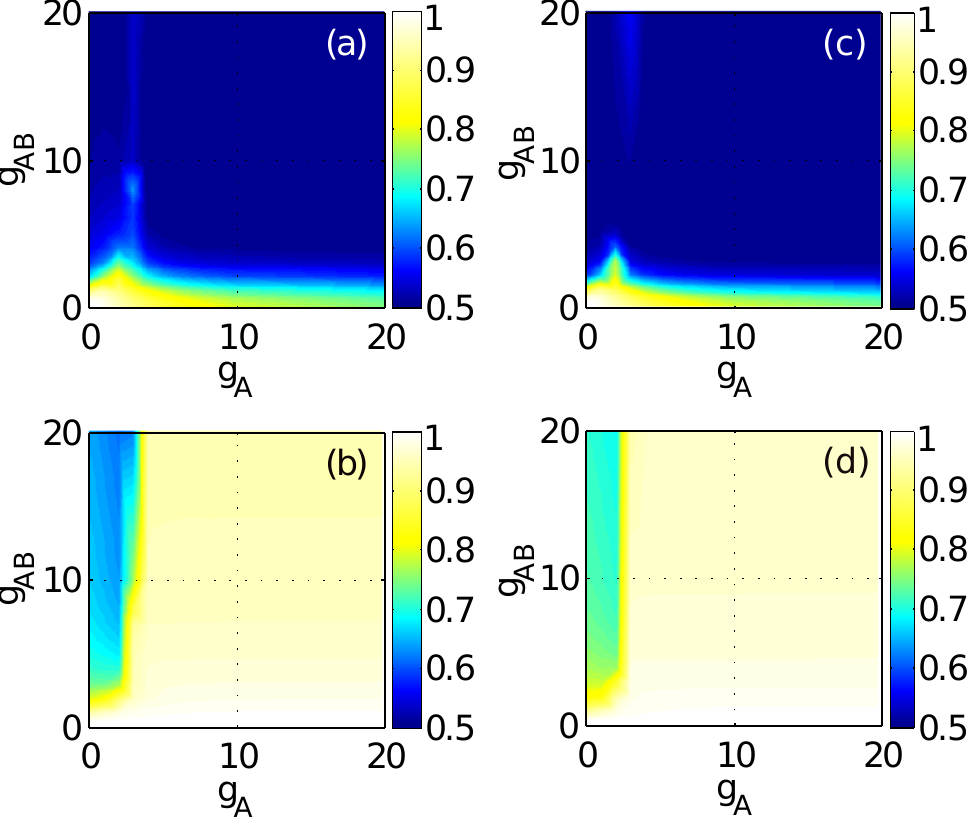}
\caption{ (a)-(b) The largest occupation numbers of the natural orbitals (A and B species, respectively) for $N_\mathrm{A}=2$  and $N_\mathrm{B}=3$ as a function of $g_\mathrm{AB}$ and   $g_\mathrm{A}$, with $g_\mathrm{B}=0$.  (c)-(d) analogue plots for the system with  $N_\mathrm{B}=4$. Figure adapted  from~\cite{2014GarciaMarchNJP}. Copyright (2014) by the IOP Publishing. \label{Fig_2014GarciaMarchNJP}}
\end{figure}

Another example occurs on the plane defined by the BEC-BEC, BEC-TG, CF, and PS limits. We discussed in section~\ref{sec:four} the unpolarized case, $N_{\rm{A}}=N_{\rm{B}}=2$~\cite{2014GarciaMarchNJP}. As shown in Fig.~\ref{Fig_CF_PS}(a) and (b) the interior of this plane shows a sharp crossover with a non-trivial dependence on the interactions (see~\cite{2013GarciaMarchPRA}). For imbalanced systems, the area in which composite fermionization persists, before the sharp crossover to phase-separation shrinks as $N_{\rm{B}}$ is increased (see Fig.~\ref{Fig_2014GarciaMarchNJP}).  For the limit when $g_{\rm{AB}} $ is large, and  $g_{\rm{A}} $ is tuned, as  $N_{\rm{B}}$ is increased the transition becomes less abrupt (see bottom panels in Fig.~\ref{Fig_2014GarciaMarchNJP}). 

Importantly, another effect occurs in the transition between BEC-TG and PS limits. For the unpolarized case, as $g_{\rm{AB}}$ is increased, keeping $g_{\rm{A}} $ large and $g_{\rm{B}}$ close to $0$, there is a smooth transition, through which the A cloud reduces its orbital occupations, due to the fact that it density separates in two pieces without coherence between them, while the B cloud occupies the center of the trap (see Fig.\ref{Fig_CF_PS}). For the polarized case, as $N_{\rm{B}}$ is increased, this transition occurs for smaller and smaller values of $g_{\rm{AB}}$ (see Fig.~\ref{Fig_2014GarciaMarchNJP}, upper panels). Indeed, as discussed in~\cite{2013GarciaMarchPRAb}, for large values of  $N_{\rm{B}}$, the transition occurs for very small values of  $g_{\rm{AB}}$. Also, the fragmented cloud of A atoms from two non-coherent TG gases, one at each side of the B atoms and each with  $N_{\rm{A}}/2$ atoms.  Indeed, for this limit and large values of $N_{\rm{B}}$, the system density admits a description in terms of a system  of two coupled non-linear Kolomeisky-like equations:
\begin{subequations}\label{KGPE}
\begin{align}
  \mu_{\rm{A}}\phi_{\rm{A}}=& -\frac{\hbar^2}{2m}\phi_{\rm{A}}^{''}
                 +\left[V(x)+\tilde g_{\rm{A}}|\phi_{\rm{A}}|^4+g_{\rm{AB}}|\phi_{\rm{B}}|^p\right]\phi_{\rm{A}}\\
  \mu_{\rm{B}}\phi_{\rm{B}}=& -\frac{\hbar^2}{2m}\phi_{\rm{B}}^{''}
                 +\left[V(x)+g_{\rm{B}}|\phi_{\rm{B}}|^2+g_{\rm{AB}}|\phi_{\rm{A}}|^p\right]\phi_{\rm{B}},
\end{align}
\end{subequations}
where $\tilde g_{\rm{A}}=\left(\pi\hbar\right)^{2}/2m$ is independent on $g_\mathrm{A}$ (see \cite{2013GarciaMarchPRAb} and \cite{2000KolomeiskyPRL}).  This system of equations has a non-linearity which is not cubic, as in the GPE equation, but quintic. The cross term modeling the inter-species interactions can have a cubic or quintic power ($p=2$ or 4), depending on the regime of interactions (small or large  $g_{\rm{AB}}$). In~\cite{2013GarciaMarchPRAb} it was shown good agreement between the density profile calculated with exact diagonalization and that calculated with Eqs.~\eqref{KGPE}. We note that when all exponents are 2 and all coupling constants are equal, these equations resemble  Manakov Equations, which are well known in the context of nonlinear optics, and are solvable via inverse scattering transform~\cite{1974ManakovJETP}.  For the case in which all coupling constants are large, Tanatar and Erkan~\cite{2000TanatarPRA} derived the following set of equations:
\begin{subequations}\label{KGPElarge}
\begin{align}
  \mu_{\rm{A}}\phi_{\rm{A}}=& -\frac{\hbar^2}{2m}\phi_{\rm{A}}^{''}
                 +\left[V(x)+\tilde g|\phi_{\rm{A}}|^4\right.\nonumber\\
                 & +\left. \bar g |\phi_{\rm{A}}|^2|\phi_{\rm{B}}|^2 + \tilde g_{\rm{AB}}|\phi_{\rm{B}}|^4\right]\phi_{\rm{A}},\\
  \mu_{\rm{B}}\phi_{\rm{B}}=& -\frac{\hbar^2}{2m}\phi_{\rm{B}}^{''}
                 +\left[V(x)+\tilde g |\phi_{\rm{B}}|^4\right.\nonumber\\
                 & +\left. \bar g |\phi_{\rm{A}}|^2|\phi_{\rm{B}}|^2 +\tilde g_{\rm{AB}}|\phi_{\rm{A}}|^4\right]\phi_{\rm{B}},
\end{align}
\end{subequations}
with $\tilde g=\left(\pi\hbar\right)^{2}/2m$, $\bar g=\left(\pi\hbar\right)^{2}/3m$ and $\tilde g_{\rm{AB}}=\left(\pi\hbar\right)^{2}/6m$. It is important to emphasize that Eqs.~\eqref{KGPE} and~\eqref{KGPElarge} can be used only to calculate density profiles. Importantly, their time dependent versions (since there are two conserved quantities, {\it i.e.}, number of atoms in each condensate, naively one writes $\phi_{\sigma}(x,t)=\phi_{\sigma}(x)\exp[-i\mu_\sigma t/\hbar]$) cannot be used to calculate dynamical properties of the system. As shown in~\cite{2000GirardeauPRL} such a utilization of the time dependent single-component Kolomeisky equation is incorrect. It was illustrated with the interference between split condensates that are recombined. The Kolomeisky approach predicts strong interference fringes, while in fact they are very shallow as shown within the exact many-body treatment. The regime of validity of Eqs.~\eqref{KGPE} (also applicable to~\eqref{KGPElarge}) was discussed in a comment by Girardeau~\cite{2000GirardeauComment}.

 For the polarized case, the limiting situation is when one of the species has only one atom. This turns to be an impurity problem, linked to the Bose polaron problem in the large $N_{\rm{B}}$ atom limit, with $g_{\rm{B}}$ and $g_{\rm{AB}}$ small. Here, we consider a few-atom limit, often with strong interactions in a 1D trap. The smallest system of this type is the three-atom case discussed in Section~\ref{sec:three}. It is clear that this system can only occur in the BEC-BEC, BEC-TG, CP, and FF limits. An ingredient which is often included in this system is the mass imbalance between the impurity atom and the bosons. With a pair-correlated wave function approach, it has been shown that through the transition from non-interacting case (BEC-BEC, $g_{\rm{B}}=g_{\rm{AB}}=0$) to the CP limit ($g_{\rm{B}}=0 $ and $g_{\rm{AB}}\to\infty$), the impurity particle A tends to localize in the edges of the majority species B, when $N_{\rm{B}}=2,3$  atoms~\cite{2016BarfknechtJPB}. Then, it has been shown that one can force the impurity particle to again localize in the center of the trap for more massive impurities when $N_{\rm{B}}=2$ to 4  atoms~\cite{2016GarciaMarchJPB} (for the 3+1 system see also appendix in~\cite{2016DehkharghaniJPB}). The limit when $g_{\rm{B}}\to\infty$ has been studied in~\cite{2017VolosnievFBS}, showing that when also $g_{\rm{AB}}\to\infty$ the particles cannot exchange their initial ordering, while when  $g_{\rm{AB}}$ is large but not infinity, the system maps into the spin chain Hamiltonian.   Then, it is shown that when the mass of the impurity is larger than the majority species mass, the impurity particle tends to localize in the center of the trap~\cite{2017VolosnievFBS}, as in the case when $g_{\rm{B}}=0$. We note that a large body of theory is based on the effective spin-chain Hamiltonian description of a mixture of few bosons. We discuss this approach in depth in section~\ref{Sec:Fermions}, together with the fermionic case. The localization of the impurity in the center of the trap when it is more massive has also been studied numerically for up to  $N_{\rm{B}}=8$ atoms in~\cite{2015DehkharghaniPRA}, both when $g_{\rm{B}}=0$ and in the transition to $g_{\rm{B}}$ large. Indeed when $m_{\rm{A}}\to\infty$ the impurity atom behaves as a delta function in the center of the trap. This behavior of the impurity also has effects on the coherence properties of the majority atoms, as the largest occupation of a natural orbital for B decreases as $m_{\rm{A}}$ is increased (see~\cite{2016GarciaMarchJPB}). For large enough values of the mass of the impurity ($m_{\rm{A}}\to\infty$), the bosons are fragmented into two incoherent halves. The opposite limit, when the impurity atom is light and also for three and four majority atoms is treated in~\cite{2014MehtaPRA}.  The problem of an impurity in strongly interacting regimes connects with the study of the Bose polaron (see recent experiments~\cite{2012CataniPRA,2011WillPRL}) beyond the Fr\"ochlich Hamiltonian (see e.g.~\cite{2017GrusdtNJP}). Finally, we note that this problem has attracted recently a lot of interest, also in other external trapping potentials, such as double wells or optical lattices~\cite{2018BarfknechtNJP,2018KeilerNJP}  as well as in the case of attractive impurity \cite{2015MehtaPRA}.  In general, when the mass imbalance is present and for any number of particles,  certain arrangements of unequal masses make the problem solvable for limits in unsolvable in the equal mass case~\cite{2017HarshmanPRX}. Some additional particular solvable cases are identified in~\cite{2017HarshmanPRX} for systems up to five particles.

\subsection{Spinor Bose mixtures}
\label{sec:Spinor}

Let us briefly review some results related to bosonic mixtures of atoms having internal degrees of freedom -- spinor Bose gases. In the simplest case of spin-1 bosons, the first quantized Hamiltonian describing $N$ particles at zero magnetic field in one dimension reads~\cite{2008DeuretzbacherPRL} 
\begin{eqnarray} \label{H-model}
\hat{\cal H} & = & \sum_{i=1}^N \left[ -\frac{\hbar^2}{2m} \frac{\partial^2}{\partial x_i^2} + \frac{m\omega^2}{2} x_i^2 \right] \mathbb{I}_i \nonumber \\
& & + \sum_{i<j} \delta(x_i - x_j) \left[ U_0 \mathbb{I}_i\cdot \mathbb{I}_j + U_2 \vec{\mathbb{F}}_i \cdot \vec{\mathbb{F}}_j \right],
\end{eqnarray}
where $\mathbb{I}_i$ and $\vec{\mathbb{F}}_i$ are the identity and spin-1 matrices in the spin space of $i$-th atom. Here $U_0$ and $U_2$ are the effective one-dimensional coupling constants of the spin-independent and spin-dependent interactions. Similarly as in the single-component case (see \eqref{gexpress} and \eqref{aexpress}), these couplings can be expressed by appropriate three-dimensional $s$-wave scattering lengths $a_{s0}$ and $a_{s2}$. In fact, they are expressed by appropriate linear combinations of these scattering lengths $c_0=(a_{s0} + 2a_{s2}) / 3$ and $c_{2} = (a_{s2} - a_{s0}) / 3$ (see \cite{1998OlshaniiPRL} for details). In this model, $p$-wave scattering is neglected~\cite{2004GrangerPRL}. Here, the number of atoms $N_{\pm1}$ and $N_0$ corresponding to states with spin $m=+1,0,-1$, are not conserved, because the scattering between two atoms of spin $s=\pm1$ can produce two atoms  with $s=0$. The total number of atoms $N=N_{+1}+N_0+N_{-1}$ and the total magnetization (total spin in the $z$ direction) $M=N_{+1}-N_{-1}$ are conserved quantities. 

This system has more limits than those described in Fig.~\ref{Fig:GarciaMarch_Cube}. Here we only review some results in the topic. For the case in which $U_0 \to\infty$ with $U_2=0$ analytical solutions has been proposed via Bose-Fermi mapping theorem (see~\cite{2004GirardeauPRA,2008DeuretzbacherPRL,2016YangPRA}; see also~\cite{2017YangPRA} for a study on the single-particle reduced density matrix and momentum distribution in this limit and its relationship with that of hard-core anyons). In~\cite{2008DeuretzbacherPRL} an exact diagonalization study is provided with up to $N=8$ when $U_0 $ is large with $U_2=0$ or  very small and negative (ferromagnetic coupling).  It is shown that the ground state is heavily degenerate for  $U_0\to\infty  $ with   $U_2=0$ and quasi-degenerate in case of $U_0$ finite and $U_2=0$ or small. For $U_0\to\infty  $ in a  gas of $N$ distinguishable atoms the degeneracy is $N!$. This degeneracy has to be reduced due to symmetrization associated to the bosonic statistics, giving that the degeneracy equals the dimension of the $N$-particle spin space $3^N$~\cite{2008DeuretzbacherPRL}. These states, either degenerate for $U_2=0$ and $U_0\to\infty$ or quasi-degenerate for the other cases, correspond to different density profiles of the three components with spin $m=+1,0,-1$ (see~\cite{2008DeuretzbacherPRL}).  A detailed density functional study with up to $N=30$ atoms concluded that the competition between the repulsive density-density interactions, $U_0$, and the spin-exchange interactions, $U_2$, led to complicated density distributions of the three spin components, even when both are kept equal~\cite{2013WangPRA}. Note that, in that case, $U_2$ had the same sign as $U_0$, corresponding to the antiferromagnetic case. A numerical exact diagonalization study was also attempted in~\cite{2016HaoEPJD}, also for the antiferromagnetic case, and the conventional situation in which $U_2\ll U_0$, ranging $U_0$ from the small interacting case to large interactions. This study was performed for a few atoms ($N=4$). The total density showed the evolution from a Gaussian-like distribution in the small interacting limit to a TG-like distribution with $N=4$ peaks, as $U_0$ was increased from 1 to 50, always keeping $U_2=U_0/100$. It also showed that, in the small and medium interaction regimes, the three component densities overlap. On the contrary, for large $U_0$ there is a density separation. Indeed, the $m=\pm1$ components occupied the center of the trap and the $m=0$ component separates to the edges. In this paper, the authors studied that, when $U_2$ is made larger and comparable to $U_0$, the fermionization does not occur in the same way, not reaching a TG-like structure with the increase of the interactions, but a double peak structure more similar to the composite fermionization observed in two-component gases. This effect is called in this context {\it weakening of fermionization}. In~\cite{2017HaoEPJD} it was performed an exact diagonalization numerical study of fermionization as $U_0$ was increased. The author considered the antiferromagnetic case, for $U_2=U_0/100$ and $N=4$, in the sectors of the Hilbert state with different magnetization $M=\int dx [\rho_{1}(x)-\rho_{-1}(x)]/N$, $\rho_m(x)$ being the density in spin component $m$. It was shown that fermionization occurred as expected, but the denrepellingsity profiles of each component showed different scenarios, from phase separation to magnetic domains (see also~\cite{2017JenPRA} for a study in the case of $U_2=0$ up to $N=16$).   
 
\section{Fermionic mixtures} \label{Sec:Fermions}
 In this section, we review the current stage of the research devoted to systems of a few fermions in one-dimensional traps. This path of exploration accelerated recently due to a series of experiments performed mainly in the J. Selim group in Heidelberg \cite{2011SerwaneScience,2012ZurnPRL,2013ZurnPRL,2013WenzScience,2015MurmannPRL,2015MurmannPRLb}. Theoretically, the idea of confining ultracold fermions in one-dimensional traps has the same origin as in the bosonic case (see \cite{2013GuanRevModPhys} for a review). Trapping in two perpendicular directions is very deep and therefore the dynamics in these directions is frozen. Consequently, in the remaining direction, the system is effectively one-dimensional with some effective interaction strength depending on the perpendicular confinement \cite{1998OlshaniiPRL,2004GirardeauPRA,2004GrangerPRL,2012GharashiPRA}. Note however one fundamental difference between bosonic and fermionic systems. Due to the Pauli exclusion principle, particles cannot occupy only the lowest single-particle state. In consequence, when a larger number of particles is considered  (even in the noninteracting case), the most excited particles may have energy comparable with excitation energy in perpendicular directions. Therefore, to keep a one-dimensional description valid, one should assure  (similarly as in the case of strongly repelling bosons) that excitations in perpendicular directions are strongly suppressed.  This issue was one of the experimental challenges and was responsible for obtaining one-dimensional fermionic systems much later than the bosonic ones.  

\subsection{Role of the spin}
In the case of ultracold fermions, the spin degree of freedom plays a crucial role. In contrast to the bosonic case, due to the quantum statistics, $s$-wave contact forces between fermions exactly vanish whenever particles have all internal quantum numbers identical. Consequently, the most prominent contribution to interactions comes from the $s$-wave scattering between fermions belonging to different internal states. Of course, it is still possible that identical fermions do interact, for example via long-range dipolar forces. However, these interactions are typically much weaker. We review this path of exploration in Sec.~\ref{DifferentInteractions}. In fact, this particular distinguishability required by quantum statistics for interacting fermions can be realized in three different ways -- identical fermionic elements may have different spin projections (like fermionic spin-$1/2$ $^3$He atoms), they may belong to different irreducible spin representations due to different spin projections of their nucleus (for example spin-$1/2$ and spin-$3/2$ $^6$Li atoms \cite{2013WenzScience}), or particles may be fundamentally different elements of different mass (for example bi-fermionic Li-K mixture \cite{Wille6Li40K,Tiecke2010Feshbach6Li40K}). In the two latter cases, numbers of particles belonging to different components are conserved and interactions reduce to a simple density-density form.

With a few exceptions, few-fermion systems with a dynamical spin degree of freedom are not considered in the literature. However, some studies in this direction were performed. Typically, in such a scenario, one assumes that the total number of particles $N_\uparrow+N_\downarrow$ is conserved, but their distribution among components $N_\uparrow-N_\downarrow$ (in this case called magnetization) depends on external conditions controlled experimentally. For example, independently on interactions, when strong enough external magnetic field is applied, the system is forced to polarize, {\it i.e.}, all particles occupy only one selected component,  minimizing magnetic energy. The field in which the system undergoes the transition to the fully-polarized system, of course, depends on mutual interactions between particles. Moreover, it also strongly depends on external confinement. First detailed studies of this phenomenon for the uniform system were given in \cite{2008GuanPRA}. Having in hand the exact form of all the many-body eigenstates of the system from the Bethe ansatz approach, the authors were able to find different properties of the system and predict the critical value of the magnetic field in which the system becomes fully polarized. The situation is much more interesting in the case of confined systems. In this case, one observes a specific interplay between interactions, external magnetic field, and the shape of external confinement in establishing ground-state magnetization of the system. As shown in \cite{2018SowinskiCondMatt}, typically the interacting system undergoes many transitions when increasing magnetic field. At each transition, one fermion flips its spin and change the total magnetization. However, when the interacting system is confined in a double-well potential, it becomes possible that some magnetizations cannot be reached by the many-body ground-state and then, in the transition points, simultaneous spin-flip of two fermions is observed. The problem of the spontaneous spin-flip and stability of the ground state was addressed recently in \cite{2018KoutentakisARX}. This path of exploration of few-fermion systems is still open and may lead to many interesting and surprising results (see also \cite{2014BarasinskiJOSAB,2015CarvalhoPRA} for spin-1 bosonic counterparts). 

From this point, we focus on cases where spin can be treated as an additional and fixed quantum number which simply distinguishes particles belonging to different components. Consequently, the most general Hamiltonian for such a system, {\it i.e.}, system of a few fermions confined in a one-dimensional trap $V_\sigma(x)$ and interacting via contact forces has the form:
\begin{align} \label{Eq:FewFermionHam}
\hat{\cal H} &= \sum_\sigma\int\!\!\mathrm{d}x\,\hat\Psi^\dagger_\sigma(x)\left[-\frac{\hbar^2}{2m_\sigma}\frac{\mathrm{d}^2}{\mathrm{d}x^2}+V_\sigma(x)\right]\hat\Psi_\sigma(x) \nonumber \\
&+\sum_{\sigma\neq\sigma'}g_{\sigma\sigma'}\int\!\!\mathrm{d}x\, \hat\Psi^\dagger_\sigma(x)\hat\Psi^\dagger_{\sigma'}(x)\hat\Psi_{\sigma'}(x)\hat\Psi_\sigma(x),
\end{align}
where $\Psi_\sigma(x)$ is the fermionic  field operator annihilating a fermion with spin $\sigma$ at position $x$. These operators obey standard fermionic anti-commutation relations
\begin{subequations}
\begin{align}
\left\{\hat\Psi_\sigma(x),\hat\Psi_{\sigma'}^\dagger(x')\right\}&=\delta_{\sigma\sigma'}\delta(x-x'), \\
\left\{\hat\Psi_\sigma(x),\hat\Psi_{\sigma'}(x')\right\}&=0.
\end{align}
\end{subequations}
In the simplest and the most widely studied case, only two-component mixture is considered. Then the standard (pseudo) spin-1/2 notation is used ($\sigma\in\{\uparrow,\downarrow\}$). By decomposing the field operators $\hat\Psi_\sigma(x)=\sum\hat{b}_{i\sigma} \varphi_{i\sigma}(x)$, the Hamiltonian can be rewritten in the more familiar form
\begin{equation}
\hat{\cal H} = \sum_{i\sigma} E_{i\sigma} \hat{b}^\dagger_{i\sigma}\hat{b}_{i\sigma} + g \sum_{ijkl} U_{ijkl} \hat{b}^\dagger_{i\uparrow}\hat{b}^\dagger_{i\downarrow}\hat{b}_{i\downarrow}\hat{b}_{i\uparrow}.
\end{equation}
Here, the fermionic operators $\hat{b}_{i\sigma}$ anihilate a particle with spin $\sigma$ in the state $\varphi_{i\sigma}(x)$,  which is the eigenstate of the corresponding single-particle Hamiltonian
\begin{equation}
\left[-\frac{\hbar^2}{2m_\sigma}\frac{\mathrm{d}^2}{\mathrm{d}x^2}+V_\sigma(x)\right]\varphi_{i\sigma}(x) = E_{i\sigma} \varphi_{i\sigma}(x).
\end{equation}
The amplitudes $U_{ijkl}=\int\mathrm{d}x\, \varphi^*_{i\uparrow}(x)\varphi^*_{j\downarrow}(x)\varphi_{k\downarrow}(x)\varphi_{l\uparrow}(x)$ describe interaction terms between opposite-spin particles (compare with \eqref{Eq:hamExactdiag} and note the difference in the notation due to the additional bosonic factor $1/2$).

The Hamiltonian \eqref{Eq:FewFermionHam} has a few symmetries which significantly simplify the analysis of its properties. First symmetry is related to the conserved number of particles in each component $\hat{\cal N}_\sigma = \int\mathrm{d}x\,\hat\Psi_\sigma^\dagger(x)\hat\Psi_\sigma(x)$ mentioned above. It means that the Hamiltonian is block-diagonal with respect to these numbers. Next symmetry (permutation of identical particles) reflects the fundamental indistinguishability of particles belonging to the same component. If the shape of the external potential is symmetric under mirror reflection, $x\rightarrow -x$, then the Hamiltonian has an additional $Z_2$ symmetry. The symmetry is guaranteed by the interaction term which, by its construction, couples only the $4$-products of single-particle orbitals which are even (even sum $i+j+k+l$ in the harmonic oscillator convention). Finally, if an external potential is quadratic and spin-independent, then the motion of the center-of-mass decouples from the internal motion, {\it i.e.}, the Hamiltonian has an additional $U(1)$ symmetry. Detailed analysis of properties of the Hamiltonian \eqref{Eq:FewFermionHam} (in a harmonic trap) in terms of its symmetries was presented in \cite{2014HarshmanPRA}. As shown, the reduction of the system's states with respect to immanent symmetries of the Hamiltonian may be very helpful when adiabatic changes of interaction strength are considered. 

In most of the cases, the trapping potential $V_\sigma(x)$ is assumed as a simple harmonic trap independent on spin $\sigma$. In this case, the motion of the center-of-mass of the system decouples from the internal dynamics and the whole discussion can be significantly simplified. For example, it is possible to determine some well-defined experimentally accessible quantities determining internal excitations. This may be a very important question when experimental outcomes in an attractive interaction regime are analyzed, due to the large uncertainty of the center-of-mass position \cite{2017PecakFewBody}. Besides harmonic confinement, it should be pointed out that, due to recent achievements in trapping techniques, it is also possible to consider other confinements, like double-well confinements \cite{2013BugnionPRAb,2015MurmannPRLb,2016SowinskiEPL,2016YannouleasNJP,2018ErdmannPRA,2019ErdmannPRA}, uniform rings \cite{2015BarfknechPRA}, one-side-open wells \cite{2013RontaniPRA,2013ZurnPRL}, or even uniform hard-wall box traps \cite{2013GauntPRL,2016PecakPRA}, and tunable periodic potentials \cite{2017PilatiPRA} {\it etc.} In the following we will focus mostly on the harmonic confinements, however, whenever it is relevant, we will refer to appropriate discussions for other confinements.

\subsection{Inter-particle correlations}

When studying different properties of interacting few-fermion systems, in a full analogy to the bosonic cases, one focuses only on the simplest quantities which characterize the state of the system, {\it i.e.}, the single-particle reduced density matrices
\begin{equation}
\rho^{(1)}_\sigma(x,x')=\frac{1}{N_\sigma}\langle \hat{\Psi}^\dagger_\sigma(x)\hat{\Psi}_\sigma(x')\rangle
\end{equation}
and the density profiles, which are their diagonal parts $n^{(1)}_\sigma(x)=\rho^{(1)}_\sigma(x,x)$. Inter-particle correlations in a given component are encoded in the corresponding $k$-particle reduced density matrices, which have the form
\begin{multline} \label{MultiCorel}
\rho^{(k)}_{\sigma}(x_1,\ldots,x_k,x_1',\ldots,x_k')= \\ \frac{(N_\sigma-k)!}{N_\sigma!}\langle \hat{\Psi}^\dagger_\sigma(x_1)\cdots\hat{\Psi}^\dagger_\sigma(x_k)\hat{\Psi}_\sigma(x_k')\hat{\Psi}_\sigma(x_1')\rangle.
\end{multline}
However, in the case of fermionic mixtures we are mostly interested in the inter-component correlations forced by interactions and the quantum statistics. These correlations are captured dominantly by the inter-component two-particle reduced density matrix
\begin{equation}
\rho^{(2)}_{\uparrow\downarrow}(x,y,x',y')=\frac{1}{N_\uparrow N_\downarrow}\langle \hat{\Psi}^\dagger_\uparrow(x)\hat{\Psi}^\dagger_\downarrow(y)\hat{\Psi}_\downarrow(y')\hat{\Psi}_\uparrow(x')\rangle
\end{equation}
and its diagonal part $n^{(2)}_{\uparrow\downarrow}(x,y)=\rho^{(2)}_{\uparrow\downarrow}(x,y,x,y)$, 
which describes the probability distribution of simultaneous detection of opposite-spin fermions. Of course, higher correlation functions, similar to these for intra-component correlations \eqref{MultiCorel}, can also be considered. In practice, reduced density matrices with $k>2$ are very hard to be calculated (due to the numerical complexity). Therefore, they receive less attention in literature.

In the case of fermionic mixtures, direct studies of inter-particle correlations in terms of single- and two-particle density matrices is much harder than in the case of bosons. The fundamental reason lies in the additional correlations forced by Pauli exclusion principle between indistinguishable fermions. These correlations are reflected in density matrices as additionally occupied orbitals, exactly the same way as correlations forced by interactions. The latter is of course much more interesting since their properties are in principle under experimental control. To overcome this difficulty one needs to find some convenient way to distinguish these two kinds of correlations, {\it i.e.}, to ,,subtract'' trivial correlations forced by the quantum statistics and remain only with those induced by interactions. One of the simplest way to determine interaction-induced correlations is to calculate the difference between diagonal parts of the two- and single-particle density matrices, called the noise correlations in position and momentum representation \cite{2008MatheyPRL,2009MatheyPRA}:
\begin{subequations} \label{NoiseCor}
\begin{align}
G_x(x,y) &= n^{(2)}_{\uparrow\downarrow}(x,y)-n^{(1)}_\uparrow(x)n^{(1)}_\downarrow(y), \label{NoiseCora}\\
G_p(p_x,p_y) &= n^{(2)}_{\uparrow\downarrow}(p_x,p_y)-n^{(1)}_\uparrow(p_x)n^{(1)}_\downarrow(p_y).\label{NoiseCorb}
\end{align}
\end{subequations}
It is quite obvious that in the limit of vanishing interactions both correlations become equal to zero. Note that $G_x$ and $G_p$ are not connected by a simple Fourier transformation since they are calculated only from the diagonal parts of corresponding density matrices. 

Since the noise correlations \eqref{NoiseCor} are calculated directly from the single- and two-particle density profiles, they have experimental relevance and, as shown in \cite{2017BrandtPRA,2018BrandtPRA}, they may encode many interesting fundamental features of interacting fermionic systems which are ,,hidden'' in bare density matrices. It is worth mentioning that some extensions of the noise correlations to higher-order densities are also possible \cite{2019YannouleasARX}.
 
\subsection{Higher-spin mixtures}
\begin{figure}[t]
\includegraphics[width=\linewidth]{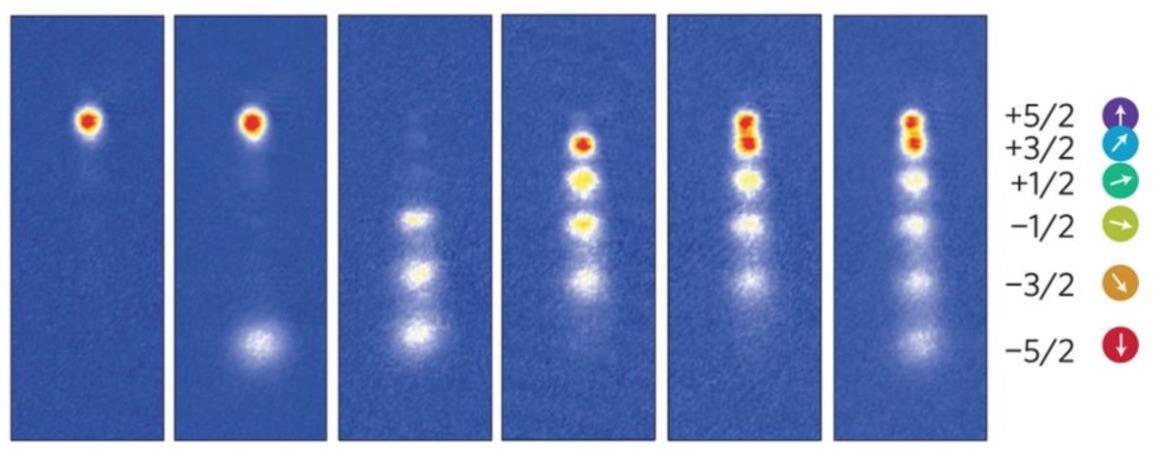}
\caption{Stern--Gerlach detection of different spin components of $^{173}$Yb atoms confined in a quasi-one-dimensional trap for different experimental scenarios resulting in different number of populated spin components. Figure adapted  from \cite{2014PaganoNatPhys}. Copyright (2014) by the Springer Nature Publishing.\label{_Fig-MultiFermions}}
\end{figure}

Importantly, it should be pointed out that few-fermion mixtures with higher spins may also be considered with a full analogy to spinor bosonic mixtures. This path of exploration was recently undertaken with more care due to the beautiful experiments on one-dimensional fermionic mixtures with internal $\mathrm{SU}(N)$ symmetry \cite{2014PaganoNatPhys} (see Fig.~\ref{_Fig-MultiFermions}) with ultracold $^{173}$Yb atoms. Due to the high total spin $5/2$, it was possible to prepare the system in an arbitrary number of spin components (up to $6$) and show how the quantum statistics together with mutual interactions influence statical and dynamical properties of the system. It was also shown that in the limit of a large number of components the system approaches a bosonic spinless liquid. 

On a theoretical footing, first discussions of multi-component one-dimensional fermionic mixtures were given in \cite{2012KuhnNJP,2012GuanPRAb} before the Florence experiment (see also \cite{2016BeverlandPRA} for a different approach to mimic $\mathrm{SU}(N)$ symmetry with atoms confined in the one-dimensional hard-wall potential). Here, by considering the system without confinement and based formally on asymptotic expansions leading to the Fredholm equations, the authors were able to obtain accurate results for the ground-state energy in a whole range of interactions and local pair-correlation function. Interestingly, they show that in the limit of infinite spin, the energy per particle becomes equal to that obtained for spinless bosons. 

Noticeably, higher-spin mixtures are less sensitive to the Pauli exclusion principle since there are much more possibilities for particles to avoid the same sets of quantum numbers. This phenomenological argument was directly observed experimentally in \cite{2014PaganoNatPhys} -- the system with a larger number of components manifests bosonic properties more clearly. This is also visible when momentum distribution is considered. For a larger number of components, the distribution becomes broadened along with slower decay for high momenta. The detailed multithreaded theoretical discussion of this behavior in terms of the Tan's contact (for a large number of particles) was given in \cite{2014DecampPRA,2017HoffmanPRA}.

The most comprehensive theoretical studies of higher-spin fermionic mixtures confined in a harmonic trap in the few-body regime were given recently in \cite{2017LairdPRA}. Using different theoretical techniques the authors study many different properties of such mixtures in a whole range of interactions. One of the prominent results for the ground-state is that with an increasing number of fermionic components, in weak as well as in strong interactions regimes, the system much faster approaches the many-body limit of a large number of particles. It should be also remarked that the dynamical properties of the uniformly confined systems, focusing mainly on collective spin-mixing dynamics, were discussed in \cite{2017XuAnPhys}.

In full analogy to the two-component mixtures, in the limit of infinite repulsions, the ground-state of the system can be obtained analytically \cite{2016DecampNJP}. Depending on the number of particles and their distribution among components, the system displays different spatial patterns and the ground manifold can be characterized in terms of the conjugacy class sums (see \cite{1962LiebPhysRev} for more details on the Lieb-Mattis theorem). A quite similar analysis for the system confined in a hard-wall potential was presented recently in 
\cite{2017PanPRB}. Based on exact solutions in the framework of Bethe ansatz it was shown that the system can be completely understood in terms of spin-exchange $\mathrm{SU}(N)$ model. 

Evidently, the path of exploration of the multi-component fermionic mixtures is still not closed. Many theoretical and experimental analysis performed previously in the many-body limit and higher dimensions are still awaiting for repetition in the regime of a few-particle sector in one spatial dimension (for a review for previous progress see \cite{2014CazalillaRPP}). In the following, we mainly focus on two-component (pseudo) spin-$1/2$ fermionic mixtures.

\subsection{Two- and three-fermion cases}
\begin{figure}[t]
\includegraphics[width=\linewidth]{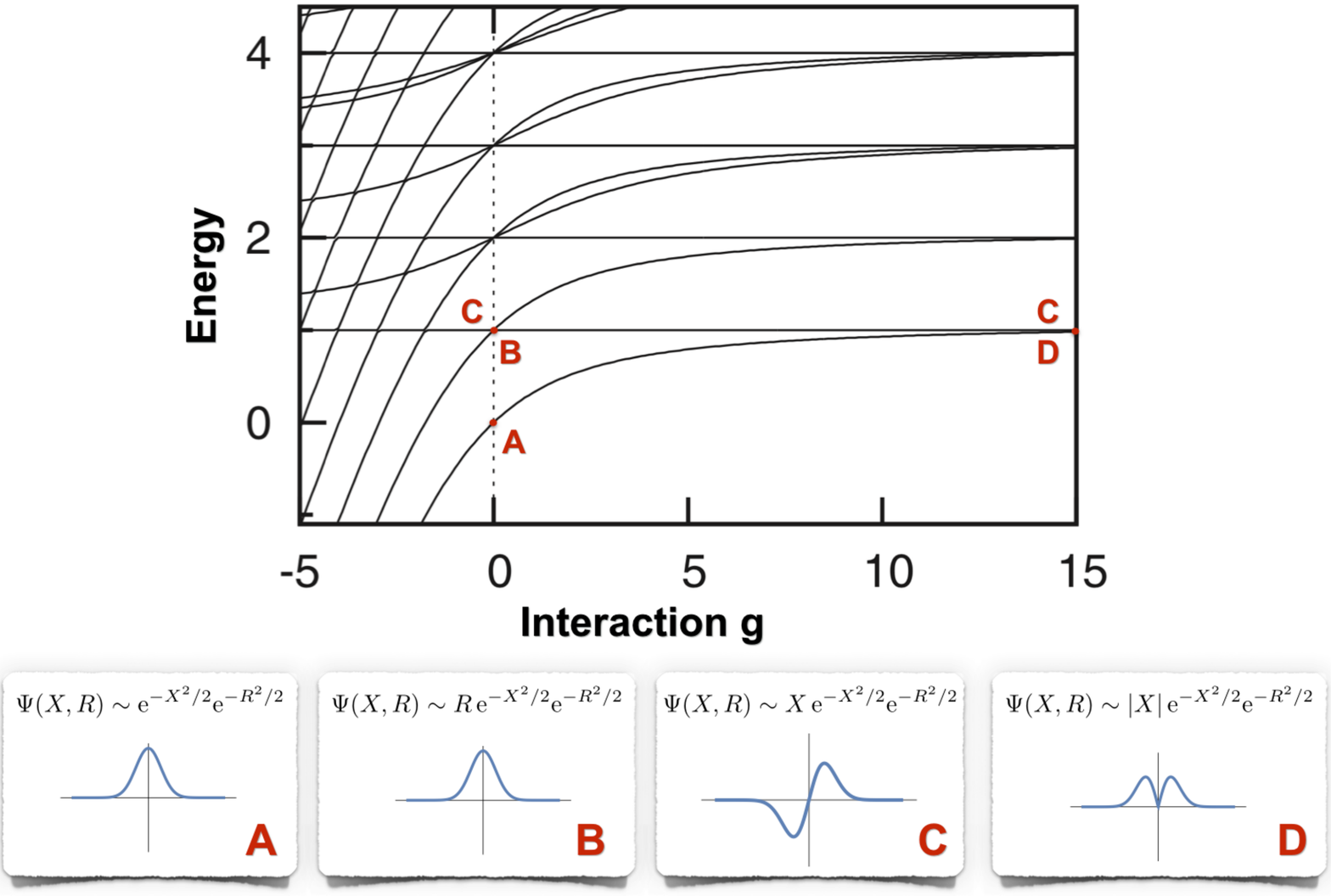}
\caption{Energy spectrum of the two-fermion system confined in a one-dimensional harmonic trap. In the bottom panel, we show the analytical expressions for the corresponding wave functions in the relative ($X$) and the center-of-mass ($R$) coordinates. Plots show shapes of relative motion wave functions (that is, as a function of $X$). For vanishing interactions ($g=0$) the ground-state wave function is isolated and the first excited state is doubly degenerated (single excitation of the relative or the center-of-mass motion). In the limit of infinite repulsions, the ground-state manifold is doubly degenerated and spanned by the symmetric and antisymmetric function of the relative motion. It is very instructive to compare this figure with Fig.~\ref{_Fig-BuschSolution} where the spectrum of relative motion of two bosons was discussed.\label{_Fig-TwoFermions}}
\end{figure} 

Due to the quantum statistics, the many-body spectrum of a few interacting fermions has many interesting features. For $N_\uparrow=N_\downarrow=1$ it is very similar to the spectrum of two interacting bosons. However, since in this case, both particles are fundamentally distinguishable, besides the symmetric states, antisymmetric combinations (which are completely insensitive to interactions) are also present \cite{2010GirardeauPRA}. Interestingly, in the limit of infinite repulsion (the TG limit), symmetric and anti-symmetric states become degenerate (see Fig.~\ref{_Fig-TwoFermions}). This degeneracy is a counterpart of fermionization described previously for bosons and can be quite easily explained by considering (rescaled) relative and the center-of-mass motion coordinates \eqref{BuschTransform}, $X=(x_1-x_2)/\sqrt{2}$ and $R=(x_1+x_2)/\sqrt{2}$. Since the first excited state (see panel (C) in Fig.~\ref{_Fig-TwoFermions}) is antisymmetric in $X$, it is insensitive to the interaction strength (for vanishing interactions it is degenerated with another state -- panel (B) in Fig.~\ref{_Fig-TwoFermions}-- which is singly excited in the center-of-mass motion coordinate). Consequently, its energy as a function of $g$ is constant. Note that its relative motion wave function is proportional to $X\mathrm{exp}(-X^2/2)$. In the limit of infinite interactions, it becomes degenerated with the symmetric ground-state (D) having relative motion wave function proportional to $|X|\mathrm{exp}(-X^2/2)$. These theoretical predictions for the system of two distinguishable fermions were carefully examined in the seminal experimental work \cite{2012ZurnPRL}. Particularly, it was shown that indeed, in the limit of strong repulsions, the system undergoes fermionization, {\it i.e.}, spatial properties of the system are exactly the same as properties of two identical non-interacting fermions.

It was argued that the ground-state degeneracy in the limit of infinite repulsions can be lifted by additional spin-changing interactions. The simplest way is to extend the model by adding the spin-orbit coupling to the single-particle part of the Hamiltonian
\begin{equation}
\hat{\cal H}_{SO} = \int\!\mathrm{d}{x}\,\hat{\boldsymbol{\Psi}}^\dagger(x)\left[\left(\frac{\hbar q_r}{im}\frac{\mathrm{d}}{\mathrm{d}x}+\frac{\delta}{2}\right)\sigma_y+\frac{\Omega}{2}\sigma_x\right]\hat{\boldsymbol{\Psi}}(x),
\end{equation}
where the algebraic vector $\hat{\boldsymbol{\Psi}}(x)=\left(\hat\Psi_\uparrow(x),\hat\Psi_\downarrow(x)\right)^\mathrm{T}$ is the two-component field operator describing both components simultaneously and $\{\sigma_x,\sigma_y,\sigma_z\}$ is a set of standard Pauli matrices acting in this two-dimensional algebraic space. Parameters $q_r$, $\delta$, and $\Omega$ encode the strengths of the different spin-orbit coupling processes (two-photon recoil momentum, two-photon detuning, and the Raman coupling strength, respectively). The resulting energy spectra in the case of two particles were carefully examined in \cite{2014GuanJPB} (see also \cite{2015SchillaciPRA} for a very nice description in a corresponding two-dimensional system of two particles). The detailed analysis of the interplay between contact interactions and spin-orbit coupling and changes of the different properties of the systems with larger number of particles (for fermions as well as for bosons) were given in \cite{2014CuiPRAb}, \cite{2014YinPRA}, \cite{2015GuanPRA}, and \cite{2016ZhuJPB}. 

As long as particles belong to different spin components, the properties of the system are not affected by the quantum statistics. In this spirit, the system of three distinguishable particles (belonging to three different components) was also studied analytically in the vicinity of infinite repulsions \cite{2014VolosnievFBS}. Since particles are completely distinguishable, in this limit the ground state is six-fold degenerated.

The simplest nontrivial system affected by the quantum statistics is a two-component mixture of $N_\uparrow=2$ and $N_\downarrow=1$ fermions. In this case, in contrast to two or three different fermions cases, the ground state is only three-fold degenerated in the limit of infinite repulsions. The first analysis of properties of the system in the vicinity of the TG limit was presented in \cite{2007KestnerPRA}. For intermediate interactions, although properties of this system cannot be fully analyzed analytically, the description becomes quite simple when appropriate Jacobi coordinates (a generalization of the two-particle relative and center-of-mass coordinates) are used. More detailed discussion along this path was given in two papers \cite{2014DAmicoJPB,2015LoftEPJD} (see also \cite{2015VolsonievEPJST} for considerations in the framework of the hyperspherical formalism) where many different features of the system were described and compared with predictions of the exact diagonalization approach. In addition, it was shown, that the method can be also applied when the additional $\downarrow$ particle has different mass (see also \cite{2014MehtaPRA} described in Sec.~\ref{Sec:Bosons} for three- and four-particle cases discussed in Born-Oppenheimer approximation and appendix~\ref{app:threeatoms}). 

\subsection{Impurity in the Fermi sea}
\begin{figure}[t]
\includegraphics[width=\linewidth]{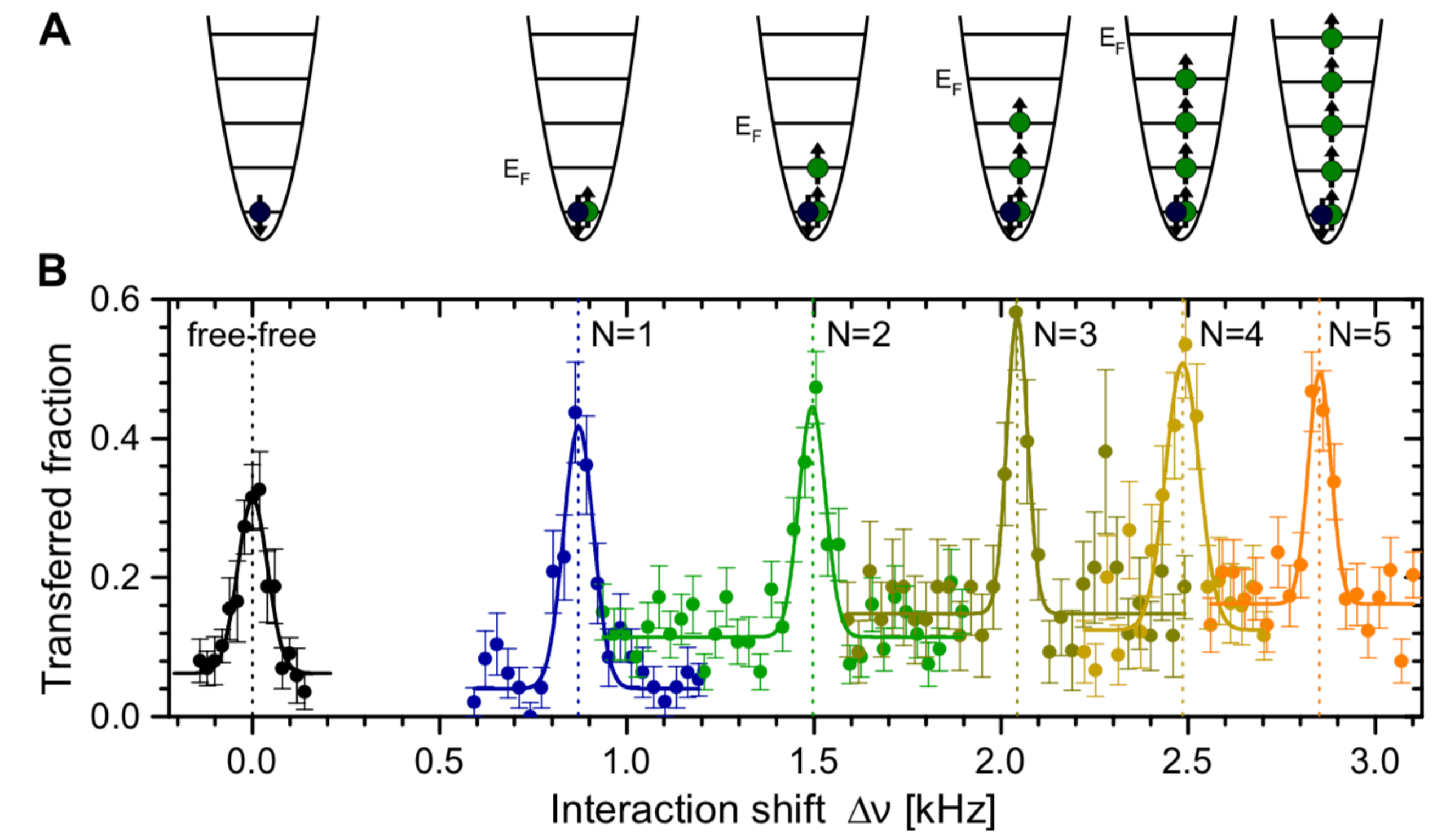}
\caption{Experimentally measured interaction shift in the system of an atom repulsively interacting ($g=2.8$) with a mesoscopic Fermi sea formed by $N=1,\ldots 5$ fermions. Figure adapted  from \cite{2013WenzScience}. Copyright (2013) by the American Association for the Advancement of Science.\label{_Fig-FermionImpurity}}
\end{figure} 

Before we discuss few-fermion systems with a larger number of particles in both components let us first mention the intermediate case of single impurity immersed in the Fermi sea composed from a larger number of fermions. The problem of a single impurity interacting with many opposite-spin fermions is not only a theoretical divagation. Recently, it was deeply studied experimentally with $^6$Li atoms \cite{2013WenzScience}. In this beautiful experiment, the interaction energy was directly measured as a function of increasing number of particles forming a Fermi sea (see Fig.~\ref{_Fig-FermionImpurity}). It was argued that, for weak interactions, along with increasing number of particles, the interaction energy rapidly converges to the value predicted by the polaron-like description \cite{1966McGuireJMP1,1966McGuireJMP2,2013AstrakharchikPRA} and surprisingly, already for $N\gtrapprox 5$ particles, the system resembles the many-body limit. Contrary, for very strong interactions, when the fermionization limit is achieved and the polaron-like approach breaks down, the system behaves like a system of $N+1$ non-interacting fermions. The transition between these two natural limits of interactions was studied theoretically with different numerical methods: the local density approximation and diffusion Monte-Carlo techniques \cite{2013AstrakharchikPRA}, perturbative expansions in the weak and strong interaction limits \cite{2015GharashiPRA}, and an effective spin model in the limit of strong repulsions \cite{2016LoftJPB}. At this point it is worth to mention that in the limit of strong repulsions many different properties of a fermionic system can be determined by considering an effective single-particle description based on the assumption that each particle is immersed in the Fermi sea (for details see \cite{2004PricoupenkoPRA}). Finally, let us also mention that some extension of the problem to the case of two impurities immersed in the few-fermion sea was addressed recently \cite{2019MistakidisNJP}.

\subsection{Mixtures close to infinite repulsions}
\begin{figure}[t]
\includegraphics{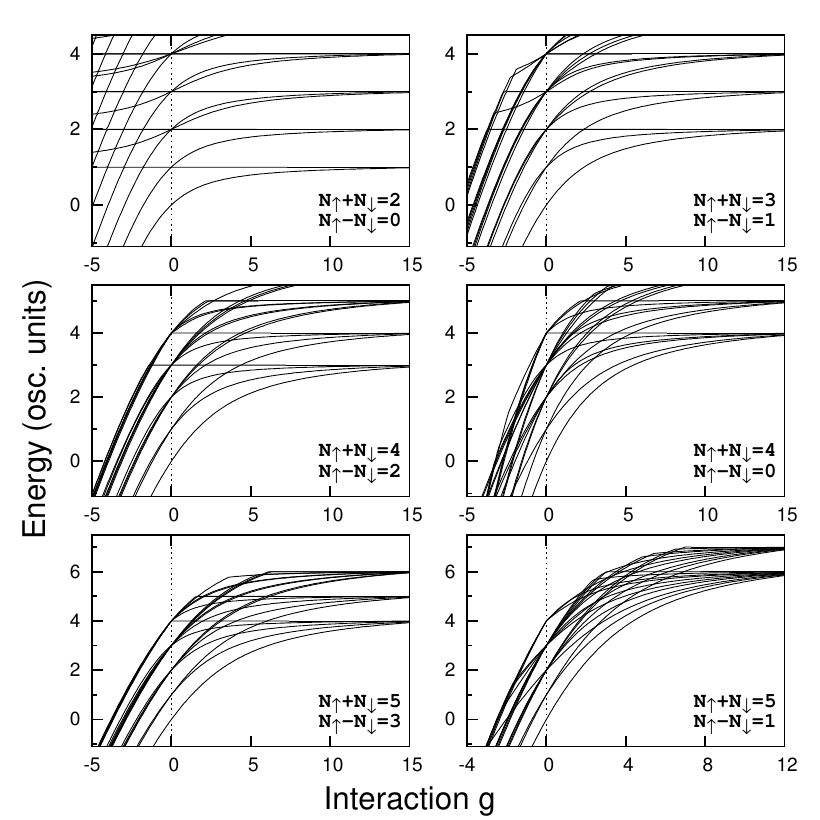}
\caption{Spectrum of the many-body Hamiltonian \eqref{Eq:FewFermionHam} for different number of equal mass fermions confined in a harmonic trap. Note the quasi-degeneracy of the spectrum in the strong repulsion regime. Figure adapted  from \cite{2013SowinskiPRA}. Copyright (2013) by the American Physical Society.\label{_Fig-FewFermionSpectrum}}
\end{figure}

For a larger number of particles, the situation becomes much more interesting \cite{2013GharashiPRL,2013SowinskiPRA,2013BugnionPRA}, as there is much more freedom to choose symmetric and antisymmetric combinations between distinguishable particles (see Fig.~\ref{_Fig-FewFermionSpectrum}). In consequence, in the limit of strong repulsions (TG limit) the degeneracy of the ground-state become much larger and can be expressed by the combinatorial factor $D=\frac{(N_\uparrow+N_\downarrow)!}{N_\uparrow! N_\downarrow!}$ (see \cite{2009GuanPRL} for details). 
It is worth to underline that all considerations about the degeneracy of the many-body spectrum in the TG limit are valid for any reasonable one-dimensional external trapping since they originate in the fundamental inter-component symmetries of the system \cite{2009YangChPL}. 

Typically, in the theoretical approaches, only the ground-state properties of the system are considered. However, this unusual quasi-degeneracy in the limit of strong repulsion may have some applications. For example, as discussed in \cite{2013SowinskiPRA,2018PlodzienPRA}, whenever such a system is in thermal equilibrium with some external thermal bath, its measurable properties are very sensitive to any changes in the temperature. Therefore, the system of a few fermions may be used for very sensitive thermometry in the nK regime. 

Similarly to the bosonic case, fermionic systems in the limit of infinite repulsions attract a lot of attention. In this limit (called full fermionization in the previous section), exact forms of all many-body eigenstates are known. The argumentation is very similar to that adopted for bosons in Sec.~\ref{Sec:Bosons}. Indeed, as shown in \cite{2009GuanPRL,2009MaJPA}, the many-body wave function in the position domain $\Psi(x_1,\sigma_1,\ldots,x_N,\sigma_N)$ has to fulfill two boundary conditions. The first condition, forced by the Pauli exclusion principle, acts in the subspace of particles having the same spin and has a form $\Psi|_{x_i=x_j;\sigma_i=\sigma_j}=0$. Additionally, the second condition caused by infinite repulsion affects pairs of opposite-spin particles and reads $\Psi|_{x_i=x_j;\sigma_i\neq\sigma_j}=0$. Both conditions lead directly to the generalized condition $\Psi|_{x_i=x_j}=0$  independently on the particles' spin. It means that the wave-function can be expressed as the Slater determinant of the corresponding single-particle states with appropriate symmetrization between components assumed \begin{equation} \label{SlaterFermion}
\Psi(x_1,\ldots,x_N) \propto {\cal S}\left(\mathrm{det}\left[\varphi_i(x_j)\right]^{i=0,\ldots,N-1}_{j=0,\ldots,N-1}\right),
\end{equation}
where ${\cal S}(.)$ is the symmetrization operator in the appropriate subspaces of distinguishable fermions of opposite spins. In the one-dimensional case, the vanishing of the wave function forced by infinite repulsions and the Pauli principle has fundamental consequences for the spatial distributions of particles. Namely, the particles cannot exchange their positions and therefore their spatial order becomes fixed. Consequently, in the limit of infinite repulsions, the ground-state manifold is spanned by the many-body states of given order -- each having appropriate modulations in the components densities \cite{2014LindgrenNJP,2016Carbonell-CoronadoNJP}. In this limit, the Hamiltonian can be written as a sum of independent Hamiltonians acting in the subspaces of a given order.

\begin{figure}[t]
\includegraphics{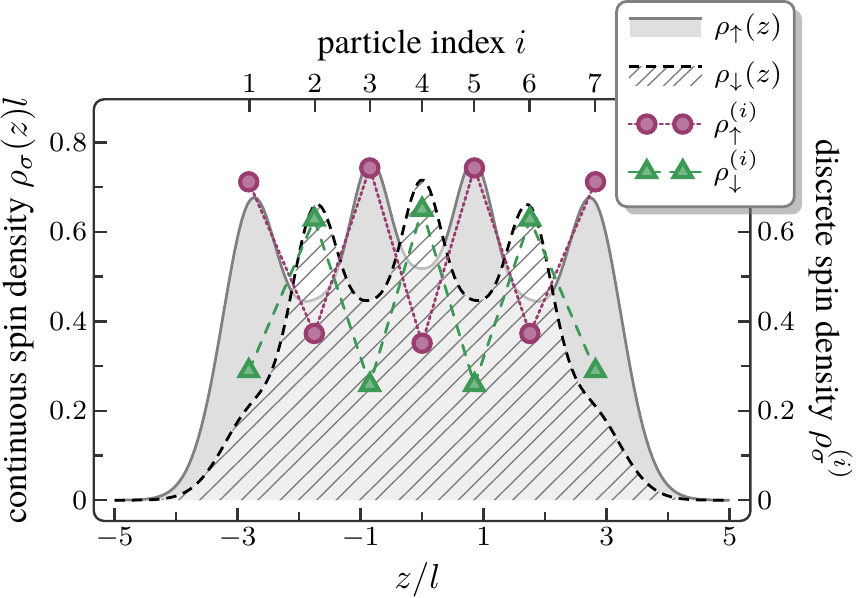}
\caption{Spin-chain representation of an infinitely repulsive system of a few fermions. Continuous lines represent density profiles $\rho_{\uparrow,\downarrow}(x)$ of the full model, while the corresponding discrete spin densities $\rho_i$ show the description in the language of the spin-chain model. Here, $N_\uparrow=4$ and $N_\downarrow=3$ are considered in the antiferromagnetic state. Figure adapted from \cite{2014DeuretzbacherPRA}. Copyright (2014) by the American Physical Society.\label{_Fig-SpinModelFermions}}
\end{figure}

The observation outlined above is essential for treating the system perturbatively for strong but finite interactions. In the series of several papers of independent theoretical groups \cite{2014DeuretzbacherPRA,2014VolosnievNatComm,2015YangPRA,2015LevinsenSciAdv} it was argued that any deviation from the infinite interactions lifts the ground-state degeneracy and the first-order corrections in $1/g$ can be written effectively (depending on a sign of $g$) as the anti-ferromagnetic or the ferromagnetic Heisenberg model with interactions proportional to $1/g$ (see Fig~\ref{_Fig-SpinModelFermions}). In the point $1/g=0$, the system undergoes the specific transition between two orderings. Importantly, the resulting model is not translationally invariant, {\it i.e.}, the exchange coefficients between effective spins depend on their positions and they are determined by the shape of an external confinement \cite{2016MarchukovEPJD}. A detailed numerical algorithm to compute these coefficients was given in \cite{2016LoftCPC}. Moreover, as shown recently in \cite{2017DeuretzbacherPRAb}, exchange coefficients in an effective spin model non-trivially depend on the transverse confinement and maybe engineer also by tuning perpendicular degrees of freedom. Generalization to higher spin representations was discussed recently in \cite{2018BarfknechtARX}.

Recently, the theoretical concept of spin-chain representation of strongly interacting fermions was examined experimentally with fermionic mixtures having up to four atoms \cite{2015MurmannPRL}. It was the first in-situ observation of the quantum magnetism forced by many-particle correlations of order higher than two. With this experimental verification, it becomes realistic to consider one-dimensional systems of strongly interacting atoms as tunable quantum simulators for different spin-chain models. Particularly, they may have an important impact in the area of state preparation and the state transfer. As shown in \cite{2015VolsonievPRA,2016LoftNJP}, by tuning the shape of the external confinement, it might be possible to transfer on demand the quantum state between the edges of the chain with almost perfect fidelity. It may shed some new light on the problems of quantum memories and quantum communication and it leads directly to the idea of another experimental possibility for constructing a quantum transistor \cite{2016MarchukovNatComm}.

For completeness, at this point we should also mention that a very similar analysis can be done for two-component bosonic mixtures \cite{2016DehkharghaniSciRep}, multi-component bosonic mixtures \cite{2017LiuPRA}, as well as Bose-Fermi mixtures (see Sec.~\ref{Sec:BoseFermi} for strongly interacting Bose-Fermi mixtures). It turns out that in all these cases, in the limit of very strong inter-component repulsions (and appropriately chosen intra-component couplings) a very similar effect of spatial separation appears in the system and, as a consequence, many system's properties can  also be well-described in the language of an effective spin model \cite{2008MatveevPRL,2015MassignanPRL}. A very nice summary of all these findings for one-dimensional systems of strongly interacting atoms having arbitrary spin was given recently in \cite{2016YangPRAb}. Additionally, the role of spin-orbit coupling in destroying the spin structure is also discussed there.

\subsection{Intermediate interactions}
On a theoretical footing one of the most challenging tasks is to determine different ground-state properties in the regime of intermediate repulsions, {\it i.e.}, when interactions are too strong to treat the system perturbatively around the non-interacting system and too weak to perform perturbation analysis around the infinitely repulsive system as described above. In this regime of interactions, if the system is confined, we do not have in our arsenal many theoretical tools to predict properties of the fermionic system. Of course, one can still try to use different methods which are suited rather for lattice systems \cite{2011SoffingPRA} or some effective approaches based on unitary transformations over known two-particle solutions \cite{2013RotureauEPJD}. However, their applicability is limited in the most interesting cases. 

\begin{table}
\begin{center}
\begin{tabular}{c||r|r}
$N_\uparrow=N_\downarrow$ & \hspace{3mm}{$\mathrm{dim}(\hat{\cal H})$}\hspace{3mm} & \hspace{3mm}{\bf Sparsity}\hspace{3mm}\hspace{0mm}\\ 
 \hline \hline
$1$ &  2025 &  $5.1 \cdot 10^{-1}$\\
$2$ &  4356 &  $5.6 \cdot 10^{-2}$\\
$3$ & 48400 & $9.3 \cdot 10^{-3}$\\
$4$ & 245025 & $2.7 \cdot 10^{-3}$\\
$5$ & 627264 & $1.3 \cdot 10^{-3}$
\end{tabular}
\end{center}
\caption{\label{TableSparse} 
The size of the cropped many-body Hilbert space and the sparsity of the Hamiltonian matrix (relative number of non-zero elements) as functions of the number of fermions $N_\uparrow=N_\downarrow$ for cut-off $M=12$. Note the tremendous expansion of the size of the Hilbert space along with decreasing sparsity of the matrix. }
\end{table}

The most straightforward method to obtain an approximate description of the system's eigenstates is to perform numerical diagonalization of the Hamiltonian in some selected many-body basis. The simplest way is to construct such a basis as a finite set of Fock states build from the $M$ lowest lying single-particle orbitals $\varphi_{i\sigma}(x)$ used in the decomposition of the field operator $\hat\Psi(x)$. Since the single-particle part of the Hamiltonian is diagonal in this basis, the only non-trivial task relays on calculating all matrix elements of the interaction Hamiltonian. Fortunately, in typical situations the resulting matrix is very sparse (see Table~\ref{TableSparse} for examples) and therefore the diagonalization can be performed for several particles and quite strong interactions with advanced Lanczos algorithm. Typically one has this scheme of diagonalization in mind when the exact diagonalization of the many-body Hamiltonian is mentioned. However, it is known that this construction of the many-body Fock basis is not the most efficient since it takes into account high-energy states in a very unsystematic way. In consequence, the convergence of the method is very slow. To solve this problem, a few different numerical approaches were introduced. One of them is based on a much more careful choice of the Fock states to construct a cropped Hilbert space in which the diagonalization is performed \cite{1998HaugsetPRA} (see also \cite{2018PlodzienARX} for a pedagogical explanation). Other is based on an observation that the single-particle orbitals of the non-interacting system $\varphi_{i\sigma}(x)$ are not well-optimized when quite strong interactions are considered. The convergence can be significantly speed-up if orbitals were slightly modified to encode shrinking (attractions) or expanding (repulsions) spatial size of the system induced by interactions. Recently, the method was used in the case of few bosons problems \cite{2018KoscikPhysLettA}. However, it can be straightforwardly adapted for fermionic systems.  At this point it should be noticed that a relatively slow convergence of the exact diagonalization approach is predominantly determined by a nature of contact interactions leading to non-analytical cusps of the relative-motion wave functions. As shown in \cite{2018JeszenszkiPRA}, by performing appropriate transformation of the many-body Hamiltonian one can remove the leading-order singularity forced by interactions. Consequently the convergence of the exact diagonalization approach is significantly improved.

\begin{figure}[t]
\includegraphics[width=\linewidth]{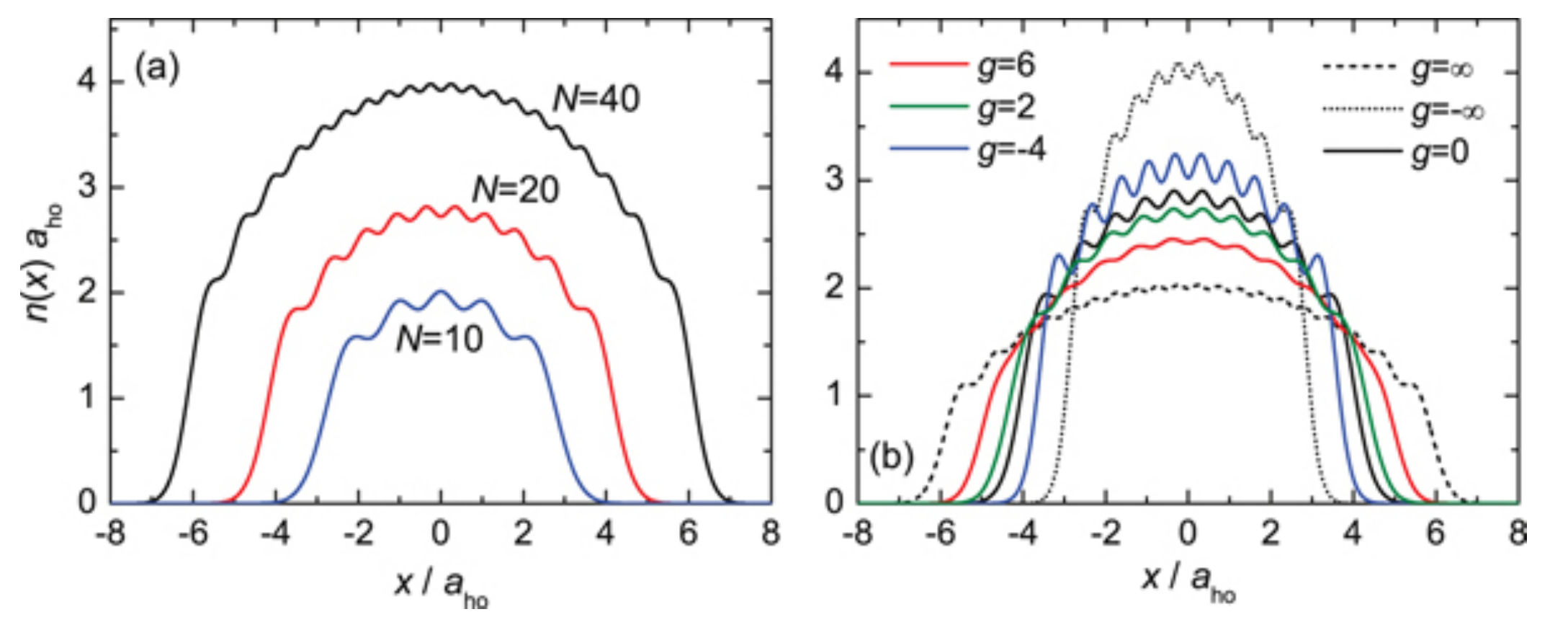}
\caption{Single-particle density profile of $N/2=N_\uparrow=N_\downarrow$ fermions confined in a harmonic trap obtained for the ground-state of an interacting system with coupled cluster method. (a) Density profiles for interaction $g=1$ and a different number of particles. (b) Density profiles for $N=10$ and different interaction strengths. Figure adapted  from \cite{2015GriningNJP}. Copyright (2015) by IOP Publishing Ltd and Deutsche Physikalische Gesellschaft.\label{_Fig-Tomza}}
\end{figure} 

Due to the mentioned tremendous expansion of the Hilbert space with the number of particles, the exact diagonalization of the many-body Hamiltonian  \eqref{Eq:FewFermionHam} can be performed numerically only for relatively small number of particles. As shown recently, other numerical {\it ab initio} methods, intensively exploited in quantum chemistry, can be adapted to study problems with a much larger number of fermions. In the introductory paper \cite{2015GriningNJP} (and later in \cite{2015GriningPRA}) the authors explain in detail many different numerical approaches, compare their accuracies, and finally use them to obtain the simplest single-particle properties of the system containing up to 80 fermions (see Fig.~\ref{_Fig-Tomza} for some examples). 

Instead of applying quantum chemistry approaches to the problem of a few interacting fermions one can consider an approach based on the famous Kohn-Sham density-functional theory \cite{1999KohnRevModPhys}. Since in this framework it is natural to express the energy functional in terms of exchange and correlation energy functionals, then one can explore more deeply the role of interactions and quantum statistics in forming bulk properties of interacting ground-state of the system \cite{2006XianlongPRA}. Of course in this approach, there are no numerical problems with a very large number of particles.  Here, we want also to mention other more sophisticated method based on the complex Langevin approach \cite{2017RammelPRD,2018RammelJPhysConf}, which in principle can be applied to many different problems of interacting fermions.

Another possible direction is to use some variational schemes. Since the ground-state of an interacting one-dimensional system is non-degenerated for any finite $g$, the variational approach is very fruitful and may give accurate predictions for the simplest measurable quantities. For example, having in hand the exact form of the ground-state for vanishing interactions $|\Psi_0\rangle$ and appropriate combination for infinite repulsions $|\Psi_\infty\rangle$ one can propose a very simple variational trial function (called interpolatory ansatz) of the form:
\begin{equation} \label{interpolatory}
|\Psi(g)\rangle = \alpha(g)|\Psi_0\rangle + \beta(g)|\Psi_\infty\rangle.
\end{equation}
Note that the extreme ground states are not orthogonal and therefore amplitudes $\alpha(g)$ and $\beta(g)$ do not fulfill standard normalization conditions. By minimizing the expectation value of the Hamiltonian one can find the approximation of the many-body ground state and calculate all quantities needed. This path of exploration was initialized in \cite{2016AndersenSciRep}, where the systems of $N_\uparrow=1$ and $N_\downarrow=1,\ldots,5$ were examined in this way. By studying different examples, also for $N_\uparrow=1, N_\downarrow=2$
 particles with different masses, it was shown that the ansatz restores the many-body spectrum in a whole range of interactions with very good accuracy. It should be underlined, however, that in fact the ansatz \eqref{interpolatory} cannot be used directly and it needs some technical improvements to describe the perturbative regime $1/g\approx 0$ correctly. The improvements are also discussed in \cite{2016AndersenSciRep}. 

Initially, the accuracy of the ansatz \eqref{interpolatory} was tested only on the energy spectrum basis. However, it is known that different variational approaches may give reasonable energies but predict other quantities less accurately. To show that the ansatz appropriately determines different experimentally accessible quantities, in \cite{2017PecakPRA}, a comprehensive discussion of the interpolatory ansatz's accuracy for $N_\uparrow=N_\downarrow=2$ particles was given. By calculating different single-particle and two-particle quantities, as well as projections to the non-interacting Fock states, it was shown that the ansatz surprisingly well predicts all these quantities also for different mass systems. To some extent, it is also possible to adopt the density matrix renormalization group scheme to study properties of confined few-particle systems. The method is however limited (similarly as exact diagonalization) to not too strong repulsions. A detailed comparison between this approach, exact diagonalization, and the interpolatory ansatz variational method was given in  \cite{2017BellottiEPJD}.

In the range of intermediate interactions, the variational approach can also be realized in a more standard way. Simply, one postulates some family of many-body states $|\Psi_{\boldsymbol{\alpha}}\rangle$ parametrized by a variational set of parameters $\boldsymbol{\alpha}=(\alpha_1,\alpha_2,\ldots)$ and then minimize the energy functional
\begin{equation}
E(\boldsymbol{\alpha}) = \frac{\langle\Psi_{\boldsymbol{\alpha}}|\hat{\cal H}|\Psi_{\boldsymbol{\alpha}}\rangle}{\langle\Psi_{\boldsymbol{\alpha}}|\Psi_{\boldsymbol{\alpha}}\rangle}
\end{equation}
to find the best approximation for the many-body ground-state. As discussed in \cite{2012RubeniPRA}, in the case of two particles ($N_\uparrow=N_\downarrow=1$) in a harmonic trap, the situation is quite simple since one can decouple the center-of-mass motion and adopt a variational method to the relative motion only. Of course, in the case of a larger number of particles, the problem is much more complicated since one needs to find an appropriate family of many-body functions which captures appropriately not only the correlations forced by interactions but also those which are induced by symmetrization conditions under exchange of same-spin fermions. 

One of the possible approaches is to go along the Jastrow ansatz scheme \cite{1995Jastrow} which assumes that all inter-particle correlations can be well-approximated by a product of two-particle correlations. In the case of two-component fermionic mixtures, this idea is captured by the ground-state wave function 
\begin{equation}
\Psi(\boldsymbol{r}_\uparrow,\boldsymbol{r}_\downarrow) \sim \Phi_\uparrow(\boldsymbol{r}_\uparrow)\Phi_\downarrow(\boldsymbol{r}_\downarrow)\prod_{i,j}\varphi(x_i^\uparrow-x_j^\downarrow),
\end{equation}
where $\boldsymbol{r}_\sigma=(x_{1\sigma},\ldots,x_{N\sigma})$ are algebraic vectors composed from the positions of all fermions of given spin $\sigma$, $\Phi_\sigma$ are many-body wave functions for different components $\sigma$  (they encode appropriately symmetry conditions under exchange of indistinguishable fermions), and $\varphi$ is the pair-correlation function encoding the inter-component two-body correlations. In general, all these functions can be treated as many-parameter variational probe functions. However, in the case of one-dimensional fermionic mixtures confined in a harmonic trap (where the analytical solution of the two-particle problem is known) one can propose a quite simple family of functions $\Phi_\sigma$ and $\varphi$ which results in the ansatz having the following properties: (i) in the case of $N_\uparrow=N_\downarrow=1$ the ansatz restores the exact solution for any interaction strength, (ii) for arbitrary number of particles the ansatz restores the solution in $g=0$ and $g\rightarrow\infty$. A detailed description of this construction is given in \cite{2013BrouzosPRA}. It is also worth noting that recently in \cite{2018KoscikEPL}, it was argued that the modified Jastrow approach having also these two properties can be significantly simplified and also nicely extended to other trapping potentials. 

\subsection{Attractive forces}

In the case of interacting fermions, one can consider also the attractive branch of inter-particle forces ($g<0$). It is justified since, due to artificial repulsions caused by fermionic statistics, the system  (more likely than bosonic system) avoids the density collapse and its ground-state remains stable. Quantum simulation of fermionic mixtures with attractive forces is quite interesting from both theoretical and experimental points of view since the problem directly corresponds to the very old idea of Cooper pairing \cite{1956CooperPR} -- the fundamental block of the theory of superconductivity. Moreover, for imbalanced systems, it may shed some light to different mechanisms of unconventional superconductivity. In the case of one-dimensional confined systems, the situation is even more interesting since standard approaches to the superconductivity assume homogenous and three-dimensional arrangement. From the perspective of few-body problems a natural question which can be addressed in this context is related to the number of particles needed to force the system to form strongly correlated pairs in a collective manner (see \cite{2010LiuPRA} for considerations for three fermions case). 

One of the first experiments on attractive ultracold fermionic mixtures confined in an array of one-dimensional tubes was performed with $^6$Li atoms in 2010 \cite{2010LiaoNATURE}. In this experiment, it was shown that, for spin-imbalanced mixtures, inter-particle correlations are in accordance with theoretical predictions, {\it i.e.}, the results may serve as some indirect proof for unconventional superconductivity. Two years later, the few-body regime was achieved with the same atoms in J. Selim's group \cite{2012ZurnPRL,2013ZurnPRL}. The experiments proved that attractive one-dimensional systems of a few fermions can be precisely prepared. By probing tunneling dynamics of an interacting system it was shown that attractions lead to substantial changes in observed probability distributions forced by pairing correlations. The first theoretical description of the tunneling dynamics was given in \cite{2012RontaniPRL}.

\begin{figure}[t]
\centering
\includegraphics[width=\linewidth]{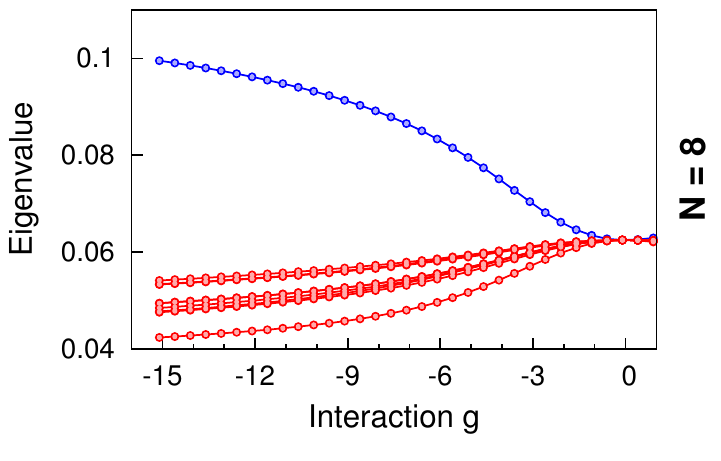}
\includegraphics[width=\linewidth]{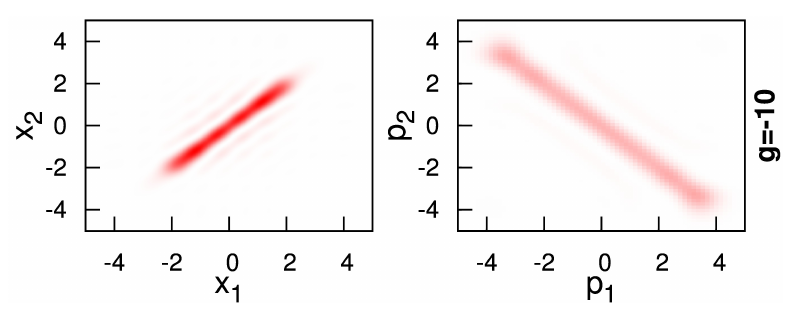}
\caption{(Upper panel) The largest eigenvalues of the two-particle reduced density matrix $\rho_{\uparrow\downarrow}^{(2)}$ as function of attractive interactions strength $g$. Note that, for increasing attractions, one of the eigenvalues strongly dominates in the system. (Bottom panel) Spatial (left) and momentum (right) density distribution of the dominant orbital in the strong attraction regime $g=-10$. Note the strong correlations in positions and strong anti-correlations in momenta. Results obtained for $N_\uparrow=N_\downarrow=N/2=4$ fermions. Figure adapted  from \cite{2015SowinskiEPL}. Copyright (2015) by EPLA. \label{_Fig-Attractive}}
\end{figure}

On theoretical footing, these general questions on pairing in few-fermion ultracold systems were addressed in two independent theoretical papers \cite{2015SowinskiEPL,2015DamicoPRA}. Based on the exact diagonalization of the Hamiltonian \eqref{Eq:FewFermionHam} (with equal masses and harmonic confinement) both groups found correlations between opposite spin fermions which emerge when attractive forces are present. In \cite{2015DamicoPRA} these correlations are quantified in terms of the conditional probability of finding opposite-spin fermions at given positions. By integrating out the center-of-mass of the appropriate pair-correlation function is determined. The paper focuses also on the {\it pairing gap}, understood as an energy loss or gain when additional unpaired fermion is present or absent in the mixture. These predictions for even-odd effect are in accordance with the corresponding experimental results in \cite{2013ZurnPRL}.

The theoretical strategy proposed in \cite{2015SowinskiEPL} is different and it is based on the properties of the reduced two-particle density matrix of opposite fermions $\rho^{(2)}_{\uparrow\downarrow}(x,y,x',y')$. It is shown that the fraction of Cooper-paired fermions, defined as the value of the dominant eigenvalue of this matrix, rapidly increases with increasing attractions in the system. At the same time, clearly visible correlations in positions and anti-correlations in momenta appear in the corresponding dominant orbital (see Fig.~\ref{_Fig-Attractive}). With this strategy, it is quite easy to determine properties of correlated fermions when the system is prepared not in the ground state, but in the thermal mixed state of many-body eigenstates. It was shown that the fraction of paired fermions decreases with temperature in accordance with predictions of the theory of superconductivity.

Due to experimental progress \cite{2013ZurnPRL}, some larger attention was also dedicated to the parity problem in attractive systems. As mentioned above, the first theoretical considerations in this direction were given in \cite{2015DamicoPRA}, where the fundamental energy gap is predicted to have a characteristic alternating sign with respect to the parity of the number of particles. This non-monotonic behavior of the gap is a direct consequence of the many-body ground-state energy which is not a convex function of the particle number. It shows that attractive forces lead to very non-trivial correlations between particles lowering the energy. This parity effect was addressed in a more comprehensive way in the recent work \cite{2016HofmannPRA}. Based on the perturbation theory and path integral approaches, the authors calculate the parity parameter \cite{1997MatveevPRL} in the weak attraction regime.  Treating this quantity as an appropriate order parameter, they showed that its behavior for a few-fermion system is fundamentally different when compared to systems with a macroscopic number of particles. 

Quite a different approach to imbalanced fermionic mixtures was presented in \cite{2015SowinskiFBS}, where systems with additional unpaired fermion were analyzed. Based on a description in the language of the reduced two-particle density matrix, it was shown that in the case studied, the pairing between fermions of opposite spins has completely different features that in the balanced case. Although for small attractive forces the two-particle orbital of correlated pairs indeed dominates in the many-body ground-state of the system, for strong interactions (far from the perturbative regime) another two-particle orbital, which does not manifest any significant pairing correlations, has the major contribution. It may suggest that a relevant description of the system in terms of the Cooper mechanism can be appropriate only in the perturbative regime of weak attractions. In a similar context, inhomogeneous superfluid pairing analogous to the Larkin-Ovchinnikov state was examined in 
\cite{2013BugnionPRL}. 

Complementary studies for attractively interacting system confined in a box potential were presented in
\cite{2015BergerPRA,2016McKenneyJPB}. Here, based on the Monte Carlo calculations in the real space, the authors explored different properties of the system. It was shown that the hard-walls lead directly to specific oscillations in density profiles (Fiedel oscillations) which in the presence of attractions change their pattern. This behavior is understood as a manifestation of pairing correlations. The analysis is nicely extended to the whole single-particle density matrix and adapted also to the momentum distribution. As shown, the distribution rapidly changes at the Fermi surface reflecting interaction effects. The authors analyze also short-distance behavior in terms of the Tan's contact density \cite{2008TanAnnPhys} encoding on-site pairing correlations. Let us also mention that a comprehensive description of few-fermion systems confined in finite-box with periodic boundary conditions  (in terms of one- and two-body reduced density matrices) was presented in \cite{2017RammelPRA}. 

Finally, a very nice extension of the attractively interacting fermions in one dimension was given in \cite{2007HaoPRA}, where instead of $s$-wave contact forces $p$-wave interactions were considered. Since this work is closely related to other scenarios exploiting $p$-wave interactions, we briefly discuss these results in the appropriate context (see subsection \ref{DifferentInteractions}). 

\subsection{Different mass fermions}
\begin{figure*}
\centering
\includegraphics{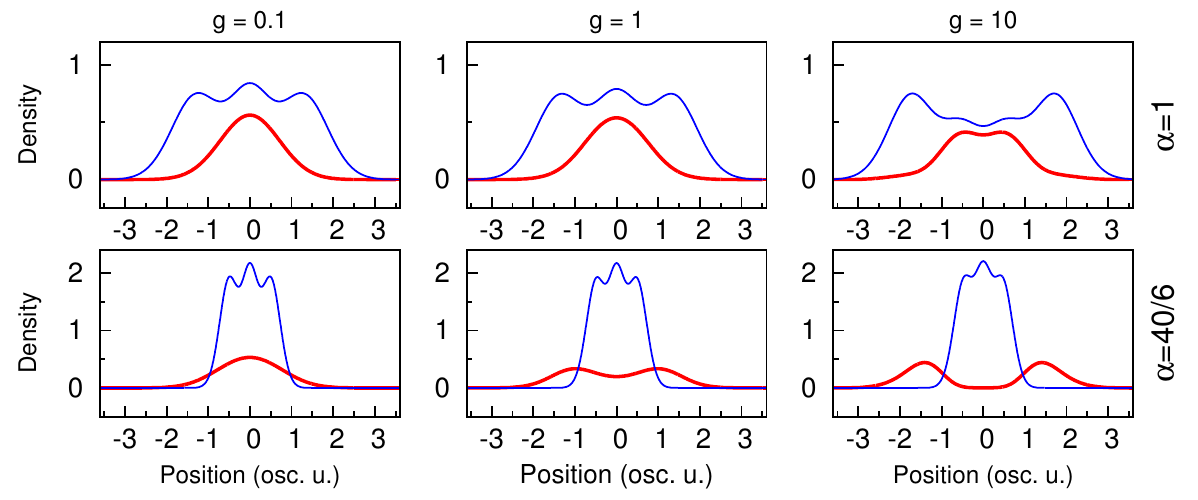}
\caption{Single-particle densities calculated in the ground state of four fermions (for $N_\uparrow =3$ and $N_\downarrow=1$) confined in a harmonic trap for different interactions and mass ratios $\alpha$ . For equal masses, the separation occurs in the species with the larger number of atoms (blue thin line). For the mass-imbalanced system ($\alpha = 40/6$), always the lighter component (red thick line) undergoes separation. Figure adapted from \cite{2016PecakNJP}. Copyright (2016) by IOP Publishing Ltd and Deutsche Physikalische Gesellschaft.  \label{_Fig-DifferentMassSeparation}}
\end{figure*}

Due to a strong experimental motivation \cite{Wille6Li40K,Tiecke2010Feshbach6Li40K}, the properties of fermionic mixtures of atoms having different masses are also widely studied. First, it is known that, in the case of a homogenous system, as shown in \cite{2013CuiPRL,2014CuiPRA}, a specific spatial separation in the single-particle density profiles occurs whenever the mass imbalance is sufficiently large. This behavior of the system strongly depends on the dimensionality and, in the case of one-dimensional systems, it has completely different properties than its three-dimensional counterpart. Along with this path,  using second-order perturbation theory, a deep analysis of such systems was performed, the thermodynamic equation of state was derived, and canonical phase diagrams for various mass ratios were determined in \cite{2014FratiniPRA}.

Although the mentioned results were obtained for systems in the thermodynamic limit, they motivated for similar research in the context of few-fermion systems. The very first analysis of the role of a mass difference (and also trap imbalance) was performed in the case of a two-component mixture of a few fermions confined in a three-dimensional spherical harmonic trap \cite{2008BlumePRA}. In the case of a one-dimensional harmonic oscillator, a comprehensive analysis was done in \cite{2016PecakNJP}. It was predicted that, independently on the number of particles in particular components, for strong enough repulsions the density profile of the lighter component is split to two domains which are pushed out from the center of the trap. At the same time, the density profile of the heavier component remains localized in the center. In contrast to the separation driven by the imbalance in the number of equal-mass fermions predicted in \cite{2014LindgrenNJP}, this behavior is found to be very robust to imperfections of the state preparation with domains having borders clearly visible (see Fig.~\ref{_Fig-DifferentMassSeparation}).

The properties of the separation of the density profiles induced by the mass difference between atoms belonging to different spin-components strongly depend on the shape of the external potential. This fact is a direct manifestation of the specific competition between the single-particle part of the Hamiltonian and the interaction terms which have different properties in different confinements. Note that, in the case of harmonic confinement, single-particle excitations are completely insensitive to the particle's mass (as long as both components are confined in the same-frequency trap). The only difference is present in the shapes of the single-particle orbitals which lead to different interaction integrals. In contrast, in the hard-wall box potential, orbitals' shapes remain unchanged but the single-particle spectrum is affected by the mass difference. As argued in  \cite{2016PecakPRA}, these observations are directly responsible for separations of different components in different shapes of external confinement. Moreover, when the shape of the trapping potential is adiabatically changed, the system may undergo a specific transition between different orderings. This transition has many interesting features which can be understood in the language of the critical transition phenomena.  

 Finally, let us note that the phase separation in binary mixtures induced by a different mass of atoms is a very general phenomena and it is present also for bosonic or mixed statistics. Some preliminary comparison of the effect for few-atom systems confined in a one-dimensional flat box potential, with special attention dedicated to spatial structures of density profiles, was given recently in \cite{2019ParajuliARX}.

A mass difference between atoms belonging to different components plays also a crucial role when attractive interactions are taken into account. It is directly related to the problem of pairing described in the previous subsection, which for different mass fermions has fundamentally different properties and is considered as one of the models leading to unconventional superconductivity \cite{2008BaranovPRA} (see also \cite{1984StewartRMP,2018KinnunenRPP} for specific reviews). In the one-dimensional few-body regime appropriate theoretical predictions were presented in \cite{2018PecakARX} for mixtures confined in a harmonic trap. The whole discussion was based on the inter-particle correlations encoded in the momentum noise correlation, Eq.~\eqref{NoiseCorb}. It was shown that along with increasing mass ratio $\mu$, the pairing correlations are strongly suppressed and the components become almost uncorrelated, {\it i.e.}, some kind of phase separation occurs in the system. Moreover, for the unexpectedly small mass ratio (around $\mu\approx 2$) the ground state can be well approximated by the product of the non-interacting ground state of the heavier component and some well-defined many-body state of lighter fermions. 

 \section{Bose-Fermi mixtures} \label{Sec:BoseFermi}

After discussing bosonic and fermionic few-body systems, here we focus on few-body mixtures of mixed statistics, {\it i.e.}, situations when particles belonging to different component have different statistics -- the Bose-Fermi mixtures (for current experimental situation see \cite{2016OnofrioUspPhys}). These mixtures form a specific bridge between purely bosonic and purely fermionic mixtures. A deep understanding of their properties forced by interactions and the specific role of the quantum statistics may be very helpful in many different contexts. 

 In the simplest case one considers a two-component Bose-Fermi mixture, {\it i.e.}, the system composed from $N_\mathrm{F}$ fermions and $N_\mathrm{B}$ bosons. The Hamiltonian of such system can be written in the form
\begin{align} \label{Eq:FewBosonFermionHam}
\hat{\cal H} &= \int\!\!\mathrm{d}x\,\hat\Psi^\dagger(x)\left[-\frac{\hbar^2}{2m_\mathrm{F}}\frac{\mathrm{d}^2}{\mathrm{d}x^2}+V_\mathrm{F}(x)\right]\hat\Psi(x) \nonumber \\
&+\int\!\!\mathrm{d}x\,\hat\Phi^\dagger(x)\left[-\frac{\hbar^2}{2m_\mathrm{B}}\frac{\mathrm{d}^2}{\mathrm{d}x^2}+V_\mathrm{B}(x)\right]\hat\Phi(x) \nonumber \\
&+\int\!\!\mathrm{d}x\,\hat\Phi^\dagger(x)\left[g_{BF}\hat\Psi^\dagger(x)\hat\Psi(x)+\frac{g_{BB}}{2}\hat\Phi^\dagger(x)\hat\Phi(x)\right]\hat\Phi(x)
\end{align}
where $m_\mathrm{F}$ ($m_\mathrm{B}$) denotes the mass of a fermionic (bosonic) atom, while $\hat\Psi(x)$ and $\hat\Phi(x)$ are fermionic and bosonic field operators obeying standard (anti-)commutation relations
\begin{subequations}
\begin{align}
\left\{\hat\Psi(x),\hat\Psi^\dagger(x')\right\} =
\left[\hat\Phi(x),\hat\Phi^\dagger(x')\right]&=\delta(x-x'), \\
\left\{\hat\Psi(x),\hat\Psi(x')\right\} =
\left[\hat\Phi(x),\hat\Phi(x')\right]&=0.
\end{align}
\end{subequations}
Inter-particle interactions are controlled by two parameters $g_{\mathrm{BB}}$ and $g_{\mathrm{BF}}$ describing the strength of the contact intra-component boson-boson and inter-component boson-fermion forces, respectively. Of course the Hamiltonian commutes independently with the operators of the total number of fermions $\hat{\cal N}_\mathrm{F}=\int\mathrm{d}x\,\hat\Psi^\dagger(x)\hat\Psi(x)$ and bosons $\hat{\cal N}_\mathrm{B}=\int\mathrm{d}x\,\hat\Phi^\dagger(x)\hat\Phi(x)$. Therefore, its properties can be analyzed in the subspaces of fixed $N_\mathrm{B}$ and $N_\mathrm{F}$.  Note that, in the simplest case, one neglects intra-component interactions for fermions from the same reasons they were neglected for pure fermionic mixtures in Sec.~\ref{Sec:Fermions}. At this point we want to point out that in many theoretical divagations on Bose-Fermi mixtures two very important simplifications are assumed. First, typically it is assumed that bosons and fermions have exactly the same mass, $m_\mathrm{B}=m_\mathrm{F}$. From the experimental point of view, one should note that, in fact, it is very challenging to find such mixtures since only isobars of different elements may have this property. In practice, equal mass assumption is fulfilled only approximately. For example in the case of $^6$Li-$^7$Li mixture \cite{2001SchreckPRL,2001TruscottScience} the deviation is quite large, $m_\mathrm{B}/m_\mathrm{F}\approx 7/6$, while in the case of heavy mixture $^{173}$Yb-$^{174}$Yb \cite{2009FukuharaPRA} one finds almost perfect equality, $m_\mathrm{B}/m_\mathrm{F}-1< 0.6\%$. For intermediate cases $^{40}$K-$^{41}$K \cite{2011WuPRA} or $^{86}$Rb-$^{87}$Rb deviations are less than $5\%$. Of course the assumption of equal mass is completely not justified in other Bose-Fermi mixtures like $^{40}$K-$^{87}$Rb \cite{2002RoatiPRL}, $^6$Li-$^{23}$Na \cite{2002HadzibabicPRL}, or $^6$Li-$^{133}$Cs \cite{2017DeSalvoPRL}. Second, it is assumed that both components experience the same external trapping potential. In principle, different elements response differently to optical pokes and introducing effectively the same external trapping for both components may be very challenging. Both simplifications should be treated  carefully when precise calculations for experimental setups are performed, since they introduce additional non-trivial symmetries to the Hamiltonian which are not present in realistic systems \cite{2005ImambekovAPhys}. 

\begin{figure}
\centering
\includegraphics[width=\linewidth]{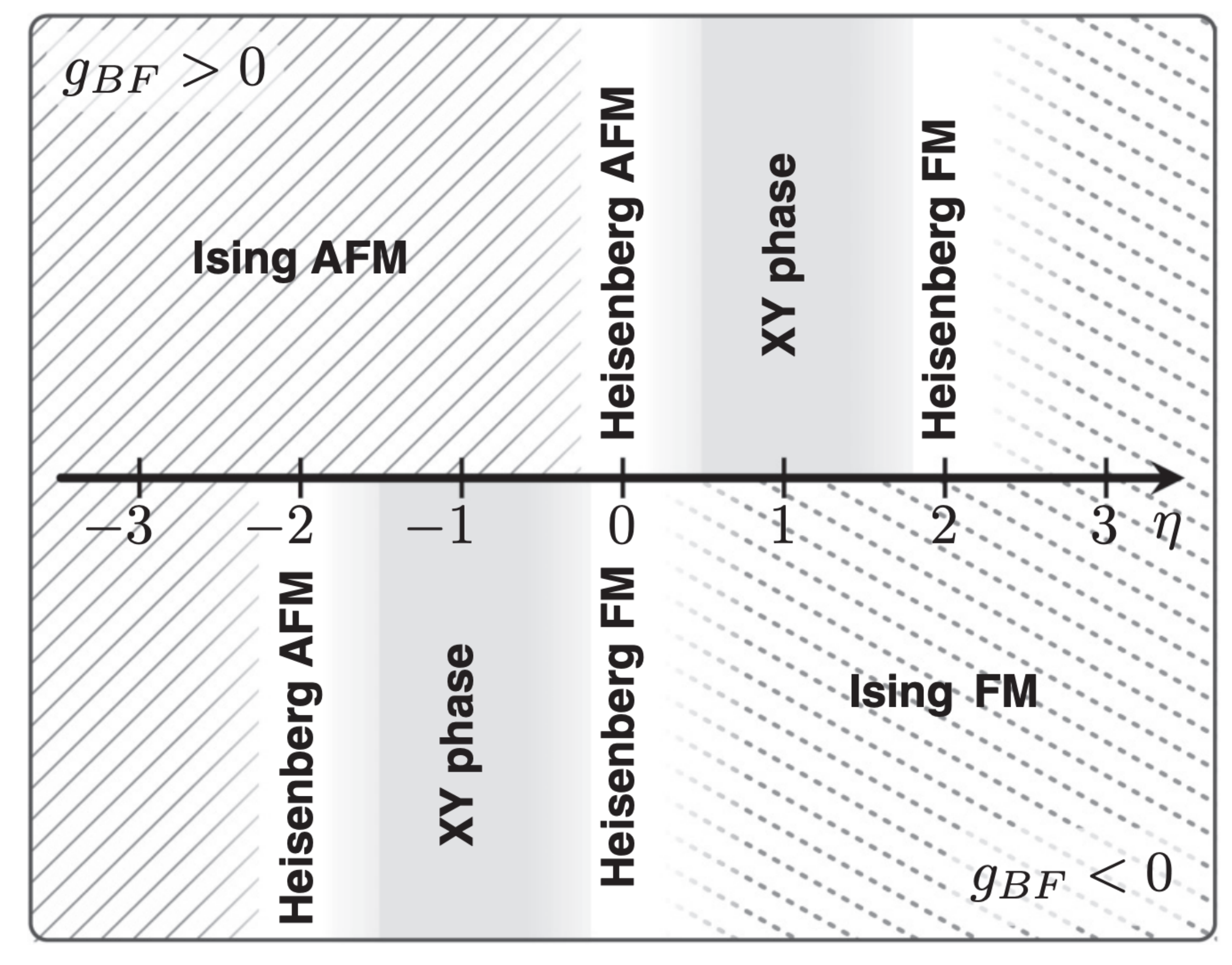}
\caption{Schematic phase diagram of the Bose-Fermi mixture in the limit of strong repulsion. Depending on the ratio $\eta=|g_{BF}|/g_{BB}$ and the sign of the boson-fermion interaction strength $g_{BF}$, the system is effectively described by different spin-chain models. Figure adapted from \cite{2017DeuretzbacherPRA}. Copyright (2017) by the American Physical Society. \label{_Fig-BoseFermiChain}}
\end{figure}

First, let us discuss the properties of the few-body Bose-Fermi mixtures in the limit of very strong interactions, {\it i.e.}, when interactions $|g_{BB}|$ and $|g_{BF}|$ tend to infinity. To formally consider this limit, let us assume that all particles have the same mass $m=m_\mathrm{B}=m_\mathrm{F}$ and they experience the same harmonic potential $V_\mathrm{F}(x)=V_\mathrm{B}(x)=m\omega^2x^2/2$. Then we can express all quantities in natural units of the harmonic oscillator and remain with only two dimensionless interaction couplings (expressed in units of $\hbar\sqrt{\hbar\omega/m}$) which can be considered as close to infinity. Consequently, according to the classification done in section~\ref{Sec:Bosons}, we can consider the four limits:
\begin{itemize}
 \item the ideal gas limit, when $g_{\mathrm{BB}}=g_{\mathrm{BF}}=0$, 
 \item TG limit, when $g_{\mathrm{BB}}\to\infty$ and $g_{\mathrm{BF}}=0$, 
 \item the phase separation limit (PS), when $g_{\mathrm{BB}}=0$ and $g_{\mathrm{BF}}\to\infty$,
 \item the full fermionization limit (FF) when both  $g_{\mathrm{BB}}\to\infty$ and $g_{\mathrm{BF}}\to\infty$. 
\end{itemize}

The ideal gas limit is trivial, and the wave function of the many-body ground state of the system can be easily written as a product of wave functions of $N_\mathrm{B}$ non-interacting  bosons and $N_\mathrm{F}$ non-interacting  fermions
\begin{equation}
\psi(\boldsymbol{r}_\mathrm{B},\boldsymbol{r}_\mathrm{F}) = \Phi_0(\boldsymbol{r}_\mathrm{B})\Psi_0(\boldsymbol{r}_\mathrm{F})
\end{equation}
where $\boldsymbol{r}_\mathrm{B}=(x_1,\ldots,x_{N_\mathrm{B}})$ and $\boldsymbol{r}_\mathrm{F}=(y_1,\ldots,y_{N_\mathrm{F}})$ are position vectors of bosons and fermions respectively, while $\Phi_0(\boldsymbol{r}_\mathrm{B})\propto\mathrm{exp}(-\boldsymbol{r}^2_\mathrm{B}/2)$ and 
$\Psi_0(\boldsymbol{r}_\mathrm{F})$ is given by \eqref{SlaterFermion}. The TG limit is also trivial since in this case the system is a composition of two completely independent gases -- a non-interacting fermionic gas and a single-component bosonic gas in the TG limit.

In the PS limit, the many-body ground state manifests a specific phase separation in the ground-state density profiles \cite{2017DeuretzbacherPRA}. Namely, the ideal gas of bosons occupies the vicinity of the center of the trap, while the density profile of the non-interacting fermionic component is split into two parts,  which are pushed out to the edges of the system to assure minimization of the interaction energy. This behavior of the system is a spectacular manifestation of the quantum statistics showing that, in the case of two mutually interacting ideal gasses, different scenarios are attributed to bosonic and fermionic components.

In the case of the full fermionization limit of Bose-Fermi mixtures, the first theoretical analysis in a harmonic trap was given in~\cite{2007GirardeauPRL},~\cite{2009FangPRA} and~\cite{2011FangPRA}. In this limit, due to infinite repulsion between bosons the system's wave function need to fulfill the condition $\psi(\boldsymbol{r}_{\mathrm{B}},\boldsymbol{r}_{\mathrm{F}})=0$ whenever $x_i=x_j$. Similarly, due to infinite repulsion between bosons and fermions, $\psi(\boldsymbol{r}_{\mathrm{B}},\boldsymbol{r}_{\mathrm{F}})=0$ whenever $x_i=y_j$. Finally, due to the fermionic statistics, $\psi(\boldsymbol{r}_{\mathrm{B}},\boldsymbol{r}_{\mathrm{F}})=0$ whenever $y_i=y_j$. All this means that, in this particular limit, the total wave function can be constructed from the purely fermionic wave function (the Slater determinant) for all $N=N_\mathrm{B}+N_\mathrm{F}$ particles provided that the appropriate symmetrization within bosonic positions will be applied, {\it i.e.}
\begin{equation}
\psi(\boldsymbol{r}_\mathrm{B},\boldsymbol{r}_\mathrm{F}) = {\cal S}_\mathrm{B}\left(\mathrm{det}\left[\varphi_i(r_j)\right]^{i=0,\ldots,N-1}_{j=0,\ldots,N-1}\right),
\end{equation}
where ${\cal S}_\mathrm{B}$ is the symmetrization operator imposing the positive sign over the wave function whenever two bosons have interchanged positions. One should point out that there is not any symmetrization condition when the position of a boson is interchanged with the position of a fermion in the determinant. Therefore, in the FF limit, the ground state has a large degeneracy, equal to $N!/(N_{\mathrm{B}}N_{\mathrm{F}})$. A complete classification of these states from the immanent symmetries of the system is given in~\cite{2017DecampNJP} while in~\cite{2016HuNJP} it is discussed how the degeneracy of the ground manifold is lifted when the interactions are large but finite. 

As discussed above, the system is controlled by two independent dimensionless quantities $\eta=|g_{BF}|/g_{BB}$ and the sign of $g_{BF}$. Therefore, one can approach the limit of strong repulsion in many different ways. It is clearly visible when the description of the system is carried out in the appropriate spin-chain representation. In fact, in the vicinity of infinite interactions, similarly to the case of fermionic mixtures, any Bose-Fermi system can be effectively described in this language~\cite{2016DeuretzbacherPRA,2017DeuretzbacherPRA}. In this case, however, the type of the resulting spin-model strongly depends on the two parameters mentioned (see Fig.~\ref{_Fig-BoseFermiChain}).

The particular case of $N_{\mathrm{B}}=N_{\mathrm{F}}=2$ particles is studied in~\cite{2011FangPRA} from the spatial densities point of view and linked to different orderings of bosons and fermions in the simplest case of $\eta=1$. The same case is also numerically studied in~\cite{2017DehkharghaniJPB}. Here, the authors find again strong indications of density separation, and differences in the odd-even number of atoms case (an effect already noted in~\cite{2011FangPRA}). 

A detailed study of the transition from the limit of ideal gases ($g_{BB}=g_{BF}=0$) to the FF limit (with $\eta=1$) is provided in~\cite{2012WangPRA} with the density functional theory, up to $N_{\mathrm{B}}=N_{\mathrm{F}}=10$ atoms. Interestingly, the authors show how the majority of bosons become concentrated in the center of the trap while fermions are pushed out to the edges as interactions are increased. Comprehensive analysis of the system's properties for intermediate interactions was performed in~\cite{2018ChenPRA} where interesting crossovers between different limits were explored. Additionally, the consequences of different masses of particles in individual components are studied. 
As argued in~\cite{2018ChenPRL}, for not too strong interactions, the properties of the individual components can be well understood in the framework of the simplified model obtained by taking into account effective intra-component interactions induced by the inter-component correlations.

Finally, although it is beyond the scope of this review, let us mention that one-dimensional few-body Bose-Fermi mixtures confined in different external potentials were also considered in the theoretical analysis. We mention here double-well traps~\cite{2009KelasPRA}, split traps~\cite{2010LuPRA}, or optical lattices~\cite{2010ChenPRA}.

\section{Other extensions} \label{Sec:OtherExt}
In this section, we would like to bring attention to different, non-standard extensions of the models described above. Additionally, our aim is to display some still undiscovered lines of possible explorations which may be very fruitful and may bring very interesting results. 

\subsection{Beyond s-wave interactions} \label{DifferentInteractions}

In the majority of works about fermionic mixtures, inter-particle interactions between fermions are assumed to be $s$-wave contact. In the one-dimensional scenario, they are modeled by a simple $\delta$-like potential (in higher dimensions $\delta(r)$ is not self-adjoint operator and some specific regularization is needed \cite{1998BuschFoundPhys}). Consequently, as mentioned before, this leads directly to vanishing interactions between particles belonging to the same component and only inter-component forces are present. The first step towards taking into account interactions between identical fermions is to include higher partial waves of the scattering process. Going along this line the simplest extension is to take into account $p$-wave scattering potential of the form \cite{1986SebaRMP,2004GirardeauPRA,2003SenJPhysA}
\begin{equation} \label{Eq:p-wavePot}
V_p(r) = g_p \overleftarrow{\frac{\partial}{\partial r}}\delta(r)\overrightarrow{\frac{\partial}{\partial r}},
\end{equation}
where $r=x_i-x_j$ is a distance between interacting particles and $g_p$ is $p$-wave interaction strength. A mining of the the non-symmetric derivative in \eqref{Eq:p-wavePot} is clarified when matrix elements of the the $p$-wave potential are calculated. If $\phi_1(r)$ and $\phi_2(r)$ are some wave functions of the relative motion of two particles then one has:
\begin{equation}
\int\mathrm{d}r \phi_1^*(r)V_p(r)\phi_2(r)= g_p\left.\frac{\partial \phi_1^*(r)}{\partial r}\right|_{r=0}\left.\frac{\partial \phi_2(r)}{\partial r}\right|_{r=0}
\end{equation}
,{\it i.e.}, the first derivative acts to the left and the second to the right \cite{2004KanjilalPRA}. In the second quantization formalism the $p$-wave interaction between fermionic particles can be written equivalently as
\begin{equation}
\hat{\cal H}_p = g_p\int\!\!\mathrm{d}x\,\frac{\mathrm{d}\hat{\Psi}^\dagger(x)}{\mathrm{d} x}\hat{\Psi}^\dagger(x)\hat{\Psi}(x)\frac{\mathrm{d}\hat{\Psi}(x)}{\mathrm{d} x}.
\end{equation}

In the context of a one-dimensional few-particle system, the opening result was presented in \cite{199CheonPRL} where a specific duality between $s$-wave interacting bosons and $p$-wave interacting fermions was displayed. It was proven that the many-body ground-state wave function of $N$, $p$-wave fermions, $\psi_\mathrm{F}(x_1,\ldots,x_N)$, can be constructed directly from the ground-state wave function of $N$, $s$-wave interacting bosons, $\psi_\mathrm{B}(x_1,\ldots,x_N)$,  as follows:
\begin{equation}
\psi_\mathrm{F}(x_1,\ldots,x_N) = \left[\prod_{i<j}^{N}\mathrm{sgn}(x_i-x_j)\right]\psi_\mathrm{B}(x_1,\ldots,x_N)
\end{equation}
provided that $g_p=-\hbar^4/(\mu^2 g)$. Here,  $\mu$ is the reduced mass of the relative motion of two particles (see also \cite{2003GirardeauARX}). This interesting mapping was pointed out also directly in a very pedagogical paper \cite{2004KanjilalPRA}, where the exact analytical solution (counterpart of the Busch {\it et. al} solution) for the two-particle problem with $p$-wave interactions in a harmonic trap was obtained (see also \cite{2016CuiPRA} for the solution obtained with different method). Based on this solution, in \cite{2006SunPRA} the first analysis of inter-particle quantum correlations in the interacting two-body ground state was studied and compared with those induced by $s$-wave interactions. In this framework, single-particle densities in the position and momentum domains were obtained. Deep analysis of the $p$-wave interacting fermions confined in a harmonic trap from the correlations point of view was presented in \cite{2005BenderPRL}.
Finally, the semi-analytical solution based on the Bethe ansatz approach for a system of a few fermions confined in a hard-wall potential was given in \cite{2007HaoPRA}. In the same external arrangement, an appropriate mapping from continuous to discretized model was analyzed in \cite{2010MuthPRA}.  As shown recently in \cite{2014ZhangPRA}, an analysis in the opposite direction is also possible, {\it i.e.}, properties of $p$-wave interacting fermionic systems can be exploited (via mentioned specific Bose-Fermi mapping) to study bosonic system being close to TG limit of infinite interactions.

A very nice extension of the $p$-wave interactions problem to the case of two-component Fermi-Fermi and Bose-Fermi mixtures was presented in \cite{2016HuPRA}. In these cases, the systems studied are described with the standard Hamiltonians \eqref{Eq:FewFermionHam} (Fermi-Fermi) and \eqref{Eq:FewBosonFermionHam} (Bose-Fermi) extended by an additional $p$-wave interaction \eqref{Eq:p-wavePot} in the fermionic components. The authors focus on the problem of strongly repelling region and they show that additional $p$-wave forces substantially change the description in the language of effective spin-chain Heisenberg model. In consequence,  different additional magnetic phases can be reached in the ground-state of the system. Going along a similar line for Fermi-Fermi mixture, the problem of experimental engineering of these new behaviors was addressed in \cite{2016YangPRAc}. 

All these studies on contact limits of  finite-range interactions seem to be still not fully uncovered. Although the direct mapping of the eigenstates to the bosonic problem is clarified, further explorations in this directions may bring us much closer to our understanding of the  zero-range limit of interactions in the case of a mesoscopic number of particles.
 
\subsection{Dipolar interactions}
One possible extension when ultracold mixtures are considered is to take into account some long-range part of the inter-particle interactions. In the context of atomic physics, the most natural long-range interaction between neutral atoms is the interaction immanently present between magnetic or electric atomic dipoles, {\it i.e.}, dipolar forces. In view of the recent ground-breaking experiments with ultracold two-component mixtures of magnetic atoms \cite{2018TrautmannPRL}, this direction for extending previous results in the context of few-body systems seems to be very promising. In contrast to contact interactions, dipolar forces are not isotropic and therefore their one-dimensional reduction strongly depends on mutual orientation between interacting dipoles. For example, if interacting dipoles are oriented parallelly in $z$ direction and their relative position is $\boldsymbol{r}=\boldsymbol{r}_1-\boldsymbol{r}_2=(x,y,z)$, then the interaction potential has the form
\begin{equation}
V_d(\boldsymbol{r}) = \frac{\gamma}{|\boldsymbol{r}|^3}\left(1-\frac{3 z^2}{|\boldsymbol{r}|^2}\right),
\end{equation}
where $\gamma$ measures the strength of the dipolar interaction. To obtain an effective one-dimensional interaction potential between particles (in $x$ direction) one assumes that the confinement in perpendicular directions is very strong and particles occupy only the lowest state in these directions. If the transverse confinement is harmonic with frequency $\Omega_\perp$, one can write this wave function as $(\kappa/\sqrt{\pi})\,\mathrm{exp}(-\rho^2/2\kappa^2)$, where $\rho=\sqrt{y^2+z^2}$ and $\kappa=\sqrt{\hbar/m\Omega_\perp}$. After integrating out these frozen degrees of freedom in the interaction integral, one finally obtains an effective one-dimensional interaction potential of the form (see \cite{2007SinhaPRL,2010DeuretzbacherPRA} for a detailed derivation and generalizations to other dipoles orientations):
\begin{align} \label{dipolar1d}
&V_d(\xi)= \\
& V_0\left[ -2\xi + \sqrt{2\pi}\left(1+\xi^2\right)\mathrm{e}^{\,\xi^2/2}\mathrm{erfc}\left(\xi/\sqrt{2}\right)-\frac{8}{3}\delta(\xi)\right],\nonumber
\end{align}
where $\xi=|x|/\kappa$ is a dimensionless distance between dipoles and $V_0=-\gamma/(8\kappa^3)$ is an effective dipolar interaction strength controlled by the transverse width. At large distances ($\xi\gg 1$) the potential \eqref{dipolar1d} can be expanded as
\begin{equation*}
V_d(\xi) = \frac{4V_0}{\xi^3}\left(1-\frac{6}{\xi^2}+\frac{45}{\xi^4}-\ldots\right)
\end{equation*}
and the leading term has the typical form for dipolar interactions, $1/\xi^3$. This observation opens a route to a simplified description of the system in the crystallization limit (similarly as done for ultracold ions \cite{1998JamesAPhysB}), {\it i.e.}, when particles are spatially well-separated \cite{2015KoscikPLA,2017KoscikFBS,2018BeraARX}. Note that the general form of the one-dimensional dipolar potential \eqref{dipolar1d} can be very unstable from the numerical point of view if calculated straightforwardly. However, appropriate numerical routines approximating its form with very well-behaving expressions are known \cite{1968OldhamMatComp}. 

Up to now, in the few-body regime, most of the attention was devoted to the problem of bosonic mixtures of polar atoms. Let us mention the most important milestones in building our understanding of such systems in one-dimensional traps. First studies, in the case of harmonic confinement, were deeply conducted in \cite{2010DeuretzbacherPRA}, where different single-particle properties of the system were presented. Later, different properties of a few dipolar bosons confined in a double- \cite{2011AbadEPL}, triple- \cite{2013GallemiPRA}, and four-well \cite{2015MazzarellaPhysB}, as well as in periodic lattice \cite{2012SowinskiPRL} were studied. In all these cases, it was argued that the interaction-induced tunnelings may have fundamental importance for the ground-state properties of the interacting particles. Due to the long-range form of the dipolar forces, one can also consider the problem of coupled bosonic systems confined in independent quasi-one-dimensional tubes. This path of exploration was presented recently in \cite{2013VolosnievNJP} and \cite{2018BjerlinPRA}, where an effective description of the system in the language of weakly interacting bound states was proposed. 
 
In the case of a sole fermionic component and Bose-Fermi mixtures only a few (introductory) results were presented already \cite{2013DeuretzbacherPRA,2015GrassPRA}. Therefore, we see a large field to be uncovered with theoretical as well as experimental approaches in the area of few-body mixtures of ultracold polar particles. 

\subsection{Artificial quantum statistics}
In the case of one-dimensional few-body systems, some attention was also dedicated to the case of artificial anyonic statistics. Although anyonic systems seem to be very exotic, they become widely studied in the context of topological computations based on precise quantum control (for comprehensive reviews see \cite{2008NayakRMP,2012PachosBook}). It seems also possible that anyonic statistics can be engineered in ultracold atomic systems \cite{2008AguadoPRL,2011KeilmannNatCom}. In the simplest case of the one-dimensional problem, one considers a single-component system of $N$ particles described by a Hamiltonian very similar to the single-component bosonic Hamiltonian of the form
\begin{align} \label{HamAnyons}
\hat{\cal H} &= \int\!\!\mathrm{d}x\,\hat\Upsilon^\dagger(x)\left[-\frac{\hbar^2}{2m}\frac{\mathrm{d}^2}{\mathrm{d}x^2}+V(x)\right]\hat\Upsilon(x) \nonumber \\
&+g\int\!\!\mathrm{d}x\, \hat\Upsilon^\dagger(x)\hat\Upsilon^\dagger(x)\hat\Upsilon(x)\hat\Upsilon(x).
\end{align}
However, in this case, one imposes the anyonic commutation relations to the field operator $\hat\Upsilon(x)$. In one-dimensional cases, these relations read
\begin{subequations}
\begin{align}
\hat\Upsilon(x)\hat\Upsilon^\dagger(x')-\mathrm{e}^{i\kappa\,\sigma(x-x')}\hat\Upsilon^\dagger(x')\hat\Upsilon(x)&=\delta(x-x'), \\
\hat\Upsilon(x)\hat\Upsilon(x')-\mathrm{e}^{i\kappa\,\sigma(x-x')}\hat\Upsilon(x')\hat\Upsilon(x)&=0,
\end{align}
\end{subequations}
where the signum function $\sigma(x)=-1,0,+1$ for $x<0$, $x=0$, and $x>0$, respectively, while the anyonic parameter $\kappa\in[0;\pi]$ determines the statistics and for $\kappa=0$ ($\kappa=\pi$) it corresponds to standard bosonic (fermionic) statistics. The analysis of anyonic problems in the context of ultracold systems in one spatial dimension is started in \cite{2006GirardeauPRL} with an observation that strongly repelling anyons can be one-to-one mapped to non-interacting fermions with almost the same arguments as in the case of bosons. In this way, it may be applied to the different exact solutions known. For example, in the case of a uniform box potential, the Hamiltonian \eqref{HamAnyons} takes the form of the standard Lieb--Liniger Hamiltonian (except the quantum statistics) and therefore it can be diagonalized analytically in terms of Bethe ansatz. First attempts to find ground-state properties of the system were given in \cite{2008HaoPRA} where the single-particle momentum distributions for different statistics were compared for different interactions strengths. This analysis can also be extended by adding an additional two-body velocity-dependent interaction term \cite{2008BatchelorJPhysA}.  First analysis of the few-anyon systems from the momentum distribution point of view  in a hard-core limit was provided in \cite{2008CampoPRA}. In the case of non-uniform external confinements, and in the strong repulsion limit, a comprehensive discussion was given in \cite{2015ZinnerPRA} and properties of the system based on different particle configurations were presented. Finally, as pointed out in \cite{2018ColcelliEPL}, the off-diagonal long-range order encoded in the reduced single-particle density matrix strongly depends on the quantum statistics, {\it i.e.}, its dominant eigenvalue $\lambda_0$ scales with the number of particles as $\lambda_0\sim N^C$, where $C$ strongly depends on interactions strength $g$ and the anyonic parameter $\kappa$.

 Finally, let us mention that studying anyonic systems can be very helpful when properties of the standard statistics systems are considered. For example, in \cite{1999KunduPRL} it was shown that a one-dimensional model of bosons interacting via two- and three-body contact forces is exactly equivalent to the problem of anyonic particles interacting via only two-body forces. Therefore, the initial model, recognized previously as unsolvable, can be fully solved with the Bethe ansatz approach.

\subsection{Tunneling to the open space}
\begin{figure}[t] 
\includegraphics[width=\linewidth]{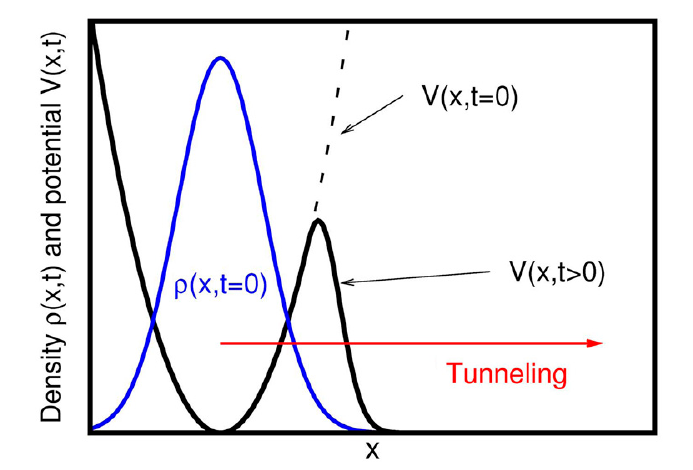}
\caption{Typical scenario of the tunneling process to the open space. At initial moment particles are confined in a trap (black dashed line) and they occupy the many-body ground state (blue line). Then at $t=0$ the trap is suddenly opened and particles may tunnel through a potential well (black solid line). Figure adapted from \cite{2012LodePNAS}. Copyright (2012) by the National Academy of Sciences.\label{_Fig-TunnelingLode}}
\end{figure}

Although in this review we do not discuss dynamical problems in one-dimensional few-body systems, here we would like to mention one of the possible directions which seems to be very challenging and opening new possibilities. It is related to the experimental schemes in which one measures static properties of interacting systems indirectly by capturing escaping particles after a sudden opening of a trap. In a typical standard scenario, one assumes that initially, the interacting particles occupy the many-body ground state $|\Psi_{ini}\rangle$ of the initial Hamiltonian (\eqref{Eq:Hamiltonian2ndQTS} and \eqref{Eq:FewFermionHam} for bosons and fermions, respectively) and then the trapping potential becomes suddenly opened and the system evolves in time (see Fig.~\ref{_Fig-TunnelingLode}). One of the simplest quantity encoding properties of the system is the survival probability ${\cal P}(t)=\langle\Psi_{ini}|\Psi(t)\rangle$, {\it i.e.}, the probability that after time $t$ the system remains in the initial state. Inspired by experimental results from \cite{2005ChuuPRL}, the first theoretical description of a decaying few-body system was given in \cite{2006CampoPRA}. Assuming that particles are initially confined in a hard-wall trap and starting from $t=0$ they may tunnel through a delta-like barrier, it was shown that, in the case of strongly repelling bosons (or non-interacting fermions), inter-particle correlations lead to violations in the exponential decay predicted for a single-particle case \cite{2011CampoPRA}.  

The particular case of tunneling ultracold bosons was studied theoretically in several papers. First in \cite{2009LodeJPhysB}, by solving time-dependent many-body Schr\"odinger equation straightforwardly, it was shown that even for a weakly interacting few-boson system, the decay process is not captured by a mean-field description and may substantially depend on the number of particles. This path of exploration was continued in \cite{2012LodePNAS}, where it was argued that the tunneling process of the initially coherent cloud of bosons can be viewed as a superposition of individual single-particle decays from sources having a different number of particles. In consequence, the system loses its coherence and becomes fragmented. This tunneling process can be controlled by tuning the shape of the external confinement or the inter-particle interactions \cite{2014LodePRA}. In this context, the interested reader can find a quite comprehensive description of decaying few-bosons system in a recent monograph \cite{2014LodeBook}.

The case of two interacting ultracold particles was studied theoretically with increased attention. In \cite{2011KimJPhysB} the decay through the delta-like barrier was studied. Then, due to the spectacular experiments with decaying two $^6$Li atoms \cite{2012ZurnPRL,2013ZurnPRL}, the tunneling through the barrier of the form 
\begin{equation}
V(\xi) = V_0\left[1-(1+\xi)^{-1}\right] - \alpha\xi,
\end{equation}
where $V_0$ and $\alpha$ are controlled experimentally while $\xi$ is the dimensionless position, was deeply studied in \cite{2015GharashiPRAb} and \cite{2015LundmarkPRA}. 

Finally, the decay problem for attractively interacting bosons was analyzed recently in \cite{2018DobrzynieckiPRAb,2019DobrzynieckiARX}. In these works, it was shown that, depending on the attraction strength, different tunneling processes dominate in the dynamics of the system. For strong attractions, particles tunnel as one composite system, while for weak attractions particles are emitted sequentially one by one. For intermediate interactions, depending on the number of particles in the initial state, the dynamics is dominated by different collective few-boson decay. 

In this general context of decaying one-dimensional systems of ultracold atoms we want to point out that, until now, all studies were devoted  almost only to bosons or two distinguishable particles (for fermionic cases see \cite{2011CampoPRA,2012PonsPRA}). Therefore a whole area of problems related to two-component mixtures of fermionic particles (or Bose-Fermi mixtures) is still  almost undiscovered. For sure, theoretical and experimental research in these directions may bring interesting results and uncover the role of the quantum statistics in the decay problems.

\section{Summary and Final remarks} \label{Sec:Final}

One-dimensional systems of bosonic, fermionic, and Bose-Fermi mixtures of a few ultracold atoms confined in parabolic traps is a topic that has attracted a considerable research effort in recent years. This intense theoretical work was triggered by the experiments that realized the Tonks-Girardeau gas~\cite{paredes2004tonks,2004KinoshitaScience} and more recently, by the striking experiments on few trapped fermions~\cite{2011SerwaneScience,2013WenzScience,2015MurmannPRL,2015MurmannPRLb}, or the experiments in one-dimensional multicomponent fermionic mixtures~\cite{2014PaganoNatPhys}, to mention few.    

For bosonic mixtures,  after short discussion of attractive forces, we first described the four conceptual pillars, coming from the thermodynamic, large $N$ limit,  that help build the developments in the few atom trapped case: the Tonks-Girardeau gas, the composite fermionized gas, the full-fermionized mixture, and the phase separation in bosonic mixtures. With these concepts at hand, we described eight limiting cases of interactions, and we use this classification to structure the description of the theoretical description from the literature of systems of three and four atoms and the extensions to larger numbers of atoms. We also reported on the studies on the transitions between limits. We remark here that, whilst these transitions are quite well studied in systems of three and four atoms, there is still room for future research progress in the cases of intermediate interactions, particularly for polarized cases, {\it i.e.}, when the number of atoms in each component is different. We also described the mass imbalanced systems.  We pay special attention to impurity systems, where one of the species consist only in one atom. Intense research in impurity problems in the few atoms and large atom cases is attracting a lot of interest recently, and many future studies are expected in the few-atom and strongly interacting cases.  We also described spinor Bose gases with a small number of atoms. Here, while the limiting cases are already described, transitions between limits remain open, among other open questions, such as mass imbalanced systems.  We also gave some remarks from the study of the symmetries in a few atom systems that guided theoretical developments in this area. Anyhow, the detailed study of symmetries in these systems is too long and out of the scope of this review. Here,  we only gave some description of these concepts, only to the extent that it may help understand the systems at hand.  
     
For fermionic mixtures, we started with a detailed description of the role of the spin in these systems. Then, we briefly outlined the results on multicomponent systems, {\it i.e.} higher-spin mixtures. From here on, we focused on two-component mixtures, and described in detail the two- and three-atom cases and the impurity problem in a small Fermi sea. Importantly, in the case of very strong repulsion, thus close to infinite repulsion, we detailed the proposed solutions and the descriptions based on an effective spin chain Hamiltonian. We henceforth described the numerical and theoretical efforts made to study intermediate interaction cases, which, as in the bosonic case, remain more elusive to theoretical descriptions, leaving room for further future research. We end the few-atom fermionic mixtures discussion with two important topics: mass imbalance and attractive interactions. Both are of crucial importance for fermionic mixtures due to their experimental and conceptual implications,  for example, in studies on Cooper-pairing or unconventional superconductivity. 

Using a similar conceptual structure based on the limiting cases used for bosonic mixtures, we describe the recent results on Bose-Fermi mixtures. Here, most of the analytical results belong to the strongly repulsive case, {\it i.e.}, when inter-component or intra-component (for the bosonic atoms), tend to infinity.  

We end our report by briefly discussing a subjectively selected collection of topics which we believe will attract future research interest and give rise to important scientific advancements. This includes more exotic interactions (beyond $s$-wave) such as $p$-wave or dipolar interactions, artificial quantum statistics, that is systems of anyons, and tunneling to open space in systems of a few atoms.

\acknowledgments
TS acknowledge financial support from the (Polish) National Science Center with grant No. 2016/22/E/ST2/00555. MAGM acknowledges funding from  the Spanish Ministry MINECO (National Plan15 Grant: FISICATEAMO No. FIS2016-79508-P, SEVERO OCHOA No. SEV-2015-0522, FPI), European Social Fund, Fundaci\'o  Cellex, Generalitat de Catalunya (AGAUR Grant No. 2017 SGR 1341 and CERCA/Program), ERC AdG OSYRIS, EU FETPRO QUIC, and the National Science Centre, Poland-Symfonia Grant No. 2016/20/W/ST4/00314.

\appendix

 \section{Symmetries in three atom systems} 
 \label{app:threeatoms}
 
 \begin{figure}[t]
\includegraphics[width=0.98\columnwidth]{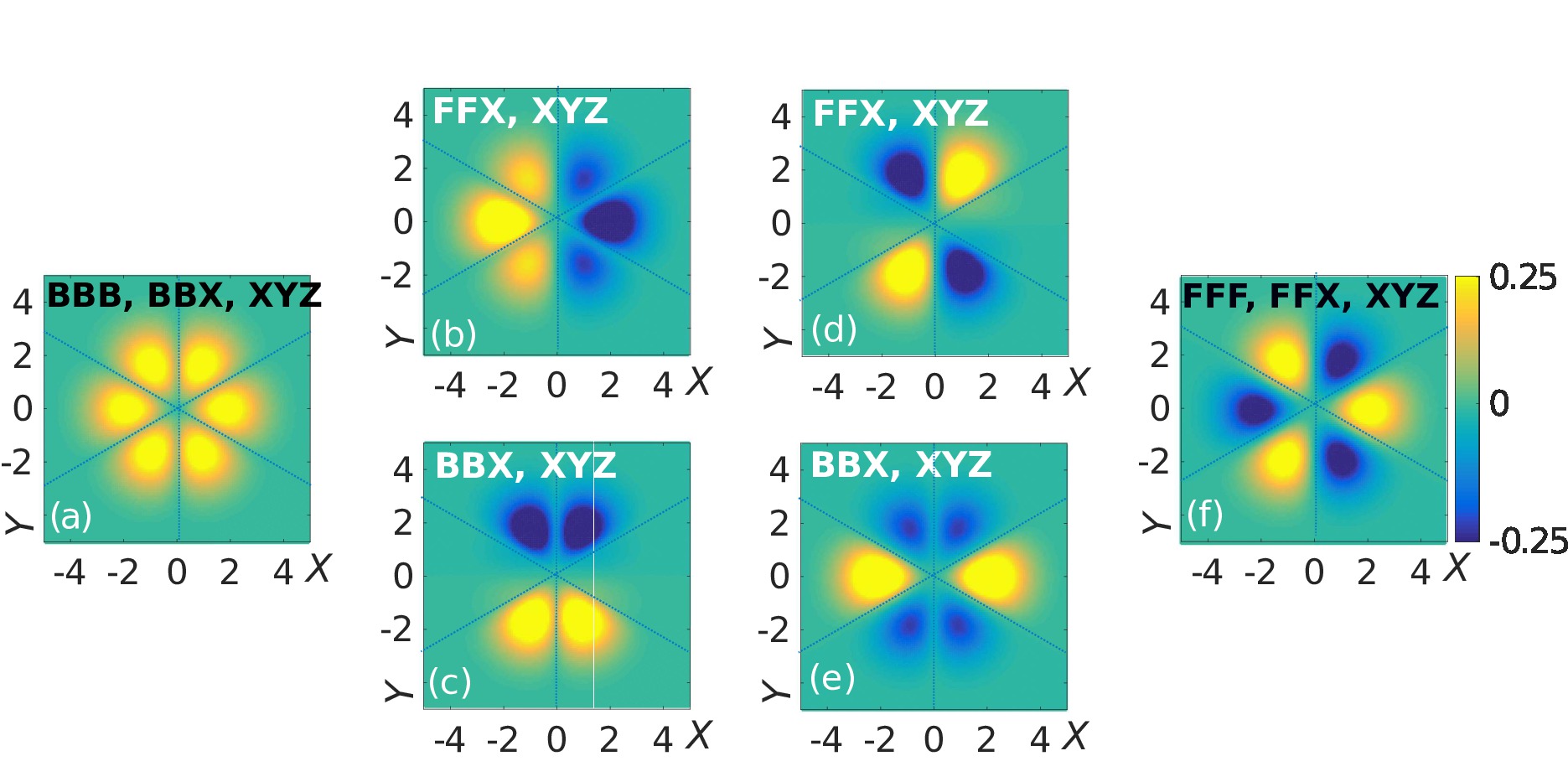}
\caption{Eigenfunctions for the three particle case taking $g_{12}=g_{23}=g_{13}=10$ .  (a) A state that occurs in the BBB, BBX, or XYZ systems,  belonging to $\mathcal{A}_1$, with the lowest energy if $g$ is finite.  (b) Corresponds to    BBX and XYZ and (c) to    FFX and XYZ; these two are degenerated and are the first excited doublet for finite $g$; These states have a reduced $\mathcal{C}_{2\nu}$ symmetry. As discussed in the text,  their combinations states generate the two (vortex like) eigenstates associated with the  $\mathcal{E}_1$  irreducible representation.  (d) corresponds to    BBX and XYZ and (e) to    FFX and XYZ;   these two are degenerated and are the second excited doublet for finite $g$; Similarly as the (B) and (c), these two states generate the (vortex-like) states that belong to  $\mathcal{E}_2$;    (f) corresponds to FFF,  FFX or XYZ, belonging to $\mathcal{B}_1$; it is the lowest energy state for the FFF system and an excited state for FFX and XYZ.  Figure adapted  from~\cite{2019GarciaMarchArxiv}. \label{Fig_distinguishability1_app}}
\end{figure}

 \begin{figure}[t]
\includegraphics[width=0.98\columnwidth]{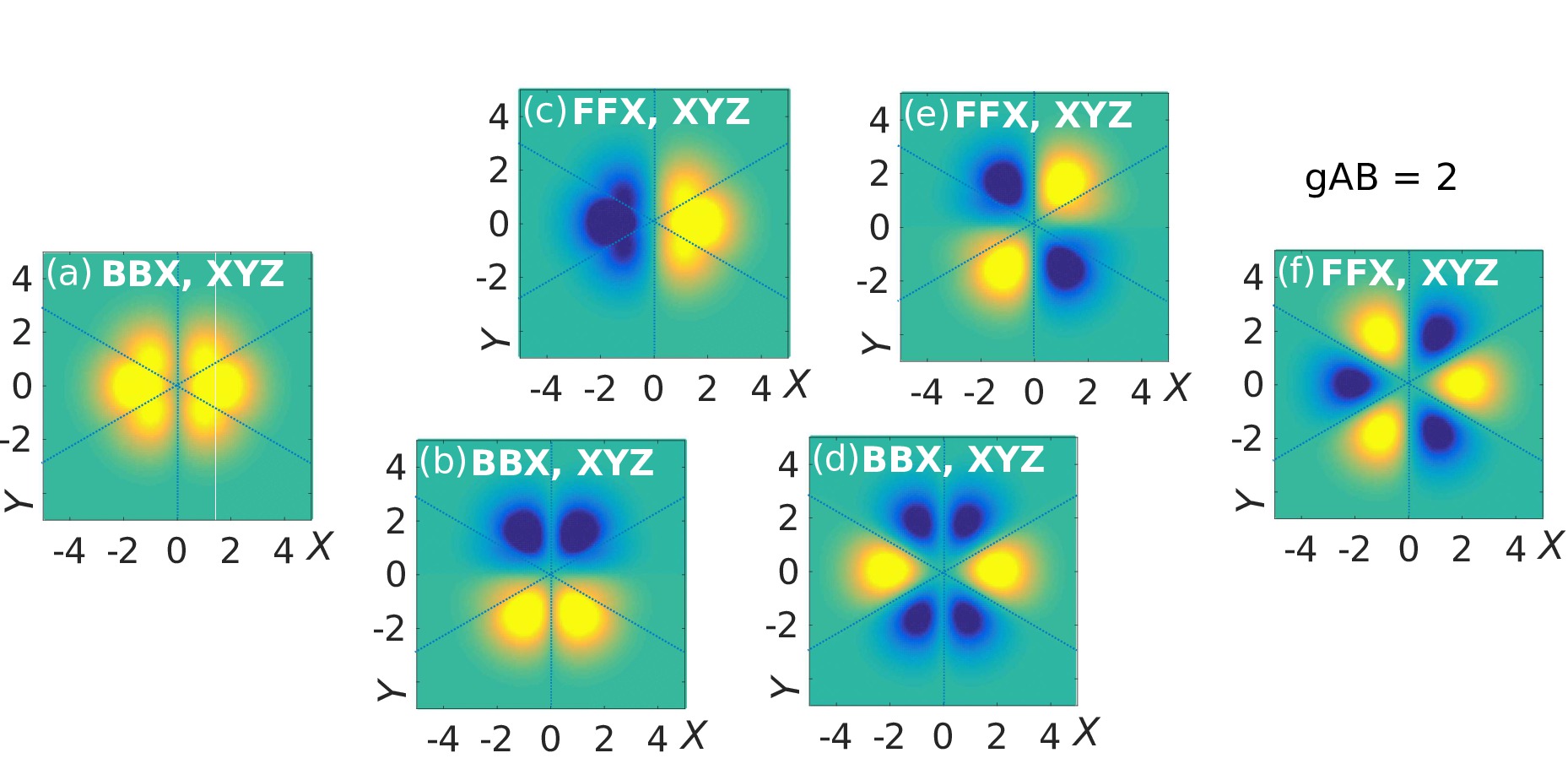}
\caption{Eigenfunctions for the three particle case taking $g_{13}=g_{23}=2$ and $g_{12}=10$.   \label{Fig_distinguishability3_app}}\end{figure}

 To illustrate the symmetry analysis in systems of few atoms, let us first  consider the three-body Hamiltonian with all quantities expressed in natural units of the harmonic oscillator
\begin{equation}
    \label{model31}
    \hat{\cal H}_{0}= \sum_{i=1}^3 \left[ -\frac{1}{2}\frac{\partial^2}{\partial x_i ^2}+\frac{1}{2}x_i^2 \right] + \sum_{i=1}^3\sum_{j=2}^3 V_{ij}\left(x_i - x_j\right),
\end{equation}
where $x_i$ are the dimensionless positions of the three particles and $V_{ij}(x)$ is the two-body interaction potential between particles $i$ and $j$. Performing the Jacobi transformation~\cite{2012HarshmanPRA,2014GarciaMarchPRA}
\begin{subequations}
\begin{align}
    \label{model33b}
    R&=\frac{x_1+ x_2 + x_3}{3}\,, \\
    X&=\frac{x_1-x_2}{\sqrt{2}}\,,\\
    Y&=\frac{x_1+x_2}{\sqrt{6}}-\sqrt{\frac{2}{3}}x_3\,
    \end{align}
\end{subequations}
the full Hamiltonian separates as $\hat{\cal H}=\hat{\cal H}_\mathrm{CM}+\hat{\cal H}_\mathrm{rel}$ with
\begin{subequations}
\begin{align}
\label{model34}
\hat{\cal H}_\mathrm{CM}&=-\frac{1}{2}\frac{\mathrm{d}^2}{\mathrm{d}R^2}+\frac{1}{2} R^2,\\
\hat{\cal H}_\mathrm{rel}&=\frac{1}{2}\left(-\frac{\partial^2}{\partial X^2}-\frac{\partial^2}{\partial Y^2}+X^2+Y^2\right) + V_\mathrm{int}(X,Y),
\end{align}
\end{subequations}
where
\begin{multline}
V_\mathrm{int}(X,Y) = V_{12}\left(\left|\sqrt{2} X\right|\right) \\
+V_{13}\left(\frac{\left|X + \sqrt{3}Y\right|}{\sqrt{2}}\right)
+ V_{23}\left(\frac{\left|X-\sqrt{3}Y\right|}{\sqrt{2}}\right).
\end{multline}
The center-of-mass Hamiltonian $\hat{\cal H}_\mathrm{CM}$ has a form of standard harmonic oscillator Hamiltonian and therefore can be diagonalized straightforwardly. If we focus only on the case of zero-range contact interactions ($V_ij(x)=g_{ij}\delta(x)$) then the interaction part of the relative motion Hamiltonian $V_\mathrm{int}$ has a form
\begin{multline}\label{eq:contact}
V_\mathrm{int}(X,Y) = 
\frac{g_{12}}{\sqrt{2}}\,\delta(X)  \\ \left.+\sqrt{2}g_{13}\,\delta\left(X+\sqrt{3}Y\right) + \sqrt{2}g_{23}\,\delta\left(X- \sqrt{3}Y\right)\right].
\end{multline}
The Hamiltonian with delta-interactions is exactly solvable when zero, one, two or three of the two-body interaction strength parameters $g_{ij}$ are infinite and the rest are zero. For intermediate values, there is no analytic solutions. 
When all $g_{ij}$ are finite and equal the relative interacting Hamiltonian $H_\mathrm{rel}$ \eqref{model34} is symmetric under the finite group of transformations of configuration space $C_{6v}$, {\it i.e.}, the rotation and reflection symmetries of a hexagon~\cite{2012HarshmanPRA}. The twelve elements of $C_{6v}$ are shown in the first column of in Table~\ref{tab:elements}. The second column of this table represents the realization of these elements of  $C_{6v}$ as transformations of relative configuration space are summarized, that is permutation of particles (represented as $\hat\sigma_{ij}$ for two particles or $\hat\sigma_{ijk}$ for the permutation of three particles) and  parity inversion (represented as $\hat\pi$). Here one can observe the main question in this approach: the fact that the system consist of three indistinguishable bosons, fermions, two bosons or two fermions plus an additional atom poses conditions on the way the wave function can transform under permutation of particles combined with parity inversion about the minimum of the harmonic trapping potential. 

\begin{table}
\centering
\begin{tabular}{|c|c|c|c|}
\hline
$g\in C_{6v}$ & $g\in S_3 \times Z_2$ & $\varphi\rightarrow \varphi'$ \\
\hline
$E$ & $\hat{e}$  & $\varphi$  \\
$\sigma_{v}$ & $\hat{\sigma}_{12}$ & $-\varphi + \pi$  \\
$\sigma_{v'} $ & $\hat{\sigma}_{23} $ & $-\varphi + \frac{\pi}{3}$  \\
$\sigma_{v''} $ & $\hat{\sigma}_{31} $ & $-\varphi - \frac{\pi}{3}$  \\
$C_3^{-1}$ & $\hat{\sigma}_{231}$ & $\varphi - \frac{2\pi}{3}$    \\
$C_3$ & $\hat{\sigma}_{312}$ & $\varphi + \frac{2\pi}{3}$    \\
$C_2$ & $\hat{\pi}$  & $\varphi+\pi$    \\
$\sigma_{d}$ & $\hat{\pi}\hat{\sigma}_{12}$ & $-\varphi$   \\
$\sigma_{d'} $ & $\hat{\pi}\hat{\sigma}_{23} $ & $-\varphi - \frac{2\pi}{3}$    \\
$\sigma_{d''} $ & $\hat{\pi}\hat{\sigma}_{31} $ & $-\varphi + \frac{2\pi}{3}$   \\
$C_6$ & $\hat{\pi}\hat{\sigma}_{231}$ & $\varphi + \frac{\pi}{3}$    \\
$C_6^{-1}$ & $\hat{\pi}\hat{\sigma}_{312}$ & $\varphi - \frac{\pi}{3}$    \\
\hline
\end{tabular}

\caption{The first column is the symmetry transformation designated by the cor\-re\-spond\-ing element of the point symmetry group of the regular hexagon permutation group $C_{6v}$.  The second column is the same transformation expressed as the corresponding element of $S_3 \times Z_2$.  We use the notation for $S_3$ permutation group elements such that $e$ is the identity, the 2-cycle $\hat{\sigma}_{ij}$ switches particles  $i$ and $j$, and 3-cycle $\hat{\sigma}_{ijk}$ switches $1$ to $i$, $2$ to $j$ and $3$ to $k$.  The element $\hat{\pi}$ is parity inversion.  The third column is the  transformation in the cylindrical Jacobi coordinate $\tan\varphi=Y/X$.}
\label{tab:elements}
\end{table}

Because $C_{6\nu}$ is a symmetry of the relative interacting Hamiltonian, energy levels are associated with its irreducible representations (irreps), whose properties are summarized in Table~\ref{tab:repbasis}. There are four one-dimensional (or singlet) irreps (denoted as $A_1, A_2, B_1, $ and $B_2$; and two two-dimensional (or doublet) irreps ($E_1$ and $E_2$). This means that there will only be singly-degenerate or doubly-degenerate energy levels. The singlets have $m=0$ or $m=3$. The solutions for  $g=g_{12}=g_{13}=g_{23}$ are well approximated by the ansatz \eqref{eq:m,nrho,ntheta}, which is exact when $g\to\infty$. Thus they can be represented by the quantum numbers $m$, $n_\rho$ and $n_\theta$. We therefore denote them as $|m,n_{\rho},n_\theta\rangle$. Systems of three bosons can only have $m=0$ and of three fermions  $m=3$, as imposed by the permutation symmetry. Of course, a system of three distinguishable atoms can belong to any irrep. In Fig.~\ref{Fig_distinguishability1_app} we represent the lowest lying eigenfunctions calculated when $g=10$. The panel (a) corresponds to a system of three bosons and it has $m=0$, while (f) to three fermions and it has $m=3$.  The two dimensional representations have $m=\pm1$ or $\pm2$. But solutions with a well defined $m$ do not have any symmetry with respect to permutations, and thus can only be realized by three distinguishable atoms. For two bosons or two fermions and a distinguishable particle, the two degenerate solutions have to be combined into a solution of the subgroup $C_{2\nu}$. Thus the combinations $ |1,n_{\rho},n_\theta\rangle+ i |1,n_{\rho},n_\theta\rangle$ and $ |2,n_{\rho},n_\theta\rangle- i |2,n_{\rho},n_\theta\rangle$ give the solutions with the correct permutation property for two bosons plus a third distinguishable particle. These are represented in panels (c) and (e)  of Fig.~\ref{Fig_distinguishability1_app}.  And the combination $ |1,n_{\rho},n_\theta\rangle- i |1,n_{\rho},n_\theta\rangle$ and $ |2,n_{\rho},n_\theta\rangle+ i |2,n_{\rho},n_\theta\rangle$ give the solutions with the correct permutation property for two fermions plus a third distinguishable particle, represented in panels (b) and (d) of Fig.~\ref{Fig_distinguishability1_app}. From here, excitations are obtained for larger values of $n_\rho$ and $n_\theta$, representing radial and angular excitations. 

\begin{table}
\centering
\begin{tabular}{|cc|c|c|c|c|}
\hline
$\mu$ & $\pm$ & $C_{6v}$ & $C_{2v}$ & Possibilities & $m$  \\
\hline
0 & N.A. & $A_1$ & $A_1$ & BBB, BBX, XYZ &   0 \\
1& $+$ & $E_1$ & $B_1$ & FFX, XYZ & N.A. (from $\pm1$ subspace)\\
1& $-$ & $E_1$ & $B_2$ & BBX, XYZ & N.A. (from  $\pm1$ subspace) \\
2& $+$ & $E_2$ & $A_1$ & BBX, XYZ & N.A. (from  $\pm2$ subspace)\\
2& $-$ & $E_2$ & $A_2$ & FFX, XYZ &  N.A. (from $\pm2$ subspace)\\
3& $+$ & $B_1$ & $B_1$ & FFF, FFX, XYZ &   3\\
3& $-$ & $B_2$ & $B_2$ & BBB, BBX, XYZ &   0\\
4& $+$ & $E_2$ & $A_1$ & BBX, XYZ &  N.A. (from  $\pm1$ subspace) \\
4& $-$ & $E_2$ & $A_2$ & FFX, XYZ &  N.A. (from  $\pm1$ subspace)\\
5& $+$ & $E_1$ & $B_1$ & FFX, XYZ &  N.A. (from  $\pm2$ subspace)\\
5& $-$ & $E_1$ & $B_2$ & BBX, XYZ &  N.A. (from  $\pm2$ subspace)\\
6& $+$ & $A_1$ & $A_1$ & BBB, BBX, XYZ &   0\\
6& $-$ & $A_2$ & $A_2$ & FFF, FFX, XYZ &   3\\
\hline
\end{tabular}
\caption{This table identifies the energy eigenbasis vectors for pseudo-angular momentum $\mu=0$ and the symmetric $(+)$ and antisymmetric $(-)$ combinations of $\mu >0$ with their corresponding symmetry representations and superselection rules.  The pattern repeats for vectors with $\mu>6$.  BBB (FFF) means three identical bosons (fermions); BBX (FFX) two identical bosons (fermions) and one other particle; XYZ three distinguishable particles. }
\label{tab:repbasis}
\end{table}

For the system of two bosons or fermions plus an additional particle, the interactions can be different, {\it i.e.}, $g_{12}\neq g_{13}=g_{23}$. In such case, the symmetry is $C_{2\nu}$ (see second column in Table \ref{tab:repbasis}).  In this case there are only four one dimensional representations $A_1, A_2, B_1, $ and $B_2$. The quantum number $m$ is still a good number, but can only take the values 0 and 1. The radial quantum number $ n_{\rho}$ and angular $n_\theta$ play the same role as in the $C_{6\nu}$ case. Since there are no two-dimensional representations, the double degeneracy associated with the doublets is lost.  In Fig.~\ref{Fig_distinguishability3_app} we show an example of the lowest energy eigenfunctions for $g_{13}=g_{23}=2$ and $g_{12}=10$. The solutions for  $g_{13}=g_{23}=0$ and $g_{12}\to\infty$ can be also obtained analytically in a similar fashion as with ansatz \eqref{eq:m,nrho,ntheta}, as discussed in the main text. 

Finally, we refer the reader to ~\cite{2016HarshmanFBSa,2016HarshmanFBSb} for an extensive study on symmetries on few-body systems with more than three atoms and different trapping potentials. Also, we note that a complementary approach to the study of permutation symmetry can be performed with Young tableaux. See e.g. Ref.~\cite{2014YurovskyPRL} for a study in general spinor quantum gases with this approach.  

\bibliography{_BibTotal}
    
\end{document}